\documentclass{article}
\usepackage{epsfig}
\usepackage{graphics}

\renewcommand{\arraystretch}{1.3}
\makeatletter
\newdimen\normalarrayskip              
\newdimen\minarrayskip                 
\normalarrayskip\baselineskip \minarrayskip\jot
\newif\ifold             \oldtrue            \def\new{\oldfalse}
\def\arraymode{\ifold\relax\else\displaystyle\fi} 
\def\eqnumphantom{\phantom{(\theequation)}}     
\def\@arrayskip{\ifold\baselineskip\z@\lineskip\z@
     \else
     \baselineskip\minarrayskip\lineskip2\minarrayskip\fi}
\def\@arrayclassz{\ifcase \@lastchclass \@acolampacol \or
\@ampacol \or \or \or \@addamp \or
   \@acolampacol \or \@firstampfalse \@acol \fi
\edef\@preamble{\@preamble
  \ifcase \@chnum
     \hfil$\relax\arraymode\@sharp$\hfil
     \or $\relax\arraymode\@sharp$\hfil
     \or \hfil$\relax\arraymode\@sharp$\fi}}
\def\@array[#1]#2{\setbox\@arstrutbox=\hbox{\vrule
     height\arraystretch \ht\strutbox
     depth\arraystretch \dp\strutbox
     width\z@}\@mkpream{#2}\edef\@preamble{\halign
\noexpand\@halignto
\bgroup \tabskip\z@ \@arstrut \@preamble \tabskip\z@ \cr}%
\let\@startpbox\@@startpbox \let\@endpbox\@@endpbox
  \if #1t\vtop \else \if#1b\vbox \else \vcenter \fi\fi
  \bgroup \let\par\relax
  \let\@sharp##\let\protect\relax
  \@arrayskip\@preamble}
\def\eqnarray{\stepcounter{equation}%
              \let\@currentlabel=\theequation
              \global\@eqnswtrue
              \global\@eqcnt\z@
              \tabskip\@centering
              \let\\=\@eqncr
 \halign to \displaywidth\bgroup
    \eqnumphantom\@eqnsel\hskip\@centering
    $\displaystyle \tabskip\z@ {##}$%
    \global\@eqcnt\@ne \hskip 2\arraycolsep
         $\displaystyle\arraymode{##}$\hfil
    \global\@eqcnt\tw@ \hskip 2\arraycolsep
         $\displaystyle\tabskip\z@{##}$\hfil
         \tabskip\@centering
    &{##}\tabskip\z@\cr}
\begingroup\ifx\undefined\newsymbol \else\def\input#1 {\endgroup}\fi

\catcode`\@=11
\def\marginnote#1{}

\newcount\hour
\newcount\minute
\newtoks\amorpm
\hour=\time\divide\hour by60 \minute=\time{\multiply\hour by60
\global\advance\minute by-\hour}
\edef\standardtime{{\ifnum\hour<12 \global\amorpm={am}%
        \else\global\amorpm={pm}\advance\hour by-12 \fi
        \ifnum\hour=0 \hour=12 \fi
        \number\hour:\ifnum\minute<10 0\fi\number\minute\the\amorpm}}
\edef\militarytime{\number\hour:\ifnum\minute<10 0\fi\number\minute}

\def\draftlabel#1{{\@bsphack\if@filesw {\let\thepage\relax
      \xdef\@gtempa{\write\@auxout{\string
          \newlabel{#1}{{\@currentlabel}{\thepage}}}}}\@gtempa \if@nobreak
    \ifvmode\nobreak\fi\fi\fi\@esphack} \gdef\@eqnlabel{#1}}
    \def\@eqnlabel{}
\def\@vacuum{}
\def\draftmarginnote#1{\marginpar{\raggedright\scriptsize\tt#1}}

\def\draft{
  \oddsidemargin -.5truein
  \def\@oddfoot{\footnotesize \sl preliminary draft \hfil
    \rm\thepage\hfil\sl\today\quad\militarytime}
  \let\@evenfoot\@oddfoot \overfullrule 3pt
    \let\label=\draftlabel
    \let\marginnote=\draftmarginnote
  \def\@eqnnum{(\theequation)\rlap{\kern\marginparsep\tt\@eqnlabel}%
    \global\let\@eqnlabel\@vacuum}

  }

\def\be{\begin{eqnarray}}
\def\ee{\end{eqnarray}}
\def\nn{\nonumber}

\def\beq{\begin{equation}}
\def\eeq{\end{equation}}
\def\ba{\beq\new\begin{array}{c}}
\def\ea{\end{array}\eeq}
\def\be{\ba}
\def\ee{\ea}

\newfont{\alef}{msbm10 at 12pt}
\newfont {\goth}{eufm10 at 11pt}
\def\mathbb#1{\hbox{{\alef #1}}}

\let\@@savethanks\thanks
\def\thanks#1{\gdef\thefootnote{\alph{footnote}}\@@savethanks{#1}}

\def\theequation{\arabic{section}.\arabic{equation}}

\title{{\bf On $n$-point Amplitudes in $N=4$ SYM
} \vspace{.5cm}}
\author{{\bf A. Mironov}\footnote{E-mail: \ mironov@itep.ru; mironov@lpi.ru}
\date{ } \\
{\small {\it Lebedev Physics Institute}
and {\it ITEP, Moscow, Russia}}\\ \\
{\bf A. Morozov}\thanks{E-mail: \ morozov@itep.ru}
\date{ } \\ {\small {\it ITEP, Moscow, Russia}}
\\ \\
{\bf T.N. Tomaras}\thanks{E-mail: \ tomaras@physics.uoc.gr}
\date{ } \\ {\small {\it Department of Physics and Institute of Plasma Pysics, University of Crete, Heraklion; Greece}}
}

\begin{document}

\maketitle

\vspace{-10.5cm}

\begin{center}
\hfill FIAN/TD-15/07\\
\hfill ITEP/TH-30/07\\
\end{center}

\vspace{9.0cm}

\begin{abstract}
\noindent The computation of n-point planar amplitudes in N=4 SYM at
strong coupling is known to be reduced to the search for solutions
of the integrable 2d SO(4,2) $\sigma$-model with growing asymptotics
on the world-sheet and to the study of their Whitham deformations
induced by an $\epsilon$-regularization, which breaks both
integrability and SO(4,2) symmetry. A multi-parameter (moduli) family
of such solutions is constructed for $n=4$. They all correspond to
the same $s$ and $t$ and some are related by SO(4,2)
transformations. Nevertheless, they lead to different regularized
areas, whose minimum is the Alday-Maldacena solution. A brief review
of results on n-point amplitudes is also provided, with special
emphasis on the underlying equivalence of the above regularized
minimal area in AdS and a double contour integral along the same
boundary, two purely geometric quantities.
\end{abstract}

\bigskip

\def\thefootnote{\arabic{footnote}}

\newpage

\tableofcontents

\newpage

\section{Introduction and results}

A new significant step was taken recently \cite{AM} towards the
computation of planar $N=4$ Super-Yang-Mills (SYM) n-point
amplitudes beyond ordinary perturbation theory, using the
gauge/string correspondence.
The celebrated exponentiation BDS-hypothesis of \cite{BDS} was
verified at strong coupling for the four point amplitude. To achieve
this, the authors made several ingenious choices and educated
guesses in order to deal with the difficulties in the string side of
the AdS/CFT duality \cite{AdS/CFT,AdS/CFToth}, studied also in a
number of recent publications \cite{AMbeg}-\cite{AMend}.

In particular, in \cite{AM}

(i) the $\sigma$-model action instead of
the Nambu-Goto one was used, what allowed to perform a $T$-duality
{\it a la} \cite{TD},

(ii) a minimal surface was constructed for
$n=4$ based on previous considerations \cite{Kru},

(iii) a rather unusual dimensional regularization was employed, instead of, for instance, the one 
described in \cite{Skenderis},

(iv) a skilful handling of the resulting integrals was required, and finally

(v)  the
KLOV interpolation \cite{KLOV} between weak and strong coupling
regimes was used.

A better understanding of these and related issues
seems to be an unavoidable step, before one can generalize this
method and derive the dilogarithmic BDS formula \cite{BDS} for
$n>4$. Before we describe the little progress made in the present
paper, let us briefly summarize the basic ingredients of the AdS/CFT
correspondence.

\subsection*{Gauge theory side}
 Start with the $N=4$ SYM.

{\bf(a)} According to the conjecture of \cite{BDS}, the {\it planar} (at
least, maximally helicity violating) $n$-point amplitude ${\cal
A}_n$ in {\it $N=4$ SYM} gauge theory has the form \be {\cal A}_n =
{\cal A}_{n,tree}\times {\cal M}_n \ee where ${\cal M}_n$ does not
depend on any color and helicity factors. In the supersymmetric case,
the leading logarithm approximation (LLA) \cite{DLA} is not required
for the amplitudes to exponentiate: both the infrared divergent and
the finite parts of amplitudes are surprisingly simple exponentials.
Explicitly, its infrared divergent part is the universal exponential
\be \label{2} {\cal M}_{IR}\sim \exp\left(-{1\over
4}\sum_{l=1}^\infty\lambda^l\left[\gamma_{(l)}+2l\epsilon{g}^{(l)}\right]
I_n^{(1)}(l\epsilon)\right) \ee with the 1-loop scalar box integral
\be
I_n^{(1)}(\epsilon)={1\over\epsilon^2}\sum_{i=1}^n\left({\mu^2\over
s_{i,i+1}}\right)^{\epsilon} \ee $s_{i,i+1}$ is the square of the
sum of the $i$ and $i+1$ external momenta and the function
$\gamma(\lambda)\equiv\sum_l\gamma_{(l)}\lambda^l$ of the 't Hooft
coupling $\lambda\equiv {N_c\alpha_S\over 2\pi}\left(4\pi
e^{\Gamma'(1)}\right)^{\epsilon}$ is called soft or cusp (Wilson
line \cite{Pol1,Kor}) anomalous dimension \cite{KL}. $\mu$ and
$\epsilon$ are the standard dimensional regularization parameters.

{\bf (b)} The finite part of ${\cal M}_n$ is also expressed through its 1-loop
counterpart $F^{(1)}_n$ as: \be F_n =
\exp\Big(\frac{1}{4}\gamma(\lambda) F^{(1)}_n+C(\lambda)\Big)
\label{BDSf} \ee where $C(\lambda)$ depends neither on $n$, nor on
kinematics. If this is true, then the main non-trivial quantity
entering both (\ref{2}) and (\ref{BDSf}) is $\gamma(\lambda)$, the
anomalous dimension of twist-two operators, an eigenvalue of a not
yet fully known Bethe ansatz \cite{BA}, but with known strong 't
Hooft coupling asymptotics $\gamma(\lambda) \sim \sqrt{\lambda} +
\hbox{const} + O(1/\sqrt{\lambda})$, \cite{strcas,Kru,Mak,Ts}.

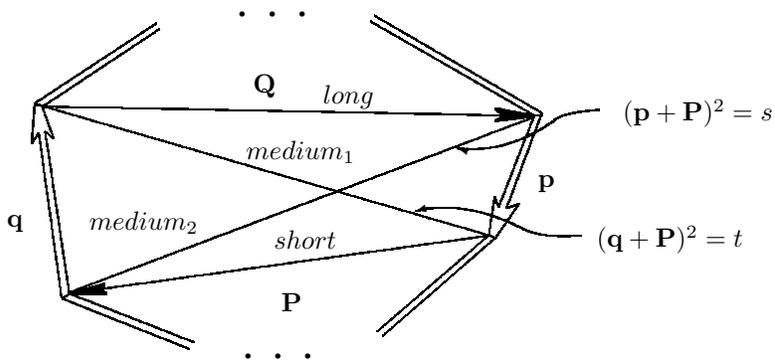
\begin{figure}
\hspace{3cm}
\unitlength 1mm 
\linethickness{0.4pt}
\ifx\plotpoint\undefined\newsavebox{\plotpoint}\fi 
\begin{picture}(96.208,51.135)(0,0)
\multiput(52.646,49.731)(.0586253369,-.0336927224){371}{\line(1,0){.0586253369}}
\put(43.622,50.74){\circle*{.791}}
\put(36.434,5.252){\circle*{.791}}
\put(35.622,50.74){\circle*{.791}}
\put(40.411,5.252){\circle*{.791}}
\put(39.622,50.74){\circle*{.791}}
\put(44.389,5.252){\circle*{.791}}
\multiput(9.217,38.501)(.0508552632,.0337401316){304}{\line(1,0){.0508552632}}
\multiput(12.52,13.548)(.086108247,-.033716495){194}{\line(1,0){.086108247}}
\multiput(68.292,21.15)(-.039047619,-.0336719577){378}{\line(-1,0){.039047619}}
\multiput(74.391,36.971)(-.08928179191,-.03371965318){692}{\line(-1,0){.08928179191}}
\multiput(8.984,38.474)(.116409002,-.0337299413){511}{\line(1,0){.116409002}}
\qbezier(58.747,24.508)(65.331,26.099)(73.154,22.033)
\qbezier(73.154,22.033)(75.938,20.84)(80.667,21.238)
\qbezier(64.227,32.817)(71.121,32.11)(77.308,36.706)
\put(64.315,32.817){\vector(-4,1){.07}}\multiput(65.641,32.64)(-.221,.0295){6}{\line(-1,0){.221}}
\put(58.747,24.508){\vector(-4,-1){.07}}\multiput(60.603,24.95)(-.1325714,-.0315714){14}{\line(-1,0){.1325714}}
\qbezier(77.307,36.705)(78.677,37.677)(83.052,37.943)
\multiput(24.186,49.433)(-.0513028391,-.0337381703){317}{\line(-1,0){.0513028391}}
\multiput(11.635,12.84)(.084855,-.033585){200}{\line(1,0){.084855}}
\multiput(53.177,50.494)(.0579326425,-.0336606218){386}{\line(1,0){.0579326425}}
\multiput(9.072,38.384)(1.76535135,-.03343243){37}{\line(1,0){1.76535135}}
\multiput(12.607,13.548)(.244618421,.033723684){228}{\line(1,0){.244618421}}
\multiput(69.264,20.884)(-.0386994819,-.0336606218){386}{\line(-1,0){.0386994819}}
\multiput(8.011,38.65)(.1915,-.0295){6}{\line(1,0){.1915}}
\multiput(12.635,13.453)(-.0536667,-.0330556){18}{\line(-1,0){.0536667}}
\multiput(11.595,12.933)(-.033494505,.241758242){91}{\line(0,1){.241758242}}
\multiput(8.547,34.933)(-.033037,-.0522963){27}{\line(0,-1){.0522963}}
\multiput(7.655,33.521)(.031,.4335){12}{\line(0,1){.4335}}
\multiput(8.027,38.723)(.138,-.031857){7}{\line(1,0){.138}}
\multiput(8.993,38.5)(.03321277,-.0917234){47}{\line(0,-1){.0917234}}
\multiput(10.554,34.189)(-.0412963,.033037){27}{\line(-1,0){.0412963}}
\multiput(9.439,35.081)(.033642105,-.226105263){95}{\line(0,-1){.226105263}}
\multiput(75.44,37.46)(-.033690647,-.090366906){139}{\line(0,-1){.090366906}}
\multiput(70.757,24.899)(.1021875,.0325){16}{\line(1,0){.1021875}}
\multiput(72.392,25.419)(-.03355914,-.048752688){93}{\line(0,-1){.048752688}}
\multiput(69.271,20.885)(-.0584286,.0318571){14}{\line(-1,0){.0584286}}
\multiput(68.453,21.331)(.0323043,.2197391){23}{\line(0,1){.2197391}}
\multiput(69.196,26.385)(.0325625,-.08825){16}{\line(0,-1){.08825}}
\multiput(69.717,24.973)(.033730496,.086446809){141}{\line(0,1){.086446809}}
\multiput(74.473,37.162)(.223,.0298){5}{\line(1,0){.223}}
\thicklines
\multiput(69.207,38.095)(.1828702,-.0335262){29}{\line(1,0){.1828702}}
\multiput(74.51,37.123)(-.2791176,-.0325637){19}{\line(-1,0){.2791176}}
\multiput(69.207,36.504)(.0744314,.0325637){19}{\line(1,0){.0744314}}
\multiput(70.621,37.123)(-.0458304,.032736){27}{\line(-1,0){.0458304}}
\multiput(69.384,38.007)(.1362637,-.0331452){24}{\line(1,0){.1362637}}
\multiput(72.654,37.211)(-.1830879,.0315669){14}{\line(-1,0){.1830879}}
\multiput(70.091,37.653)(.0757605,-.0315669){14}{\line(1,0){.0757605}}
\multiput(71.152,37.211)(-.0883872,-.0315669){14}{\line(-1,0){.0883872}}
\multiput(69.914,36.769)(.202028,.0315669){14}{\line(1,0){.202028}}
\multiput(17.412,14.761)(-.12660875,-.03344382){37}{\line(-1,0){.12660875}}
\put(12.728,13.523){\line(1,0){4.95}}
\multiput(17.677,13.523)(-.082863,.0331452){16}{\line(-1,0){.082863}}
\multiput(16.352,14.054)(.0511716,.0325637){19}{\line(1,0){.0511716}}
\multiput(17.324,14.672)(-.1141669,-.0331452){24}{\line(-1,0){.1141669}}
\multiput(14.584,13.877)(.397743,-.029462){6}{\line(1,0){.397743}}
\multiput(16.97,13.7)(-.132581,.029462){6}{\line(-1,0){.132581}}
\put(16.175,13.877){\line(-1,0){2.0329}}
\put(38.606,41.151){\makebox(0,0)[cc]{$\bf Q$}}
\put(42.202,12.367){\makebox(0,0)[cc]{$\bf P$}}
\put(5.432,22.797){\makebox(0,0)[cc]{$\bf q$}}
\put(76.05,28.161){\makebox(0,0)[cc]{$\bf p$}}
\put(96.208,37.501){\makebox(0,0)[cc]{$({\bf p}+{\bf P})^2=s$}}
\put(92.495,20.708){\makebox(0,0)[cc]{$({\bf q}+{\bf P})^2=t$}}
\put(49.749,39.75){\makebox(0,0)[cc]{$long$}}
\put(43.999,20.75){\makebox(0,0)[cc]{$short$}}
\put(43.374,32.375){\makebox(0,0)[]{$medium_1$}}
\put(22.5,23){\makebox(0,0)[cc]{$medium_2$}}
\end{picture}
\caption{{\footnotesize The $4$-cluster formed by two
non-intersecting edges of the polygon $\Pi$, it also contains four
vertices (hence the name) and four diagonals. ("Diagonals" are those
of $\Pi$, two of the four are actually {\it sides} of the
quadrilateral.) From these four diagonals one is "long", another
"short" and two "medium" -- associated respectively with ${\bf
P}^2$, ${\bf Q}^2$ (or vice versa), $s$ and $t$. "Long" and "short"
refer to the smallest number of edges of $\Pi$ in between the ends
of diagonal. The contribution of the $4$ cluster to the dilogarithmic part
of the BDS formula (\ref{Fdil}) is $\frac{1}{2}Li_2\left(1-
\frac{{\bf P}^2{\bf Q}^2}{st}\right)$ where $\log \frac{{\bf
P}^2{\bf Q}^2}{st} = \tau_{l} + \tau_{s} - \tau_{m1} - \tau_{m2}$.
Dilogarithmic contribution does not distinguish long
and short diagonals. The logarithmic one does, see
Figure \ref{degclust}.} } \label{clust}
\end{figure}

{\bf (c)} The 1-loop amplitude $F_n^{(1)}$ may be expressed as a sum
over $4$-clusters in an auxiliary polygon $\Pi$, Figures
\ref{polyg},\ref{clust}, formed by the external momenta ${\bf p}_a$
of the process at hand, which plays the central role in the
description of both sides of the AdS/CFT duality. In Section
\ref{picto} one may find a simple pictorial representation of the
BDS formula for $F_n^{(1)}$.

\begin{figure}
\hspace{2cm}
\unitlength 1mm 
\linethickness{0.4pt}
\ifx\plotpoint\undefined\newsavebox{\plotpoint}\fi 
\begin{picture}(160.874,78.25)(0,0)
\put(67.392,76.808){\circle{.867}}
\put(20.121,55.848){\circle{.867}}
\put(20.121,51.092){\circle{.867}}
\put(20.121,46.483){\circle{.867}} \put(70.96,22.7){\circle{.867}}
\put(74.081,23.145){\circle{.867}} \put(77.5,23.443){\circle{.867}}
\put(72.787,76.072){\circle{.867}}
\put(77.557,75.231){\circle{.867}}
\multiput(87.157,29.974)(-.03363,-.03994){100}{\line(0,-1){.03994}}
\multiput(83.794,25.98)(.0315,-.0946){10}{\line(0,-1){.0946}}
\multiput(84.109,25.034)(.10957447,.03355319){47}{\line(1,0){.10957447}}
\multiput(89.259,26.611)(-.1616923,.0323077){13}{\line(-1,0){.1616923}}
\multiput(87.157,27.031)(.0500823529,.0336970588){340}{\line(1,0){.0500823529}}
\multiput(104.185,38.488)(-.1131538,.0323846){13}{\line(-1,0){.1131538}}
\multiput(102.714,38.909)(-.03355319,.09391489){47}{\line(0,1){.09391489}}
\multiput(101.137,43.323)(.0663684,-.0331579){19}{\line(1,0){.0663684}}
\multiput(102.398,42.693)(-.03343939,.30577273){66}{\line(0,1){.30577273}}
\multiput(100.191,62.874)(-.08408333,.03328333){60}{\line(-1,0){.08408333}}
\multiput(95.146,64.871)(.1997,.0316){10}{\line(1,0){.1997}}
\multiput(97.143,65.187)(-.047186667,.033635556){225}{\line(-1,0){.047186667}}
\multiput(86.526,72.755)(.0334545,.0382273){22}{\line(0,1){.0382273}}
\multiput(59.026,76.323)(.0334545,.0382273){22}{\line(0,1){.0382273}}
\multiput(35.777,76.005)(.0334545,.0382273){22}{\line(0,1){.0382273}}
\multiput(22.628,35.159)(.0327407,.0373704){27}{\line(0,1){.0373704}}
\multiput(39.128,22.559)(.0327407,.0373704){27}{\line(0,1){.0373704}}
\multiput(87.262,73.596)(.045927928,-.033617117){222}{\line(1,0){.045927928}}
\put(97.458,66.133){\line(0,1){1.366}}
\multiput(97.353,67.499)(.033525862,-.035336207){116}{\line(0,-1){.035336207}}
\multiput(101.242,63.4)(.03336508,-.32868254){63}{\line(0,-1){.32868254}}
\multiput(103.344,42.693)(.03364,.06304){25}{\line(0,1){.06304}}
\put(104.185,44.269){\line(0,-1){5.781}}
\multiput(100.086,63.084)(.0889231,.0323846){13}{\line(1,0){.0889231}}
\put(86.83,73.23){\circle*{1.863}}
\put(20.532,63.419){\circle*{1.863}}
\put(64.681,22.54){\circle*{1.863}}
\put(59.33,76.798){\circle*{1.863}}
\put(36.081,76.48){\circle*{1.863}}
\put(22.993,35.729){\circle*{2.236}}
\put(39.493,23.129){\circle*{2.236}}
\put(100.729,63.196){\circle*{1.863}}
\put(103.479,38.743){\circle*{1.863}}
\put(83.708,25.067){\circle*{1.863}}
\multiput(20.75,62.875)(.040282392,.0336378738){301}{\line(1,0){.040282392}}
\multiput(32.875,73)(-.03125,-.34375){4}{\line(0,-1){.34375}}
\multiput(32.75,71.625)(.033602151,.044354839){93}{\line(0,1){.044354839}}
\multiput(53.875,75.875)(-.0326087,-.0326087){23}{\line(0,-1){.0326087}}
\multiput(53.125,75.125)(.16158537,.03353659){41}{\line(1,0){.16158537}}
\multiput(59.75,76.5)(-.125,.03365385){52}{\line(-1,0){.125}}
\multiput(53.25,78.25)(.03125,-.072917){12}{\line(0,-1){.072917}}
\multiput(53.625,77.375)(-2.203125,-.03125){8}{\line(-1,0){2.203125}}
\multiput(36,77.125)(-.059139785,-.033602151){93}{\line(-1,0){.059139785}}
\multiput(30.5,74)(.375,-.03125){4}{\line(1,0){.375}}
\multiput(32,73.875)(-.0409556314,-.0337030717){293}{\line(-1,0){.0409556314}}
\multiput(20,64)(.0326087,-.048913){23}{\line(0,-1){.048913}}
\multiput(44.796,21.803)(.03333333,-.0375){36}{\line(0,-1){.0375}}
\multiput(45.996,20.453)(-.14333333,.03333333){45}{\line(-1,0){.14333333}}
\multiput(39.546,21.953)(-.0443877551,.0336734694){294}{\line(-1,0){.0443877551}}
\put(26.496,31.853){\line(0,-1){1.95}}
\put(26.496,29.903){\line(-3,4){4.05}}
\put(160.874,29.25){\line(0,1){0}}
\multiput(26.946,32.903)(.0453736655,-.0336298932){281}{\line(1,0){.0453736655}}
\put(39.696,23.453){\line(5,1){6.75}}
\multiput(46.446,24.803)(-.03367347,-.03367347){49}{\line(0,-1){.03367347}}
\put(44.796,23.153){\line(1,0){19.35}}
\put(64.146,23.153){\line(0,-1){1.35}}
\multiput(26.796,32.903)(.15,.0321429){14}{\line(1,0){.15}}
\multiput(28.896,33.353)(-.06851852,.03333333){81}{\line(-1,0){.06851852}}
\multiput(36.125,75.875)(2.21875,.03125){8}{\line(1,0){2.21875}}
\multiput(87.125,29.875)(-.03125,-.145833){12}{\line(0,-1){.145833}}
\multiput(86.75,28.125)(.0503968254,.0337301587){315}{\line(1,0){.0503968254}}
\multiput(44.796,21.953)(3.87,.03){5}{\line(1,0){3.87}}
\put(95.433,73.285){\makebox(0,0)[cc]{${\bf p}_n$}}
\put(106.136,53.365){\makebox(0,0)[cc]{${\bf p}_1$}}
\put(97.217,30.027){\makebox(0,0)[cc]{${\bf p}_2$}}
\end{picture}
\caption{\footnotesize A formal polygon $\Pi$ made from
external momentum $p_a$ (one can call it a Wilson loop in the dual
momentum space). It plays a surprisingly important role in the
description of both perturbative and strong coupling sides of
gauge/string duality. } \label{polyg0}
\end{figure}
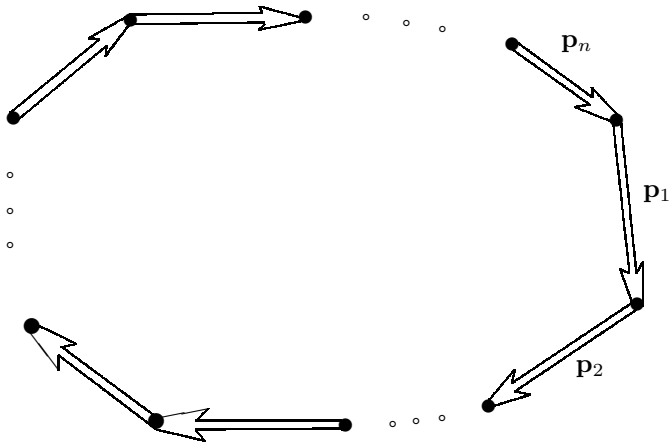

\begin{figure}
\hspace{2cm}
\unitlength 1mm 
\linethickness{0.4pt}
\ifx\plotpoint\undefined\newsavebox{\plotpoint}\fi 
\begin{picture}(110.125,50.375)(0,0)
\put(20,21.875){\line(1,0){20}} \put(82,21.875){\line(1,0){20}}
\put(40,21.875){\line(0,1){20}} \put(102,21.875){\line(0,1){20}}
\put(40,41.875){\line(-1,0){20}} \put(102,41.875){\line(-1,0){20}}
\put(20,41.875){\line(0,-1){20}} \put(82,41.875){\line(0,-1){20}}
\multiput(20.125,41.875)(-.0336734694,.0336734694){245}{\line(0,1){.0336734694}}
\multiput(82.125,41.875)(-.0336734694,.0336734694){245}{\line(0,1){.0336734694}}
\multiput(20,21.875)(-.033613445,-.034138655){238}{\line(0,-1){.034138655}}
\multiput(82,21.875)(-.033613445,-.034138655){238}{\line(0,-1){.034138655}}
\multiput(40.125,21.75)(.033695652,-.03423913){230}{\line(0,-1){.03423913}}
\multiput(102.125,21.75)(.033695652,-.03423913){230}{\line(0,-1){.03423913}}
\multiput(40,41.75)(.034138655,.033613445){238}{\line(1,0){.034138655}}
\multiput(102,41.75)(.034138655,.033613445){238}{\line(1,0){.034138655}}
\thicklines
\multiput(20,41.625)(-.0336345382,.0336345382){249}{\line(-1,0){.0336345382}}
\multiput(20.25,42)(-.0336734694,.0341836735){245}{\line(0,1){.0341836735}}
\multiput(20,41.75)(-.033713693,.03526971){241}{\line(0,1){.03526971}}
\multiput(39.837,21.863)(.033675325,-.034580087){231}{\line(0,-1){.034580087}}
\multiput(40.048,22.074)(.033653509,-.034118421){228}{\line(0,-1){.034118421}}
\multiput(47.616,14.19)(-.033635556,.034568889){225}{\line(0,1){.034568889}}
\put(73.999,49.823){\line(1,-1){7.989}}
\put(74.209,50.138){\line(1,-1){8.094}}
\put(82.093,42.045){\line(-1,1){8.094}}
\multiput(82.303,21.863)(-.033708861,-.034147679){237}{\line(0,-1){.034147679}}
\multiput(81.987,22.074)(-.033704641,-.034151899){237}{\line(0,-1){.034151899}}
\multiput(82.198,21.863)(-.034141026,-.033688034){234}{\line(-1,0){.034141026}}
\put(16.397,50.349){\makebox(0,0)[cc]{$\bf Q$}}
\put(41.414,50.138){\makebox(0,0)[cc]{$\bf p$}}
\put(16.503,14.085){\makebox(0,0)[cc]{$\bf q$}}
\put(40.363,16.082){\makebox(0,0)[cc]{$\bf P$}}
\put(79.78,49.403){\makebox(0,0)[cc]{$\bf P$}}
\put(79.78,15.662){\makebox(0,0)[cc]{$\bf Q$}}
\put(103.746,49.192){\makebox(0,0)[cc]{$\bf p$}}
\put(101.748,16.503){\makebox(0,0)[cc]{$\bf q$}}
\put(30.062,8.514){\makebox(0,0)[cc]{A}}
\put(92.393,9.355){\makebox(0,0)[cc]{B}}
\end{picture}
\caption{\footnotesize Ordinary box Feynman diagrams for a
massless scalar field in $4+2\epsilon$ dimensions with four external
momenta. Two external momenta are on-shell, two are off-shell: ${\bf
p}^2={\bf q}^2=0$, ${\bf P}^2\neq 0$, ${\bf Q}^2\neq 0$. In "easy" (or 2-mass easy)
box (A) off-shell momenta are at opposite corners, in "heavy" box
(B) they are adjacent. Only "easy" boxes contribute to $F_n^{(1)}$
in (\ref{2me}). } \label{boxdi}
\end{figure}
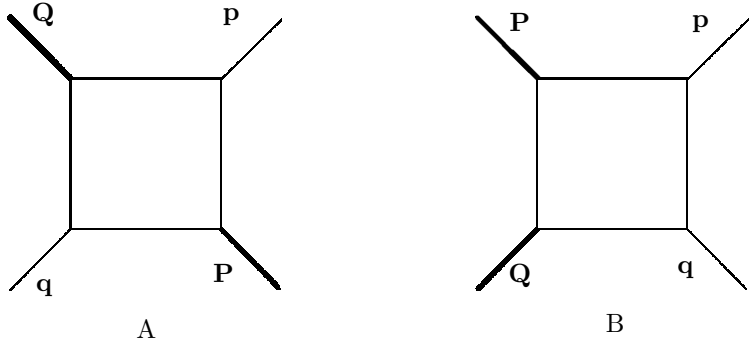

Furthermore, in $N=4$ SYM $F^{(1)}_n$ is given to leading order
\cite{BDDK,BST2} as a sum of contributions $F^{2me}$ from "2-mass
easy" (2me) square diagrams \cite{BDK}, i.e. square diagrams with
two external legs at opposite corners on-shell and the other two
off-shell, Figure \ref{boxdi}. Formally, \be F^{(1)}_n = \sum_{a<b}
F^{2me}({\bf p}_a,{\bf P}_{ab},{\bf p}_b,{\bf P}_{ba}) \label{2me}
\ee Here ${\bf p}_a$ are the $n$ external momenta\footnote{Throughout
the paper, we use the bold font for $4d$ vectors, while arrows are
used for $2d$ vectors.} and ${\bf P}_{ab} = \sum_{c=a+1}^{b-1} {\bf
p}_c$, where we assume that ${\bf p}_{a+n}\equiv {\bf p}_a$. The two
lower case arguments ${\bf p}_a$ of $F^{2me}$ are on-shell (${\bf
p}_a^2=0$), while the other two are in general off-shell.

Each $F^{2me}({\bf p},{\bf P},{\bf q},{\bf Q})$ can be expressed
through dilogarithmic functions of four invariant scalars $s=({\bf
p}+{\bf P})^2$, $t=({\bf p}+{\bf Q})^2$, ${\bf P}^2$, ${\bf Q}^2$. A
particularly interesting dilogarithimic representation is \cite{DN}:
\be F^{2me}({\bf p},{\bf P},{\bf q},{\bf Q}) =
\frac{1}{\epsilon^2(-s)^\epsilon} +
\frac{1}{\epsilon^2(-t)^\epsilon}  + \frac{1}{\epsilon^2(-{\bf
P}^2)^\epsilon}+ \frac{1}{\epsilon^2(-{\bf Q}^2)^\epsilon} + \ee \be
+Li_2(1-as) + Li_2(1-at) -  Li_2(1-a{\bf P}^2) - Li_2(1-a{\bf Q}^2)
\ee with \be a = \frac{s+t-{\bf P}^2-{\bf Q}^2}{st-{\bf P}^2{\bf
Q}^2}. \ee

{\bf (d)} As shown recently in \cite{BHT}, the above sum of dilogarithms
is actually a double contour integral (the leading contribution to
the Wilson-loop average \cite{Pol1}), only in $T$-dual coordinates
in the target space along a polygon, formed by the external momenta
${\bf p}_a$ of the scattering process: \be F^{(1)}_n =
\oint_\Pi\oint_\Pi \frac{dy^\mu dy'_\mu} {(y-y')^{2+\epsilon}}
\label{loopin} \ee Equation (\ref{loopin}) constitutes a purely
geometric formulation of the BDS formula for $F^{(1)}_n$. Together
with a Bethe-ansatz description \cite{BA} of the function
$\gamma(\lambda)$ it should provide a particularly satisfactory
solution of planar $N=4$ SYM.

\subsection*{String Theory side}

The situation on the String Theory side of the AdS/CFT
correspondence looks at the moment less optimistic. Let us briefly
review the current understanding.

{\bf (e)} According to the AdS/CFT duality \cite{AdS/CFT,AdS/CFToth}
the geometric integral (\ref{loopin}) should in fact coincide with
another geometric quantity: an area of a minimal surface in AdS 
\be F^{(1)}_n = {\rm Minimal\
area} \ee with boundary defined by the external momenta. After a
$T$-duality transformation of the $n$-point function the boundary
conditions become Dirichlet and state that the boundary of the
surface is the same polygon $\Pi$, Figure \ref{polyg0}, formed by
the external momenta ${\bf p}_a$ and located on the boundary of AdS \cite{AM}.

Classically, the {\it minimal} area
(defined by the Nambu-Goto string action) can be rewritten as the
classical action of the AdS $\sigma$-model in conformal gauge \cite{AM}.
The world-sheet $\sigma$-model equations of motion are \be
\Delta z = zL, \nn \\
\Delta {\bf v} = {\bf v}L \nn \\
z^2L-(\partial z)^2 = (z\partial {\bf v} - {\bf v}\partial z)^2 \ee
where $\Delta\equiv \partial^2/\partial u_1^2+\partial^2/\partial
u_2^2\equiv \partial^2$ is the Laplacian on the world-sheet,
described by the coordinates ${\vec u}=(u_1, u_2), -\infty < u_i <
+\infty $, while $z$ and ${\bf v}$ are coordinates of AdS, related
to the  $T$-dual $(y^\mu, r)$ and the embedding ones $({\bf Y},Y_\pm)$ by
\be {\bf v}
= \frac{\bf y}{r}={\bf Y}, \ \ \ z = \frac{1}{r} = Y_+, \ \ \
\frac{r^2-{\bf y}^2}{r}= Y_- \ee

{\bf (f)} Let us concentrate on the $n=4$ case, a case in which
considerable progress has been made in \cite{AM}. As we demonstrate in Section 3, in this case there is
a whole class of solutions with constant $L$, some of which are related by
$SO(4,2)$ transforms to the Alday-Maldacena solution \cite{AM}. They are \be z =
\sum_{a=1}^n z_a e^{\vec k_a\vec u}\ \ \ \ \ \ {\bf v} =
\sum_{a=1}^n {\bf v}_a e^{\vec k_a\vec u} \label{exponsol} \ee The
four $2$-vectors $\vec k_a$
all have the same length $\vec k_a^2=L$ and are directed along the
diagonals of a rectangular. The
four $4$-vectors ${\bf v}_a$ are related to the external momenta by
\be {\bf p}_a = \frac{{\bf v}_{a+1}}{z_{a+1}}-\frac{{\bf v}_a}{z_a}
\ee This relation can be considered as defining ${\bf v}_a$ for
given ${\bf p}_a$ and $z_a$, while the only remaining constraint
imposed by the equations of motion on the four parameters $z_a$ is
\be z_1z_3 s +z_2z_4 t = 1
 \label{consa}
\ee

With constant $L$, the  action (minimal surface area) $\int L d^2u$
looks independent of external momenta, but actually the integral
diverges and requires regularization. According to \cite{AM}, a
special dimensional regularization defines properly the surface area
and reproduces the expression of $F^{(1)}_n$ conjectured in
\cite{BDS}, but it breaks the AdS-structure of the model, together
with its symmetries and integrability, perhaps in an unnecessarily
violent manner. See \cite{KRTT} for an extension of \cite{AM} to
$1/\sqrt{\lambda}$ corrections (at this stage only divergent terms,
excluded from $F^{(1)}_n$, are examined) and discussion of potential
problems beyond one loop caused by this version of
$\epsilon$-regularization.

Nevertheless, in Section 4 the regularization prescription of
\cite{AM} is naively applied to our solutions as well. It leads to a
regularized minimal area in the form of an integral, which depends
only on $z$. This means that solutions with different $\{z_a\}$'s,
even if they are $SO(4,2)$ transforms of each other, can give rise to
different "areas" after regularization: what may be considered as a
new kind of {\it anomaly}. Then, as usual in anomalous theories, we
minimized the resulting expression over the moduli space of
solutions under the constraint (\ref{consa}). The area of the
minimal surface obtained depends, of course, on $s$ and $t$ because
of the constraint and remarkably enough, it reproduces exactly the
Alday-Maldacena result \cite{AM}.


\bigskip

A few general remarks and speculations about the present program are
in order here.

First, for $n>4$ the Lagrangian density $L$ for configurations of
the form (\ref{exponsol}) is no longer constant, so that the latter
may at best be considered an approximate solution -- a "trial
function" -- which might provide an almost but not exactly minimal
surface. Exact solutions of the $SO(4,2)$ sigma-model
\cite{int42,int42AdS}, allowing for growing asymptotics, remain to be
found. The vast majority of studies in the field of sigma-models are
concentrated on two issues: Lax representation and finite-action
(instanton-like) solutions. By contrast, what is needed here are
solutions with infinite action to make regularization necessary.
Moreover, the target space is non-compact and in the standard
$\sigma$-model coordinates the solutions of interest are
exponentially growing. It is not a big surprise that they have not
been studied thoroughly in the literature.

Second, if the regularization scheme has to break the integrability
of the sigma model, then the Whitham theory \cite{Whith} may be
relevant, since it is well-known \cite{RG} that renormalization
group (RG) flows in the vicinity of integrable systems are well
described in terms of Whitham hierarchies. At the same time AdS
geometry itself should provide a reasonable description of the same
RG behavior \cite{AdSRG}, so that one can probably stay within the
pure integrable framework.

Third, whatever regularization is used, the resulting integral will
have the general form \be \int d^2u  L_\epsilon z^\epsilon \ee A
key step would be to recognize in the finite part of that integral
the same kind of {\it bilinear} structure that exists in its
counterpart formula (\ref{2me}) or its equivalent {\it
double}-contour integral (\ref{loopin}).

Finally, one might expect that the hoped for relation \be
\oint_{\partial S}\oint_{\partial S} \frac{dy^\mu
dy'_\mu}{(y-y')^{2+\epsilon}} ={\rm Area}_{\,\epsilon}{\rm of \ a\
minimal\ surface}\ S \label{1.15} \ee
advertized by the AdS/CFT approach, is one
between two purely geometric quantities, and should
not require any reference to quantum field theory in order to be
formulated and proved.
However, the {\it regularized} area on the r.h.s.
still needs to be defined in geometric terms.
Perhaps, the puzzling relation (\ref{1.15}) can itself be
used as a clue to such a definition.

Also, the "area" on the r.h.s. is not well
defined until a regularization prescription is clearly formulated
and the resulting "anomaly" problem is resolved.

\bigskip

This concludes our introductory remarks, as well as a brief
presentation of our results. The rest of the paper contains detailed
explanations of the statements made above and is organized as
follows. In Section 2 we expand upon the Gauge Theory side and give
a potentially useful pictorial representation of the BDS
formula for $F_n^{(1)}$. Section 3 contains the presentation of our
multi-parameter class of solutions of the AdS $\sigma$-model for
$n=4$. In Section 4 we use
the regularization prescription employed in \cite{AM} and compute
the minimal area, as a function of the parameters of the
solutions. Upon minimization with respect to the moduli we obtain
the Alday-Maldacena result. The final Section contains a brief
summary of our results, a review of open questions related to the
present work and suggestions of possible directions for further
study.

\section{Properties of $F^{(1)}$}
\setcounter{equation}{0}
\subsection{The BDS formula}

\begin{figure}
\hspace{2cm}
\unitlength 1mm 
\linethickness{0.4pt}
\ifx\plotpoint\undefined\newsavebox{\plotpoint}\fi 
\begin{picture}(160.874,79.78)(0,0)
\put(66.705,20.348){\circle{.867}}
\put(67.392,76.808){\circle{.867}}
\put(19.017,58.808){\circle{.867}}
\put(18.497,55.625){\circle{.867}}
\put(17.828,52.416){\circle{.867}}
\put(71.238,21.537){\circle{.867}}
\put(72.787,76.072){\circle{.867}}
\put(75.593,22.429){\circle{.867}}
\put(77.557,75.231){\circle{.867}}
\multiput(87.157,29.974)(-.03363,-.03994){100}{\line(0,-1){.03994}}
\multiput(83.794,25.98)(.0315,-.0946){10}{\line(0,-1){.0946}}
\multiput(84.109,25.034)(.10957447,.03355319){47}{\line(1,0){.10957447}}
\multiput(89.259,26.611)(-.1616923,.0323077){13}{\line(-1,0){.1616923}}
\multiput(87.157,27.031)(.0500823529,.0336970588){340}{\line(1,0){.0500823529}}
\multiput(104.185,38.488)(-.1131538,.0323846){13}{\line(-1,0){.1131538}}
\multiput(102.714,38.909)(-.03355319,.09391489){47}{\line(0,1){.09391489}}
\multiput(101.137,43.323)(.0663684,-.0331579){19}{\line(1,0){.0663684}}
\multiput(102.398,42.693)(-.03343939,.30577273){66}{\line(0,1){.30577273}}
\multiput(100.191,62.874)(-.08408333,.03328333){60}{\line(-1,0){.08408333}}
\multiput(95.146,64.871)(.1997,.0316){10}{\line(1,0){.1997}}
\multiput(97.143,65.187)(-.047186667,.033635556){225}{\line(-1,0){.047186667}}
\multiput(86.526,72.755)(.0334545,.0382273){22}{\line(0,1){.0382273}}
\multiput(35.777,76.005)(.0334545,.0382273){22}{\line(0,1){.0382273}}
\multiput(22.152,28.63)(.0334545,.0382273){22}{\line(0,1){.0382273}}
\multiput(35.902,18.13)(.0334545,.0382273){22}{\line(0,1){.0382273}}
\multiput(87.262,73.596)(.045927928,-.033617117){222}{\line(1,0){.045927928}}
\put(97.458,66.133){\line(0,1){1.366}}
\multiput(97.353,67.499)(.033525862,-.035336207){116}{\line(0,-1){.035336207}}
\multiput(101.242,63.4)(.03336508,-.32868254){63}{\line(0,-1){.32868254}}
\multiput(103.344,42.693)(.03364,.06304){25}{\line(0,1){.06304}}
\put(104.185,44.269){\line(0,-1){5.781}}
\multiput(100.086,63.084)(.0889231,.0323846){13}{\line(1,0){.0889231}}
\put(86.83,73.23){\circle*{1.863}}
\put(36.081,76.48){\circle*{1.863}}
\put(22.456,29.105){\circle*{1.863}}
\put(36.206,18.605){\circle*{1.863}}
\put(100.729,63.196){\circle*{1.863}}
\put(103.479,38.743){\circle*{1.863}}
\put(83.708,25.067){\circle*{1.863}}
\multiput(20.75,62.875)(.0402822096,.0336377214){301}{\line(1,0){.0402822096}}
\multiput(32.875,73)(-.03125,-.34375){4}{\line(0,-1){.34375}}
\multiput(32.75,71.625)(.033601998,.044354638){93}{\line(0,1){.044354638}}
\multiput(53.875,75.875)(-.0326085,-.0326085){23}{\line(0,-1){.0326085}}
\multiput(53.125,75.125)(.16158463,.03353643){41}{\line(1,0){.16158463}}
\multiput(59.75,76.5)(-.12499943,.03365369){52}{\line(-1,0){.12499943}}
\multiput(53.25,78.25)(.03125,-.072916){12}{\line(0,-1){.072916}}
\multiput(53.625,77.375)(-2.203115,-.03125){8}{\line(-1,0){2.203115}}
\multiput(36,77.125)(-.059139517,-.033601998){93}{\line(-1,0){.059139517}}
\multiput(30.5,74)(.375,-.03125){4}{\line(1,0){.375}}
\multiput(32,73.875)(-.0409554459,-.0337029191){293}{\line(-1,0){.0409554459}}
\multiput(20,64)(.0326085,-.0489128){23}{\line(0,-1){.0489128}}
\multiput(40.625,17.5)(.0333332,-.0374998){30}{\line(0,-1){.0374998}}
\multiput(41.625,16.375)(-.14144673,.03289459){38}{\line(-1,0){.14144673}}
\multiput(36.25,17.625)(-.0443875541,.0336733169){245}{\line(-1,0){.0443875541}}
\put(25.375,25.875){\line(0,-1){1.625}}
\multiput(25.375,24.25)(-.03341569,.044554254){101}{\line(0,1){.044554254}}
\multiput(22,28.75)(-.033601998,.139784313){93}{\line(0,1){.139784313}}
\multiput(18.875,41.75)(-.0326085,-.0489128){23}{\line(0,-1){.0489128}}
\put(18.125,40.625){\line(0,1){6.5}}
\multiput(18.125,47.125)(.03343008,-.06104624){86}{\line(0,-1){.06104624}}
\multiput(21,41.875)(-.114583,.03125){12}{\line(-1,0){.114583}}
\put(160.874,29.25){\line(0,1){0}}
\multiput(19.625,42.125)(.033601998,-.137096153){93}{\line(0,-1){.137096153}}
\multiput(25.75,26.75)(.045405777,-.033653694){234}{\line(1,0){.045405777}}
\put(36.375,18.875){\line(5,1){5.625}}
\multiput(42,20)(-.03353643,-.03353643){41}{\line(-1,0){.03353643}}
\put(40.625,18.625){\line(1,0){16.125}}
\put(56.75,18.625){\line(0,-1){1.125}}
\multiput(25.625,26.75)(.145833,.03125){12}{\line(1,0){.145833}}
\multiput(27.375,27.125)(-.06902954,.03358194){67}{\line(-1,0){.06902954}}
\multiput(36.125,75.875)(2.21874,.03125){8}{\line(1,0){2.21874}}
\multiput(87.125,29.875)(-.03125,-.145833){12}{\line(0,-1){.145833}}
\multiput(86.75,28.125)(.0503965972,.033730006){315}{\line(1,0){.0503965972}}
\multiput(40.625,17.625)(4.03123,.03125){4}{\line(1,0){4.03123}}
\put(36.25,18.875){\line(0,1){57.5}}
\multiput(36.25,76.375)(-.033665683,-.1181416097){401}{\line(0,-1){.1181416097}}
\thicklines
\multiput(34.125,70.625)(.03348199,.09151744){56}{\line(0,1){.09151744}}
\multiput(36,75.75)(-.0336537,-.2067298){26}{\line(0,-1){.2067298}}
\put(87.157,76.118){\makebox(0,0)[cc]{$n$}}
\put(95.041,71.388){\makebox(0,0)[cc]{${\bf p}_n$}}
\put(103.344,64.556){\makebox(0,0)[cc]{$1$}}
\put(105.131,52.994){\makebox(0,0)[cc]{${\bf p}_1$}}
\put(106.287,37.647){\makebox(0,0)[cc]{$2$}}
\put(96.933,29.659){\makebox(0,0)[cc]{${\bf p}_2$}}
\put(84.845,22.406){\makebox(0,0)[cc]{$3$}}
\put(6.185,60.158){\makebox(0,0)[cc]{${\bf P}_{ab}$}}
\put(70.339,50.891){\makebox(0,0)[cc]{$t_{ab}=({\bf p}_a+{\bf
P}_{ab})^2$}} \put(45.408,79.78){\makebox(0,0)[cc]{${\bf p}_b$}}
\put(33.951,79.359){\makebox(0,0)[cc]{$b$}}
\put(14.295,36.263){\makebox(0,0)[cc]{${\bf p}_{a+1}$}}
\put(17.133,28.38){\makebox(0,0)[cc]{$a+1$}}
\put(27.434,20.917){\makebox(0,0)[cc]{${\bf p}_a$}}
\put(34.161,15.557){\makebox(0,0)[cc]{$a$}}
\put(47.826,14.821){\makebox(0,0)[cc]{${\bf p}_{a-1}$}} \thinlines
\qbezier(11.142,60.439)(13.77,60.649)(17.238,61.28)
\put(30.798,59.283){\vector(4,-1){.07}}\qbezier(17.238,61.28)(20.97,61.648)(30.798,59.283)
\qbezier(54.238,51.4)(49.455,51.452)(46.144,47.931)
\put(36.999,45.408){\vector(-1,0){.07}}\qbezier(46.144,47.931)(43.148,45.618)(36.999,45.408)
\thicklines
\multiput(34.897,70.74)(.0328474,.1445284){32}{\line(0,1){.1445284}}
\multiput(35.948,75.365)(-.03332806,-.09998418){41}{\line(0,-1){.09998418}}
\multiput(34.582,71.055)(.0328474,.111681){32}{\line(0,1){.111681}}
\multiput(35.633,74.629)(-.03319313,-.09681329){38}{\line(0,-1){.09681329}}
\end{picture}
 \caption{\footnotesize A formal polygon $\Pi$ made from
external momentum $p_a$, the same as in Fig.\ref{polyg0}. The
picture shows the labeling of vertices, edges and diagonals used
throughout the text. Note that diagonals appear only through their
squares $t_{ab}$. All vectors are Minkowskian, and in the
$N=4$ SYM theory, all ${\bf p}_a$ are null ${\bf p}_a^2=0$. }
\label{polyg}
\end{figure}
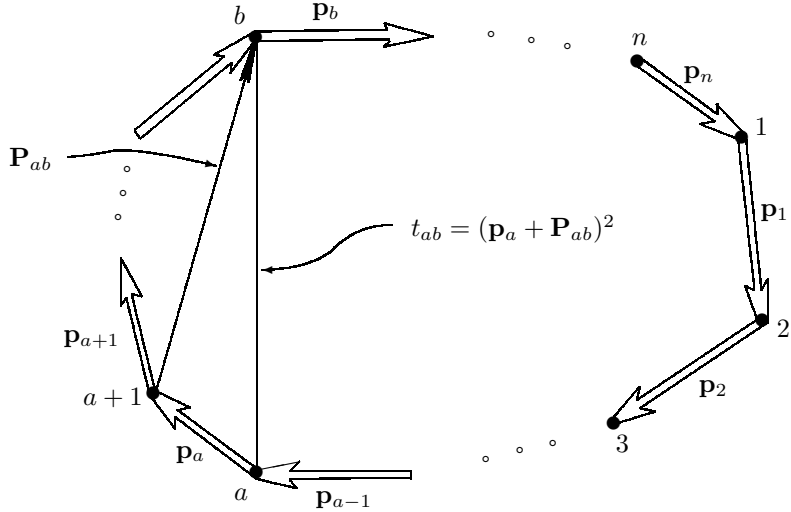

$F_n^{(1)}$ in (\ref{BDSf}) is a function of invariant variables
$t_{ab} = \left(\sum_{c=a}^{b-1} {\bf p}_c\right)^2 = ({\bf
p}_a+{\bf P}_{ab})^2$. These $t$-variables are nothing but squares
of diagonals in a polygon $\Pi$,
Figure \ref{polyg}, which is closed due to momentum conservation
$\sum_{a=1}^n {\bf p}_a=0$. Given this association of $t$-variables with
diagonals, it is also natural to denote $t_{ab} = t_a^{[b-a]}$ where
$[b-a]$ is the {\it size} of the diagonal: the number of polygon
sides that it embraces. Of course, $t_{ab}=t_{ba}$,
$t^{[2]}_a=s_{a,a+1}$ and one can restrict diagonal sizes $r$ by
$r\leq n/2$. Of all $t$ variables only $3n-4-6=3n-10$ are actually
independent ($3$ stands for three independent components of a
null-vector, $4$ constraints are imposed by 4-momentum conservation,
$\sum_{a=1}^n {\bf p}_a = 0$, $10$ is the number of Lorentz
rotations and translations in $4$ dimensions, which all act on
$p$-variables as long as $n\geq 4$), this is, however, not important
for our purposes below.

According to \cite{BDS}, in the $N=4$ supersymmetric gauge theory
$F_n^{(1)}$ decomposes into a sum of terms, each depending on only
four out of all $t$-variables. Some of these terms are dilogarithms,
while others are squares of ordinary logarithms. The BDS formula
(eqs.(4.59)-(4.63) of \cite{BDS}, in a slightly different notation)
states that
\be F_n^{(1)} = D_n + L_{n}^{(1)} + L_{n}^{(2)} + \hbox{const}
\label{Fdilog} \ee where
\be D_n =
-\frac{1}{4}\sum_{a=1}^n
\sum_{r=2}^{n-4} Li_2\left(1-\frac{t^{[r]}_at^{[r+2]}_{a+1}}
{t^{[r+1]}_at^{[ r+1]}_{a+1}}\right),  
\label{Fdil} \ee and
\be L_{n}^{(1)} = -\frac{1}{2}\sum_{a=1}^n\sum_{r=2}^{[n/2]-1}
\Big(\tau^{[r+1]}_{a}-\tau^{[r]}_a\Big)
\Big(\tau^{[r+1]}_{a}-\tau^{[r]}_{a+1}\Big) \label{Flog} \ee with
$\tau^{[r]}_a = \log (-t_{a,a+r})$.
The remaining logarithmic term is different for even and odd $n$, namely
%
\be {\rm for\ even} \ \ n=2m: \ \ \ \ \ L_{n}^{(2)} =
-\frac{1}{8}\sum_{a=1}^n \Big(\tau_a^{[m]}-\tau_{a+m+1}^{[m]}\Big)
\Big(\tau_{a+1}^{[m]}-\tau_{a+m}^{[m]}\Big) \label{Flogeven} \ee
while
%
\be {\rm for\ odd} \ \ n=2m+1: \ \ \ \ \ L_{n}^{(2)} =
-\frac{1}{4}\sum_{a=1}^n \Big(\tau_a^{[m]}-\tau_{a+m+1}^{[m]}\Big)
\Big(\tau_{a+1}^{[m]}-\tau_{a+m}^{[m]}\Big) \label{Flogodd} \ee
where $m = \big[n/2\big]$. Note that in (\ref{Flogeven}) each
$\tau_a^{[m]}$ appears twice, as $\tau_a^{[m]}$ and
$\tau_{a+m}^{[m]}$, but this is not the case in (\ref{Flogodd}) --
this explains the difference in the coefficients in front of the
sums.

The dilogarithmic part of (\ref{Fdilog}) is actually a sum over
$4$-clusters (Figure \ref{clust}) in a polygon $\Pi$, (Figure
\ref{polyg}) of the quantity
\be \frac{1}{2} \sum_{{\bf p},{\bf q}} Li_2\left(1-\frac{{\bf
P}^2{\bf Q}^2}{st}\right) \label{Fdilog2} \ee A $4$-cluster, Figure
\ref{clust}, is formed by some two non-adjacent edges ${\bf p}_a={\bf
p}$ and ${\bf p}_b={\bf q}$ in $\Pi$ and consists of the four
diagonals of $\Pi$, connecting the corresponding four vertices $a$,
$a+1$, $a+r$, $a+r+1$. Squared lengths of these four diagonals are
$t_{a,b+1}=t_a^{[r+2]}={\bf Q}^2$, $t_{ab} = t_a^{[r+1]}=s$,
$t_{a+1,b+1} = t_{a+1}^{[r+1]}=t$ and $t_{a+1,b}=t_{a+1}^{[r]}={\bf
P}^2$, where ${\bf P} = {\bf P}_{ab}$ and ${\bf Q}={\bf P}_{ba}$ and
$r=j-i-1$ is the size of the shortest diagonal in the cluster. It is
indeed a diagonal only for $r\geq 2$ and such $4$-clusters -- and
thus dilogarithmic contributions to $F_n^{(1)}$ -- exist for $n\geq
6$.

Note that, from the point of view of scattering theory,
eq.(\ref{Fdilog}) is better than (\ref{Fdilog2}) because the summation
over different appearances of each $t_{ab}$ is already performed.
However, the older formula (\ref{Fdilog2}) better suites {\it our}
purposes, since it describes more nicely the internal structures of the problem.

\subsection{A pictorial representation \label{picto}}

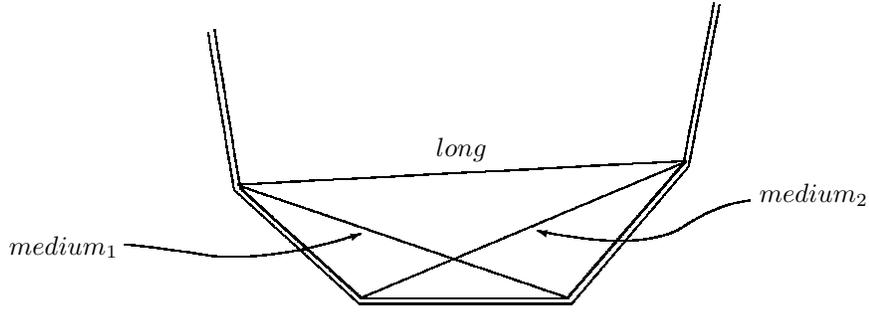
\begin{figure}
\hspace{2cm}
\unitlength 1mm 
\linethickness{0.4pt}
\ifx\plotpoint\undefined\newsavebox{\plotpoint}\fi 
\begin{picture}(109.11,59.599)(0,0)
\multiput(32.585,35.423)(.632904255,.033542553){94}{\line(1,0){.632904255}}
\multiput(32.795,35.213)(.0987065463,-.0336930023){443}{\line(1,0){.0987065463}}
\put(76.522,20.287){\line(-1,0){27.75}}
\multiput(48.772,20.287)(.0804029851,.0337294776){536}{\line(1,0){.0804029851}}
\thicklines
\multiput(92.078,38.576)(-.0336901709,-.0390790598){468}{\line(0,-1){.0390790598}}
\multiput(32.795,35.002)(.0370411899,-.0336727689){437}{\line(1,0){.0370411899}}
\thinlines
\multiput(32.795,35.213)(-.03364,.20812){100}{\line(0,1){.20812}}
\multiput(91.868,38.576)(.033563025,.176663866){119}{\line(0,1){.176663866}}
\multiput(28.59,55.709)(.033679612,-.204097087){103}{\line(0,-1){.204097087}}
\multiput(32.059,34.687)(.0369888641,-.0337104677){449}{\line(1,0){.0369888641}}
\put(48.667,19.551){\line(1,0){28.275}}
\multiput(76.942,19.551)(.0336709957,.0395865801){462}{\line(0,1){.0395865801}}
\multiput(92.498,37.84)(.033606557,.176622951){122}{\line(0,1){.176622951}}
\put(62.285,40.136){\makebox(0,0)[cc]{$long$}}
\put(9.374,27.118){\makebox(0,0)[cc]{$medium_1$}}
\put(109.11,34.041){\makebox(0,0)[cc]{$medium_2$}}
\qbezier(100.488,33.298)(96.103,32.778)(91.42,29.582)
\put(72.393,29.284){\vector(-4,1){.07}}\qbezier(91.42,29.582)(85.772,26.46)(72.393,29.284)
\put(48.877,28.906){\vector(3,1){.07}}\qbezier(28.696,25.963)(37.525,25.122)(48.877,28.906)
\qbezier(28.801,25.963)(21.601,27.119)(17.554,27.434)
\end{picture}
\caption{\footnotesize Degenerate $4$-cluster: one of
the would-be diagonals is actually an edge of $\Pi$, thus it is
null and the corresponding $t$ would be zero. Degenerate clusters
do not contribute to the dilogarithmic part of the BDS formula (\ref{Fdil}),
but do contribute to the logarithmic part (\ref{Flog}), see
Fig.\ref{clust}.} \label{degclust}
\end{figure}

The simple structure of the somewhat sophisticated formula
(\ref{Fdilog}) can be revealed in pictures. As already mentioned,
eq.(\ref{Fdilog}) represents $F_n^{(1)}$ as a sum over all
$4$-clusters of the polygon, with the single exception of peculiar
$5$-clusters, Figure \ref{longlongodd}, contributing for odd $n$.
{\it Degenerate} $4$-clusters where the smallest diagonal coincides
with an edge, Figure \ref{degclust}, do not contribute to the
dilogarithmic piece (\ref{Fdil}), but only to the logarithmic one
(\ref{Flog}). Since for $n=4$ and $n=5$ all $4$-clusters are
degenerate, there are no dilogarithms at all in $F_4^{(1)}$ and
$F_5^{(1)}$.

\begin{figure}
\hspace{3cm}
\unitlength 1mm 
\linethickness{0.4pt}
\ifx\plotpoint\undefined\newsavebox{\plotpoint}\fi 
\begin{picture}(99.26,46.323)(0,0)
\multiput(20.483,37.573)(.03370787,-.15589888){89}{\line(0,-1){.15589888}}
\multiput(23.483,23.698)(.287564767,.033678756){193}{\line(1,0){.287564767}}
\multiput(78.983,30.198)(.033602151,.173387097){93}{\line(0,1){.173387097}}
\multiput(33.733,11.323)(.112903226,-.033602151){93}{\line(1,0){.112903226}}
\put(59.608,8.073){\line(2,1){11.25}}
\multiput(78.983,30.323)(-.0793650794,-.0337301587){567}{\line(-1,0){.0793650794}}
\multiput(33.983,11.198)(.49333333,.03333333){75}{\line(1,0){.49333333}}
\multiput(70.983,13.698)(-.162116041,.0337030717){293}{\line(-1,0){.162116041}}
\multiput(23.6,23.599)(.0336419355,-.0402){310}{\line(0,-1){.0402}}
\multiput(78.931,30.229)(-.033603306,-.068301653){242}{\line(0,-1){.068301653}}
\put(48.702,7.778){\circle*{.791}}
\put(51.795,7.778){\circle*{.791}}
\put(54.801,7.866){\circle*{.791}}
\multiput(19.622,37.3)(.033630435,-.15276087){92}{\line(0,-1){.15276087}}
\put(22.716,23.246){\line(5,-6){10.607}}
\multiput(33.322,10.518)(.107525773,-.03371134){97}{\line(1,0){.107525773}}
\multiput(83.085,45.962)(-.033494737,-.168410526){95}{\line(0,-1){.168410526}}
\multiput(79.903,29.963)(-.0336396761,-.0687044534){247}{\line(0,-1){.0687044534}}
\multiput(71.594,12.993)(-.066416185,-.033722543){173}{\line(-1,0){.066416185}}
\put(49.851,29.787){\makebox(0,0)[cc]{$long$}}
\put(51.089,14.407){\makebox(0,0)[cc]{$short$}}
\put(4.243,19.799){\makebox(0,0)[cc]{$medium_1$}}
\put(99.26,25.456){\makebox(0,0)[cc]{$medium_2$}}
\qbezier(12.816,20.064)(17.103,19.887)(20.506,17.943)
\put(36.77,20.153){\vector(3,1){.07}}\qbezier(20.506,17.943)(25.986,15.777)(36.77,20.153)
\qbezier(88.3,25.544)(83.925,24.926)(79.373,23.246)
\put(61.43,22.185){\vector(-1,0){.07}}\qbezier(79.373,23.246)(72.611,21.125)(61.43,22.185)
\thicklines \put(75,33.938){\line(0,-1){2.063}}
\put(62.438,12.625){\line(0,-1){2.063}}
\put(74,32.875){\line(1,0){1.938}}
\put(61.438,11.563){\line(1,0){1.938}}
\put(72,25.938){\line(1,0){2}} \put(67,15.938){\line(1,0){2}}
\end{picture}
 \caption{\footnotesize Relative signs of the
contributions of different diagonals of a given cluster to an
argument of dilogarithm in (\ref{Fdil}): $\log \frac{{\bf P}^2{\bf
Q}^2}{st} = \tau_{l} + \tau_{s} - \tau_{m1} - \tau_{m2}$, see
Fig.\ref{clust}. } \label{dilo}
\end{figure}
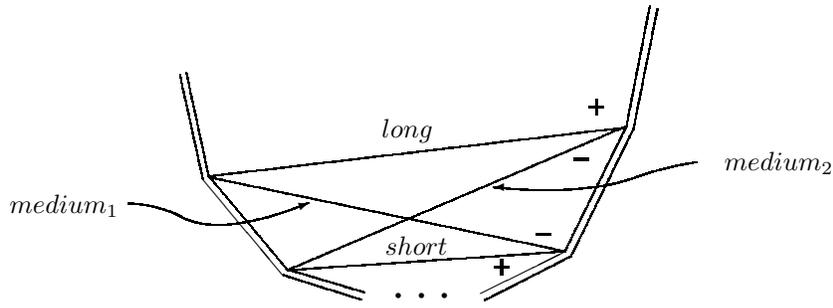

The dilogarithm is made out of the four diagonals of a
non-degenerate $4$-cluster. The four $\tau$-parameters associated
with the four diagonals are summed with the signs shown in Figure
\ref{dilo}, the sum is exponentiated to give a ratio of $t$'s, which
is further subtracted from unity and used as an argument of a
dilogarithm.

\begin{figure}
\hspace{3cm}
\unitlength 1mm 
\linethickness{0.4pt}
\ifx\plotpoint\undefined\newsavebox{\plotpoint}\fi 
\begin{picture}(99.26,46.323)(0,0)
\multiput(20.483,37.573)(.03370787,-.15589888){89}{\line(0,-1){.15589888}}
\multiput(23.483,23.698)(.287564767,.033678756){193}{\line(1,0){.287564767}}
\multiput(78.983,30.198)(.033602151,.173387097){93}{\line(0,1){.173387097}}
\multiput(33.733,11.323)(.112903226,-.033602151){93}{\line(1,0){.112903226}}
\put(59.608,8.073){\line(2,1){11.25}}
\multiput(70.983,13.698)(-.162116041,.0337030717){293}{\line(-1,0){.162116041}}
\multiput(23.6,23.599)(.0336419355,-.0402){310}{\line(0,-1){.0402}}
\multiput(78.931,30.229)(-.033603306,-.068301653){242}{\line(0,-1){.068301653}}
\put(48.702,7.778){\circle*{.791}}
\put(51.795,7.778){\circle*{.791}}
\put(54.801,7.866){\circle*{.791}}
\multiput(19.622,37.3)(.033630435,-.15276087){92}{\line(0,-1){.15276087}}
\put(22.716,23.246){\line(5,-6){10.607}}
\multiput(33.322,10.518)(.107525773,-.03371134){97}{\line(1,0){.107525773}}
\multiput(83.085,45.962)(-.033494737,-.168410526){95}{\line(0,-1){.168410526}}
\multiput(79.903,29.963)(-.0336396761,-.0687044534){247}{\line(0,-1){.0687044534}}
\multiput(71.594,12.993)(-.066416185,-.033722543){173}{\line(-1,0){.066416185}}
\put(49.851,29.787){\makebox(0,0)[cc]{$long$}}
\put(51.089,14.407){\makebox(0,0)[cc]{$short$}}
\put(4.243,19.799){\makebox(0,0)[cc]{$medium_1$}}
\put(99.26,25.456){\makebox(0,0)[cc]{$medium_2$}}
\qbezier(12.816,20.064)(17.103,19.887)(20.506,17.943)
\put(36.77,20.153){\vector(3,1){.07}}\qbezier(20.506,17.943)(25.986,15.777)(36.77,20.153)
\qbezier(88.3,25.544)(83.925,24.926)(79.373,23.246)
\put(61.43,22.185){\vector(-1,0){.07}}\qbezier(79.373,23.246)(72.611,21.125)(61.43,22.185)
\multiput(23.795,23.087)(.269519,.031785){7}{\line(1,0){.269519}}
\multiput(27.568,23.532)(.269519,.031785){7}{\line(1,0){.269519}}
\multiput(31.341,23.977)(.269519,.031785){7}{\line(1,0){.269519}}
\multiput(35.114,24.422)(.269519,.031785){7}{\line(1,0){.269519}}
\multiput(38.888,24.867)(.269519,.031785){7}{\line(1,0){.269519}}
\multiput(42.661,25.312)(.269519,.031785){7}{\line(1,0){.269519}}
\multiput(46.434,25.757)(.269519,.031785){7}{\line(1,0){.269519}}
\multiput(50.207,26.202)(.269519,.031785){7}{\line(1,0){.269519}}
\multiput(53.981,26.647)(.269519,.031785){7}{\line(1,0){.269519}}
\multiput(57.754,27.092)(.269519,.031785){7}{\line(1,0){.269519}}
\multiput(61.527,27.537)(.269519,.031785){7}{\line(1,0){.269519}}
\multiput(65.301,27.982)(.269519,.031785){7}{\line(1,0){.269519}}
\multiput(69.074,28.427)(.269519,.031785){7}{\line(1,0){.269519}}
\multiput(72.847,28.872)(.269519,.031785){7}{\line(1,0){.269519}}
\multiput(76.62,29.317)(.269519,.031785){7}{\line(1,0){.269519}}
\put(78.507,29.717){\line(-1,0){.0442}}
\multiput(78.419,29.717)(-.0773206,-.0324347){23}{\line(-1,0){.0773206}}
\multiput(74.862,28.225)(-.0773206,-.0324347){23}{\line(-1,0){.0773206}}
\multiput(71.305,26.733)(-.0773206,-.0324347){23}{\line(-1,0){.0773206}}
\multiput(67.748,25.241)(-.0773206,-.0324347){23}{\line(-1,0){.0773206}}
\multiput(64.192,23.749)(-.0773206,-.0324347){23}{\line(-1,0){.0773206}}
\multiput(60.635,22.257)(-.0773206,-.0324347){23}{\line(-1,0){.0773206}}
\multiput(57.078,20.765)(-.0773206,-.0324347){23}{\line(-1,0){.0773206}}
\multiput(53.521,19.273)(-.0773206,-.0324347){23}{\line(-1,0){.0773206}}
\multiput(49.965,17.781)(-.0773206,-.0324347){23}{\line(-1,0){.0773206}}
\multiput(46.408,16.289)(-.0773206,-.0324347){23}{\line(-1,0){.0773206}}
\multiput(42.851,14.797)(-.0773206,-.0324347){23}{\line(-1,0){.0773206}}
\multiput(39.294,13.305)(-.0773206,-.0324347){23}{\line(-1,0){.0773206}}
\multiput(35.738,11.813)(-.0773206,-.0324347){23}{\line(-1,0){.0773206}}
\multiput(34.224,11.155)(.986606,.066889){38}{{\rule{.4pt}{.4pt}}}
\end{picture}
\caption{\footnotesize Generic $4$-cluster with well
defined "long" diagonal. Its contribution to the logarithmic part of the
BDS formula (\ref{Flog}) is $\frac{1}{2}(\tau_l -
\tau_{m_1})(\tau_l-\tau_{m_2})$ and does not depend on the "short"
diagonal. Therefore this contribution is well defined even for the
degenerate clusters in Fig.\ref{degclust}. {\it Generic} $4$-cluster
is fully defined by its long diagonal, so that each long diagonal
contributes once to the logarithmic part of the BDS formula and
twice -- to its dilogarithmic part. Exceptions from this rule are
the {\it longest} (main) diagonals of $\Pi$, see
Figs.\ref{logodi}-\ref{longlongodd}.} \label{logo}
\end{figure}
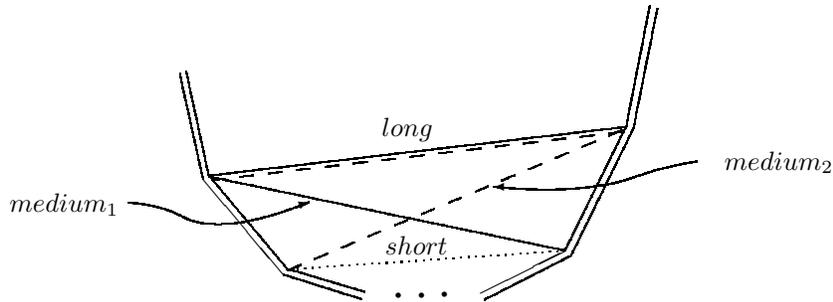

\begin{figure}
\hspace{5cm}
\unitlength 1mm 
\linethickness{0.4pt}
\ifx\plotpoint\undefined\newsavebox{\plotpoint}\fi 
\begin{picture}(46.25,42.754)(0,0)
\put(18.935,41.904){\line(1,0){18.817}}
\multiput(18.935,41.904)(-.0337275862,-.0348275862){290}{\line(0,-1){.0348275862}}
\multiput(8.41,32.336)(.0336993464,.0340457516){306}{\line(0,1){.0340457516}}
\put(18.722,42.754){\line(1,0){19.668}}
\multiput(37.525,41.835)(.033635556,-.054191111){225}{\line(0,-1){.054191111}}
\multiput(45.093,29.642)(-.033587629,-.152793814){97}{\line(0,-1){.152793814}}
\multiput(38.156,42.57)(.033725,-.053866667){240}{\line(0,-1){.053866667}}
\multiput(46.25,29.642)(-.033679612,-.147980583){103}{\line(0,-1){.147980583}}
\multiput(9.25,31.744)(.03319298,-.24157895){57}{\line(0,-1){.24157895}}
\multiput(11.142,17.974)(.066493023,-.033730233){215}{\line(1,0){.066493023}}
\put(25.438,10.722){\line(4,1){16.397}}
\multiput(8.304,32.164)(.03328333,-.24876667){60}{\line(0,-1){.24876667}}
\multiput(10.301,17.238)(.066390351,-.033653509){228}{\line(1,0){.066390351}}
\multiput(25.438,9.565)(.124913043,.033514493){138}{\line(1,0){.124913043}}
\multiput(25.438,10.616)(.0336823204,.0865303867){362}{\line(0,1){.0865303867}}
\multiput(11.177,17.799)(.0940167,.0321167){20}{\line(1,0){.0940167}}
\multiput(14.937,19.083)(.0940167,.0321167){20}{\line(1,0){.0940167}}
\multiput(18.698,20.368)(.0940167,.0321167){20}{\line(1,0){.0940167}}
\multiput(22.459,21.653)(.0940167,.0321167){20}{\line(1,0){.0940167}}
\multiput(26.219,22.937)(.0940167,.0321167){20}{\line(1,0){.0940167}}
\multiput(29.98,24.222)(.0940167,.0321167){20}{\line(1,0){.0940167}}
\multiput(33.741,25.507)(.0940167,.0321167){20}{\line(1,0){.0940167}}
\multiput(37.501,26.791)(.0940167,.0321167){20}{\line(1,0){.0940167}}
\multiput(41.262,28.076)(.0940167,.0321167){20}{\line(1,0){.0940167}}
\put(28.591,4.94){\makebox(0,0)[cc]{n odd}}
\put(17.764,35.633){\makebox(0,0)[cc]{$l_3$}}
\put(39.102,24.386){\makebox(0,0)[cc]{$l_4$}}
\put(29.957,16.398){\makebox(0,0)[cc]{$l_1$}}
\put(13.875,24.386){\makebox(0,0)[cc]{$l_2$}}
\multiput(11.142,17.974)(.03741259483,.03370102788){708}{\line(1,0){.03741259483}}
\multiput(18.85,41.764)(.031946,-.151484){12}{\line(0,-1){.151484}}
\multiput(19.616,38.128)(.031946,-.151484){12}{\line(0,-1){.151484}}
\multiput(20.383,34.493)(.031946,-.151484){12}{\line(0,-1){.151484}}
\multiput(21.15,30.857)(.031946,-.151484){12}{\line(0,-1){.151484}}
\multiput(21.917,27.222)(.031946,-.151484){12}{\line(0,-1){.151484}}
\multiput(22.683,23.586)(.031946,-.151484){12}{\line(0,-1){.151484}}
\multiput(23.45,19.95)(.031946,-.151484){12}{\line(0,-1){.151484}}
\multiput(24.217,16.315)(.031946,-.151484){12}{\line(0,-1){.151484}}
\multiput(24.983,12.679)(.031946,-.151484){12}{\line(0,-1){.151484}}
\end{picture}
 \caption{\footnotesize In the case of odd $n$ the
longest (main) diagonals enter through a $5$-cluster (involving five
vertices of $\Pi$). The contribution to the BDS formula (\ref{Flogodd})
is made from four longest diagonals and is given by $\frac{1}{2}
\Big(\tau_{l_1}-\tau_{l_2})(\tau_{l_3}-\tau_{l_4}\Big)$ -- again
very different from the one in Fig.\ref{logo}. This is the only case
where $5$-clusters contribute to the BDS formula. }
\label{longlongodd}
\end{figure}
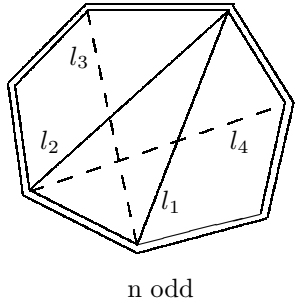

Logarithms are a little bit trickier. The smallest diagonal does not
contribute (thus degenerate clusters are allowed), while the largest
one contributes twice. The contribution of a $4$-cluster is the
product of the two differences of $\tau$ for the largest diagonal
and $\tau$'s for the two medium diagonals, Figure \ref{logo}.

Additional complications arise when the largest diagonal in the
cluster is the main diagonal of the polygon $\Pi$. The situation is
somewhat different for $n$ even and odd.

\begin{figure}
\hspace{2.5cm}
\unitlength 1mm 
\linethickness{0.4pt}
\ifx\plotpoint\undefined\newsavebox{\plotpoint}\fi 
\begin{picture}(110.359,49.056)(0,0)
\thicklines
\multiput(18.268,41.683)(.071607527,.033564516){186}{\line(1,0){.071607527}}
\multiput(104.232,40.933)(-.071607527,.033564516){186}{\line(-1,0){.071607527}}
\multiput(31.587,47.926)(.097015038,-.033639098){133}{\line(1,0){.097015038}}
\multiput(90.913,47.176)(-.097015038,-.033639098){133}{\line(-1,0){.097015038}}
\multiput(44.49,43.452)(.033624454,-.037716157){229}{\line(0,-1){.037716157}}
\multiput(78.01,42.702)(-.033624454,-.037716157){229}{\line(0,-1){.037716157}}
\multiput(18.268,41.475)(-.033546053,-.057506579){152}{\line(0,-1){.057506579}}
\multiput(104.232,40.725)(.033546053,-.057506579){152}{\line(0,-1){.057506579}}
\multiput(13.482,18.478)(.033590551,-.053251969){127}{\line(0,-1){.053251969}}
\multiput(109.018,17.728)(-.033590551,-.053251969){127}{\line(0,-1){.053251969}}
\multiput(17.748,11.715)(.093795775,-.033711268){142}{\line(1,0){.093795775}}
\multiput(104.752,10.965)(-.093795775,-.033711268){142}{\line(-1,0){.093795775}}
\multiput(30.963,6.824)(.124694215,.03353719){121}{\line(1,0){.124694215}}
\multiput(91.537,6.074)(-.124694215,.03353719){121}{\line(-1,0){.124694215}}
\multiput(46.051,10.882)(.033588608,.052031646){158}{\line(0,1){.052031646}}
\multiput(76.449,10.132)(-.033588608,.052031646){158}{\line(0,1){.052031646}}
\thinlines
\put(31.205,47.648){\line(0,-1){.9923}}
\put(31.18,45.663){\line(0,-1){.9923}}
\put(31.154,43.679){\line(0,-1){.9923}}
\put(31.129,41.694){\line(0,-1){.9923}}
\put(31.104,39.709){\line(0,-1){.9923}}
\put(31.078,37.725){\line(0,-1){.9923}}
\put(31.053,35.74){\line(0,-1){.9923}}
\put(31.028,33.755){\line(0,-1){.9923}}
\put(31.002,31.77){\line(0,-1){.9923}}
\put(30.977,29.786){\line(0,-1){.9923}}
\put(30.951,27.801){\line(0,-1){.9923}}
\put(30.926,25.816){\line(0,-1){.9923}}
\put(30.901,23.832){\line(0,-1){.9923}}
\put(30.875,21.847){\line(0,-1){.9923}}
\put(30.85,19.862){\line(0,-1){.9923}}
\put(30.825,17.878){\line(0,-1){.9923}}
\put(30.799,15.893){\line(0,-1){.9923}}
\put(30.774,13.908){\line(0,-1){.9923}}
\put(30.749,11.924){\line(0,-1){.9923}}
\put(30.723,9.939){\line(0,-1){.9923}}
\put(30.698,7.954){\line(0,-1){.9923}}
\put(91.154,46.898){\line(0,-1){.9923}}
\put(91.18,44.913){\line(0,-1){.9923}}
\put(91.205,42.929){\line(0,-1){.9923}}
\put(91.23,40.944){\line(0,-1){.9923}}
\put(91.256,38.959){\line(0,-1){.9923}}
\put(91.281,36.975){\line(0,-1){.9923}}
\put(91.307,34.99){\line(0,-1){.9923}}
\put(91.332,33.005){\line(0,-1){.9923}}
\put(91.357,31.02){\line(0,-1){.9923}}
\put(91.383,29.036){\line(0,-1){.9923}}
\put(91.408,27.051){\line(0,-1){.9923}}
\put(91.433,25.066){\line(0,-1){.9923}}
\put(91.459,23.082){\line(0,-1){.9923}}
\put(91.484,21.097){\line(0,-1){.9923}}
\put(91.509,19.112){\line(0,-1){.9923}}
\put(91.535,17.128){\line(0,-1){.9923}}
\put(91.56,15.143){\line(0,-1){.9923}}
\put(91.586,13.158){\line(0,-1){.9923}}
\put(91.611,11.174){\line(0,-1){.9923}}
\put(91.636,9.189){\line(0,-1){.9923}}
\put(91.662,7.204){\line(0,-1){.9923}}
\multiput(30.685,6.962)(.031452,.0823){11}{\line(0,1){.0823}}
\multiput(31.377,8.773)(.031452,.0823){11}{\line(0,1){.0823}}
\multiput(32.069,10.583)(.031452,.0823){11}{\line(0,1){.0823}}
\multiput(32.761,12.394)(.031452,.0823){11}{\line(0,1){.0823}}
\multiput(33.453,14.204)(.031452,.0823){11}{\line(0,1){.0823}}
\multiput(34.145,16.015)(.031452,.0823){11}{\line(0,1){.0823}}
\multiput(34.837,17.826)(.031452,.0823){11}{\line(0,1){.0823}}
\multiput(35.529,19.636)(.031452,.0823){11}{\line(0,1){.0823}}
\multiput(36.221,21.447)(.031452,.0823){11}{\line(0,1){.0823}}
\multiput(36.913,23.257)(.031452,.0823){11}{\line(0,1){.0823}}
\multiput(37.605,25.068)(.031452,.0823){11}{\line(0,1){.0823}}
\multiput(38.297,26.879)(.031452,.0823){11}{\line(0,1){.0823}}
\multiput(38.988,28.689)(.031452,.0823){11}{\line(0,1){.0823}}
\multiput(39.68,30.5)(.031452,.0823){11}{\line(0,1){.0823}}
\multiput(40.372,32.31)(.031452,.0823){11}{\line(0,1){.0823}}
\multiput(41.064,34.121)(.031452,.0823){11}{\line(0,1){.0823}}
\multiput(41.756,35.932)(.031452,.0823){11}{\line(0,1){.0823}}
\multiput(42.448,37.742)(.031452,.0823){11}{\line(0,1){.0823}}
\multiput(43.14,39.553)(.031452,.0823){11}{\line(0,1){.0823}}
\multiput(43.832,41.363)(.031452,.0823){11}{\line(0,1){.0823}}
\multiput(91.674,6.212)(-.031452,.0823){11}{\line(0,1){.0823}}
\multiput(90.982,8.023)(-.031452,.0823){11}{\line(0,1){.0823}}
\multiput(90.29,9.833)(-.031452,.0823){11}{\line(0,1){.0823}}
\multiput(89.598,11.644)(-.031452,.0823){11}{\line(0,1){.0823}}
\multiput(88.907,13.454)(-.031452,.0823){11}{\line(0,1){.0823}}
\multiput(88.215,15.265)(-.031452,.0823){11}{\line(0,1){.0823}}
\multiput(87.523,17.076)(-.031452,.0823){11}{\line(0,1){.0823}}
\multiput(86.831,18.886)(-.031452,.0823){11}{\line(0,1){.0823}}
\multiput(86.139,20.697)(-.031452,.0823){11}{\line(0,1){.0823}}
\multiput(85.447,22.507)(-.031452,.0823){11}{\line(0,1){.0823}}
\multiput(84.755,24.318)(-.031452,.0823){11}{\line(0,1){.0823}}
\multiput(84.063,26.129)(-.031452,.0823){11}{\line(0,1){.0823}}
\multiput(83.371,27.939)(-.031452,.0823){11}{\line(0,1){.0823}}
\multiput(82.679,29.75)(-.031452,.0823){11}{\line(0,1){.0823}}
\multiput(81.987,31.56)(-.031452,.0823){11}{\line(0,1){.0823}}
\multiput(81.295,33.371)(-.031452,.0823){11}{\line(0,1){.0823}}
\multiput(80.603,35.182)(-.031452,.0823){11}{\line(0,1){.0823}}
\multiput(79.911,36.992)(-.031452,.0823){11}{\line(0,1){.0823}}
\multiput(79.219,38.803)(-.031452,.0823){11}{\line(0,1){.0823}}
\multiput(78.527,40.613)(-.031452,.0823){11}{\line(0,1){.0823}}
\put(12.388,29.905){\circle*{.495}}
\put(110.112,29.155){\circle*{.495}}
\put(52.852,30.771){\circle*{.495}}
\put(69.648,30.021){\circle*{.495}}
\put(12.388,26.317){\circle*{.495}}
\put(110.112,25.567){\circle*{.495}}
\put(52.852,27.183){\circle*{.495}}
\put(69.648,26.433){\circle*{.495}}
\put(12.388,22.728){\circle*{.495}}
\put(110.112,21.978){\circle*{.495}}
\put(52.852,23.594){\circle*{.495}}
\put(69.648,22.844){\circle*{.495}}
\multiput(12.463,33.322)(.033727133,.059313233){152}{\line(0,1){.059313233}}
\multiput(110.037,32.572)(-.033727133,.059313233){152}{\line(0,1){.059313233}}
\multiput(17.589,42.338)(.070710678,.033587572){200}{\line(1,0){.070710678}}
\multiput(104.911,41.588)(-.070710678,.033587572){200}{\line(-1,0){.070710678}}
\multiput(31.731,49.056)(.092655371,-.033526615){145}{\line(1,0){.092655371}}
\multiput(90.769,48.306)(-.092655371,-.033526615){145}{\line(-1,0){.092655371}}
\multiput(45.166,44.194)(.033690626,-.038446949){223}{\line(0,-1){.038446949}}
\multiput(77.334,43.444)(-.033690626,-.038446949){223}{\line(0,-1){.038446949}}
\multiput(12.816,18.031)(.033671751,-.055418091){126}{\line(0,-1){.055418091}}
\multiput(109.684,17.281)(-.033671751,-.055418091){126}{\line(0,-1){.055418091}}
\multiput(17.059,11.049)(.092513137,-.033587572){150}{\line(1,0){.092513137}}
\multiput(105.441,10.299)(-.092513137,-.033587572){150}{\line(-1,0){.092513137}}
\multiput(30.936,6.01)(.132582521,.033707421){118}{\line(1,0){.132582521}}
\multiput(91.564,5.26)(-.132582521,.033707421){118}{\line(-1,0){.132582521}}
\multiput(46.581,9.988)(.033620108,.052599201){163}{\line(0,1){.052599201}}
\multiput(75.919,9.238)(-.033620108,.052599201){163}{\line(0,1){.052599201}}
\multiput(32.085,47.641)(-.0331456,-2.5190679){16}{\line(0,-1){2.5190679}}
\multiput(90.415,46.891)(.0331456,-2.5190679){16}{\line(0,-1){2.5190679}}
\multiput(45.962,10.872)(-.033732751,.0890288429){414}{\line(0,1){.0890288429}}
\multiput(76.538,10.122)(.033732751,.0890288429){414}{\line(0,1){.0890288429}}
\end{picture}
 \caption{\footnotesize Special $4$-clusters in which the
long diagonal is the {\it longest} (main) in $\Pi$: diameter of
$\Pi$ for $n$ even. In this picture the cluster has a single longest
diagonal. Then its contribution to the logarithmic part of the BDS
formula (\ref{Flog}) is the same as in Fig.\ref{logo}, but there are
{\it two} clusters associated with this longest diagonal --
one to its right and the other to its left. } \label{logodi}
\end{figure}
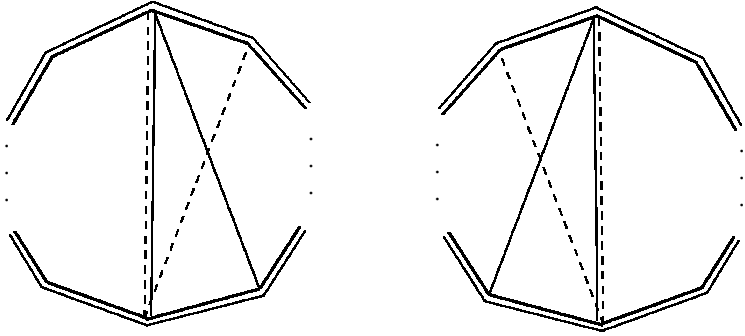

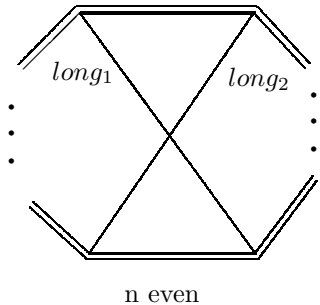
\begin{figure}
\hspace{4cm}
\unitlength 1mm 
\linethickness{0.4pt}
\ifx\plotpoint\undefined\newsavebox{\plotpoint}\fi 
\begin{picture}(59.485,52.326)(0,0)
\thicklines \put(28.079,51.397){\line(1,0){23.204}}
\put(29.12,19.452){\line(1,0){22.372}} \thinlines
\put(20.483,43.905){\line(1,1){7.492}}
\multiput(27.975,51.397)(.03373227207,-.0462301013){691}{\line(0,-1){.0462301013}}
\multiput(51.284,19.452)(.033624454,.0449869){229}{\line(0,1){.0449869}}
\multiput(57.943,44.009)(-.033661765,.036215686){204}{\line(0,1){.036215686}}
\multiput(51.076,51.397)(-.0337222222,-.0492978395){648}{\line(0,-1){.0492978395}}
\multiput(29.224,19.452)(-.03672549,.033666667){204}{\line(-1,0){.03672549}}
\put(18.886,31.76){\circle*{.553}}
\put(59.103,33.306){\circle*{.553}}
\put(18.886,35.472){\circle*{.553}}
\put(59.103,37.019){\circle*{.553}}
\put(18.886,38.937){\circle*{.553}}
\put(59.103,40.483){\circle*{.553}}
\multiput(19.799,44.548)(.03372807,.034114035){228}{\line(0,1){.034114035}}
\put(27.489,52.326){\line(1,0){24.042}}
\multiput(51.53,52.326)(.0337343,-.036719807){207}{\line(0,-1){.036719807}}
\multiput(21.302,25.544)(.037078049,-.033629268){205}{\line(1,0){.037078049}}
\put(28.903,18.65){\line(1,0){22.804}}
\multiput(51.707,18.65)(.033670996,.045151515){231}{\line(0,1){.045151515}}
\put(38.891,13.77){\makebox(0,0)[cc]{n even}}
\put(28.38,43.411){\makebox(0,0)[cc]{$long_1$}}
\put(51.82,42.57){\makebox(0,0)[cc]{$long_2$}}
\end{picture}
 \caption{\footnotesize Special $4$-clusters where the
long diagonals are the {\it longest} (main) in $\Pi$. The case of
even $n$ is shown, when {\it two} long diagonals are diameters. The
contribution of this $4$-cluster to the BDS formula
(\ref{Flog})+(\ref{Flogeven}) is
$\frac{1}{4}\Big(\tau_{l_1}-\tau_{l_2}\Big)^2$ -- very different
from the one in Fig.\ref{logo}. } \label{didi}
\end{figure}

For even $n$  the main diagonal is just the diameter of $\Pi$. When
the largest diagonal of the cluster is a diameter, it appears in two
different $4$-clusters -- to the right and to the left of the
diameter, Figure \ref{logodi}, -- and both clusters should be
included into the sum. Finally, some clusters do not have the
largest diagonal: they contain adjacent {\it main} diagonals
(diameter for even $n$), Figure \ref{didi}. The contribution of such
a cluster is just the square of differences between $\tau$'s for the
two main diagonals, the smaller diagonals do not contribute (they
could coincide with edges if such a cluster happens to be degenerate
-- which is indeed the case for $n=4$ and $n=5$).

For odd $n$  contributing (along with Figure \ref{logo}) are
peculiar $5$-clusters with four {\it main} diagonals, see Figure
\ref{longlongodd}.

\subsection{Asymptotics for nearly light-like diagonals}

\begin{figure}
\hspace{4cm}
\unitlength 1mm 
\linethickness{0.4pt}
\ifx\plotpoint\undefined\newsavebox{\plotpoint}\fi 
\begin{picture}(69.164,62.962)(0,0)
\multiput(50.623,35.752)(-.0337343,.064048309){207}{\line(0,1){.064048309}}
\multiput(43.64,49.01)(-.17481111,.03338889){90}{\line(-1,0){.17481111}}
\multiput(30.117,13.743)(-.06947907,.033711628){215}{\line(-1,0){.06947907}}
\multiput(15.179,20.991)(-.033573643,.117162791){129}{\line(0,1){.117162791}}
\multiput(30.117,13.566)(.16029661,.033711864){118}{\line(1,0){.16029661}}
\multiput(49.032,17.544)(.03314583,.37564583){48}{\line(0,1){.37564583}}
\multiput(28.261,52.811)(.17775556,-.0334){90}{\line(1,0){.17775556}}
\multiput(44.259,49.805)(.033711628,-.064130233){215}{\line(0,-1){.064130233}}
\multiput(51.507,36.017)(-.0336,-.38184){50}{\line(0,-1){.38184}}
\multiput(29.675,12.771)(-.069363229,.033690583){223}{\line(-1,0){.069363229}}
\multiput(14.207,20.284)(-.0335,.124741935){124}{\line(0,1){.124741935}}
\multiput(30.028,12.594)(.159669355,.0335){124}{\line(1,0){.159669355}}
\put(49.827,16.748){\line(0,1){.177}}
\multiput(30.117,12.417)(-.048273,.032182){11}{\line(-1,0){.048273}}
\multiput(30.117,13.566)(.033723192,.0883890274){401}{\line(0,1){.0883890274}}
\multiput(43.64,49.01)(.033563291,-.198594937){158}{\line(0,-1){.198594937}}
\multiput(48.943,17.632)(-.33764,.03359){100}{\line(-1,0){.33764}}
\multiput(15.179,20.991)(.0820600462,.0336812933){433}{\line(1,0){.0820600462}}
\thicklines
\multiput(50.711,35.575)(-.0337269737,-.0361990132){608}{\line(0,-1){.0361990132}}
\multiput(30.028,13.743)(.0337269737,.0360526316){608}{\line(0,1){.0360526316}}
\multiput(30.47,13.655)(.033735,.0362383333){600}{\line(0,1){.0362383333}}
\multiput(50.623,34.956)(-.0337291312,-.035988075){587}{\line(0,-1){.035988075}}
\put(39.642,45.884){\line(1,0){1.9325}}
\put(42.615,40.904){\line(1,0){2.007}}
\put(43.655,41.87){\line(0,-1){1.932}}
\put(16.601,23.958){\line(1,0){1.932}}
\put(21.655,21.728){\line(1,0){2.007}}
\put(22.621,22.694){\line(0,-1){1.858}}
\put(9.62,45.182){\makebox(0,0)[cc]{$t_{b-1}^{[3]}$}}
\put(21.548,62.962){\makebox(0,0)[cc]{$t_{b}^{[3]}$}}
\put(69.164,53.187){\makebox(0,0)[cc]{$t_{b+1}^{[2]}$}}
\put(68.113,19.446){\makebox(0,0)[cc]{$t_{b}^{[2]}$}}
\put(59.914,8.935){\makebox(0,0)[cc]{$t_{b-1}^{[2]}$}}
\put(9.67,18.5){\makebox(0,0)[cc]{$b-1$}}
\put(29.642,10.091){\makebox(0,0)[cc]{$b$}} \thinlines
\put(26.383,26.173){\vector(1,-4){.07}}\qbezier(14.611,44.357)(22.809,43.779)(26.383,26.173)
\put(38.051,36.579){\vector(3,-2){.07}}\qbezier(21.233,56.55)(21.022,46.354)(38.051,36.579)
\put(41.73,24.491){\vector(-3,2){.07}}\qbezier(63.488,19.866)(49.14,19.866)(41.73,24.491)
\put(40.889,17.869){\vector(-2,3){.07}}\qbezier(54.448,9.67)(44.62,11.037)(40.889,17.869)
\put(46.039,36.999){\vector(-1,0){.07}}\qbezier(65.8,47.195)(60.755,38.629)(46.039,36.999)
\end{picture}
\caption{\footnotesize The single $4$-cluster
contributing to the asymptotic (\ref{asymp}) of $F_n^{(1)}$ as
$t_n^{[2]}\longrightarrow 0$ (the corresponding diagonal is shown
by thick line and becomes light-like in the limit). This is
the only type of singularities that $F_n^{(1)}$ has at small
$t_a^{[r]}$. Other clusters can also have singularities, but
they cancel between dilogarithmic and logarithmic contributions to
(\ref{Fdilog}). $t_a^{[r]}$ with $r=2$ are distinguished because
they never appear in denominators in arguments of dilogarithms.}
\label{t2sing}
\end{figure}
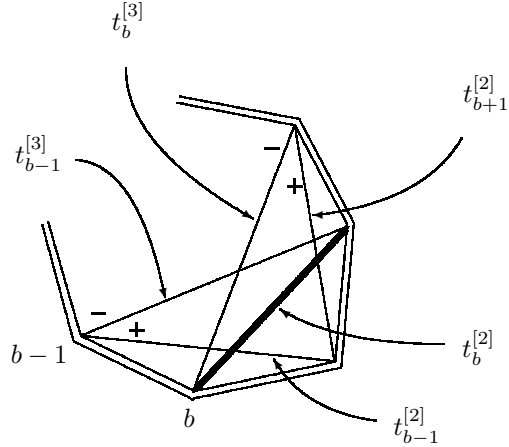

If some $t^{[r]}_b$ is much smaller than all other $t$'s, then the
arguments of the corresponding dilogarithms are large, so that the
dilogarithms become squares of logarithms and cancel against the
logarithmic terms. Indeed, let us take some $t_b^{[r]} \to 0$
 -- this means that the corresponding $\tau_b^{[r]} \to -\infty$ and see
what happens to (\ref{Fdilog}). If it appears in the numerator of the
argument of a dilogarithm, nothing happens because $Li_2(z)$ is regular
near $z=1$. However, if it appears in the denominator, the dilogarithm blows
up and gets expressed through ordinary logarithms via $Li_2(z)
\sim -\frac{1}{2}(\log z)^2$ for $|z|\rightarrow \infty$. Taking
this into account, we can write (note that our $t_b^{[r]}$ appears
in dilogarithms twice: also as $t_{b+r}^{[n-r]}$, hence the extra
factor of two in the dilogarithmic contributions in the first line of (\ref{cance}) below.) \be
\frac{2}{8}\Big(\tau_b^{[r-1]}+\tau_{b-1}^{[r+1]}-
\underline{\tau_b^{[r]}} - \tau_{b-1}^{[r]}\Big)^2 +
\frac{2}{8}\Big(\tau_{b+1}^{[r-1]}+\tau_{b}^{[r+1]}
 - \tau_{b+1}^{[r]} - \underline{\tau_b^{[r]}}\Big)^2 - \nn \\
-\frac{1}{2}\Big(\tau_b^{[r+1]}-\underline{\tau_b^{[r]}}\Big)
\Big(\tau_b^{[r+1]}-\tau_{b+1}^{[r]}\Big)
-\frac{1}{2}\Big(\underline{\tau_b^{[r]}}-\tau_{b}^{[r-1]}\Big)
\Big(\underline{\tau_b^{[r]}}-\tau_{b+1}^{[r-1]}\Big) -
\frac{1}{2}\Big(\tau_{b-1}^{[r+1]}-\tau_{b-1}^{[r]}\Big)
\Big(\tau_{b-1}^{[r+1]} - \underline{\tau_b^{[r]}}\Big)
\label{cance} \ee
 It is easy to see that all terms with the
underlined quantity cancel. As a result, this $t^{[r]}_b$ completely
drops out from $F_n^{(1)}$ for $r\geq 3$. For $r=2$ one should make
a separate calculation (since we ignored the restriction $r>2$ in
(\ref{cance}) and kept terms with $r-1$), and it is easy to see
that, in contrast to $t_b^{[r]}$ with $r>2$, the small $t_b^{[2]}$
leads to a singularity in $F_n^{(1)}$ of the form, Figure
\ref{t2sing}:
\be F_n^{(1)} = -{1\over 2}
\underline{\tau_b^{[2]}}\Big(\tau_{b-1}^{[2]}+\tau_{b+1}^{[2]}
-\tau_{b-1}^{[3]}-\tau_b^{[3]}\Big) + {\rm \ terms\ finite\ as\ }
\tau_b^{[r]} \rightarrow -\infty \label{asymp} \ee Eq.(\ref{asymp})
is not directly applicable also when $t_b^{[r]} \rightarrow 0$ with
maximal $r=\big[n/2\big]$, such $t^{[r]}$s enter (\ref{Fdilog}) in a
more sophisticated way; moreover, they are different for even and
odd $n$, see Figures \ref{didi}-\ref{longlongodd}. Still, it is easy
to demonstrate that no singularity in $F_n^{(1)}$ occurs when
$t^{[r]} \rightarrow 0$ with $r=\big[n/2\big]$ -- unless
$\big[n/2\big] = 2$, i.e. $n=4$ or $n=5$. Note that at $n=4$
eq.(\ref{asymp}) also requires a correction: \be F_4^{(1)} =
\frac{1}{4}\Big(\log\frac{s}{t}\Big)^2 = \frac{1}{4} (\log s)^2 -
\frac{1}{2} {\log s} \ {\log t} \ +\  {\rm terms\ finite\ as\ } \log s
\rightarrow -\infty \ee coming from Figure \ref{didi}.

To analyse the singularities at large $t$ one has to take into account all relations among different $t$'s and is
beyond the scope of this paper.

\subsection{Boxes: two dilogarithmic representations}

We now turn to the analysis of individual box contributions to
$F_n^{(1)}$. Denoting ${\bf p}^2={\bf q}^2=0$, $s=({\bf p}+{\bf
P})^2$, $t=({\bf q}+{\bf P})^2$ for a particular box, associated
with the $4$-cluster in Figure \ref{clust}, we obtain for the
easy-box Feynman diagram, Figure \ref{boxdi}.A, with a massless
scalar field in the loop:
\be \label{2Fme} F^{2me}(s,t;{\bf P}^2,{\bf Q}^2) \sim \int_0^1
\frac{d\beta_1d\beta_2d\beta_3d\beta_4
\delta(1-\beta_1-\beta_2-\beta_3-\beta_4)} {(-s\beta_1\beta_3 -{\bf
P}^2\beta_3\beta_4 - t\beta_2\beta_4 - {\bf
Q}^2\beta_1\beta_2)^{2+\epsilon}} \ee The calculation of this
integral is somewhat tedious. As a result, there are two essentially
different formulas for this quantity: the first one \cite{BDK} is
convenient to make contact with (\ref{Fdilog2}), the second one
\cite{DN} with (\ref{contdilog}), which will be studied in the
next section \ref{contin} and, furthermore, with the contour integral
(\ref{loopin}). It is actually here, at (\ref{2Fme}), that the roads
split: one leads to the BDS formula in the form (\ref{Fdilog}) and the
other to its geometric representation (\ref{loopin}).

The BDK formula (eq.(4.44) of \cite{BDK}) states that
$$
F^{2me}(s,t;{\bf P}^2,{\bf Q}^2) = \frac{2i}{4\pi^2}
\frac{\Gamma(1-\epsilon)\Gamma^2(1+\epsilon)}
{\Gamma(1-2\epsilon)}\frac{1} {st-{\bf P}^2{\bf Q}^2}
\left\{\frac{1}{\epsilon^2}\left[ {1\over (-s)^\epsilon} +{1\over (
-t)^\epsilon} - {1\over (-{\bf P}^2)^\epsilon} - {1\over (-{\bf
Q}^2)^\epsilon}\right] + \right. $$
\be \left. + Li_2\left(1-\frac{{\bf P}^2{\bf Q}^2}{st}\right) -
Li_2\left(1-\frac{{\bf P}^2}{s}\right) - Li_2\left(1-\frac{{\bf
P}^2}{t}\right) - Li_2\left(1-\frac{{\bf Q}^2}{s}\right) -
Li_2\left(1-\frac{{\bf Q}^2}{t}\right) \right\} \label{BDK} \ee The
first dilogarithm is exactly the same as in (\ref{Fdilog2}), other
dilogarithms cancel non-trivially between different 4-clusters in
the sum (\ref{2me}). For $n=4$  when both ${\bf P}^2={\bf Q}^2=0$
and $n=5$ when either ${\bf P}^2=0$ or ${\bf Q}^2=0$ the first term
disappears and there are no dilogarithms in the answer at all (for
$n=5$ this still involves non-trivial cancelation of other
dilogarithms between different 4-clusters).

The DN formula (eq.(71) of \cite{DN}) states that
$$
F^{2me}(s,t;{\bf P}^2,{\bf Q}^2) = \frac{2i}{4\pi^2}
\frac{\Gamma(1-\epsilon)\Gamma^2(1+\epsilon)}
{\Gamma(1-2\epsilon)}\frac{1}{st-{\bf P}^2{\bf Q}^2}\cdot
$$ $$ \cdot
\left\{\frac{1}{\epsilon^2}\left[
\left(\frac{-s-i\varepsilon}{4\pi\mu^2}\right)^{-\epsilon} +
\left(\frac{-t-i\varepsilon}{4\pi\mu^2}\right)^{-\epsilon} -
\left(\frac{-{\bf P}^2-i\varepsilon}{4\pi\mu^2}\right)^{-\epsilon} -
\left(\frac{-{\bf
Q}^2-i\varepsilon}{4\pi\mu^2}\right)^{-\epsilon}\right] + \right. $$
\be \left. + Li_2\big(1-a(s+i\varepsilon)\big) +
Li_2\big(1-a(t+i\varepsilon)\big) - Li_2\big(1-a({\bf
P}^2+i\varepsilon)\big) - Li_2\big(1-a({\bf
Q}^2+i\varepsilon)\big)\right\} \label{DN} \ee with
$$a = \frac{s+t-{\bf P}^2-{\bf Q}^2}{st-{\bf P}^2{\bf Q}^2}$$

The equivalence of (\ref{BDK}) and (\ref{DN}) shown in \cite{DN}, is based on basic
dilogarithmic identities such as
\be Li_2(z) = \sum_{k=1}^\infty \frac{z^k}{k^2}, \ \ \ Li_2(0) = 0 ,
\ \ \
Li_2(1) = \zeta_2 = \frac{\pi^2}{6}, \nn \\
Li_2(z) + Li_2(1-z)  = -\log (1-z)\log z - \frac{\pi^2}{6},\nn \\
Li_2(z) + Li_2(1/z) = -\frac{1}{2}\big(\log (-z)\big)^2 -
\frac{\pi^2}{6} \ee The rather lengthy proof is presented in
Appendix A of \cite{DN}. The equivalence of (\ref{BDK}) and
(\ref{Fdilog}) is one of the subjects of \cite{BDS}. We skip these
derivations here.

\subsection{Double integral \cite{Pol1} along a polygon \cite{BHT}
\label{contin}}

As shown in \cite{BHT}, the sum (\ref{2me}) of the easy-box
diagrams, if represented in the form (\ref{DN}), is nothing but
the double contour integral (\ref{loopin}) along the auxiliary
polygon $\Pi$: \be F_n^{(1)} = (\ref{2me}) \
\stackrel{(\ref{DN})}{=}\ \oint_{\Pi}\oint_{\Pi} \frac{dy^\mu
dy'_\mu}{({\bf y}-{\bf y}')^{2+\epsilon}} \ee Indeed, (\ref{loopin})
is a sum of contributions coming from pairs of segments in $\Pi$.
There are three different types of pairs, which we briefly consider
following \cite{BHT}. In this section $\tau_p$ and $\tau_q$
parameterize the segments ${\bf p}$ and ${\bf q}$.

\subsubsection{One null-segment p}

No contribution because $dydy' = {\bf p}^2 d\tau d\tau'$ in the
numerator vanishes for ${\bf p}^2=0$.

\subsubsection{Two adjacent null-segments p and q}

\be \int_0^1\int_0^1 \frac{({\bf pq}) d\tau_p d\tau_q} {(\tau_p {\bf
p} +\tau_q {\bf q})^{2+\epsilon}} = ({\bf pq})^{\epsilon/2}
\int_0^1\int_0^1 \frac{d\tau_p
d\tau_q}{(\tau_p\tau_q)^{1+\epsilon/2}} = \frac{({\bf
pq})^{\epsilon/2}}{\epsilon^2} \ee

\subsubsection{Two non-adjacent null-segments p and q}

$2{\bf pq} = u = {\bf P}^2+{\bf Q}^2-s-t$
$$
\int_0^1\int_0^1 \frac{({\bf pq}) d\tau_p d\tau_q} {\Big(\tau_p {\bf
p} + {\bf P} + (1-\tau_q){\bf q}\Big)^{2+\epsilon}} \
\stackrel{\tau_q \rightarrow 1-\tau_q}{=}\
$$
$$
= \frac{1}{2}\int_0^1\int_0^1 \frac{({\bf P}^2+{\bf Q}^2-s-t)\
d\tau_p d\tau_q} {\Big({\bf P}^2+(s-{\bf P}^2)\tau_p + (t-{\bf
P}^2)\tau_q + ({\bf P}^2+{\bf
Q}^2-s-t)\tau_p\tau_q\Big)^{1+\frac{\epsilon}{2}}} =
$$
$$
= \frac{1}{2}\int_0^1 \frac{({\bf P}^2+{\bf Q}^2-s-t)\ d\tau_p}
{({\bf P}^2+{\bf Q}^2-s-t)\tau_p + t -{\bf P}^2}\ \log \frac{t +
({\bf Q}^2-t)\tau_p}{{\bf P}^2+(s-{\bf P}^2)\tau_p}+ O(\epsilon) \
=
$$
\be =Li_2(1-as) + Li_2(1-at) -  Li_2(1-a{\bf P}^2) - Li_2(1-a{\bf
Q}^2) + O(\epsilon) \label{contdilog} \ee with \be a = \frac{{\bf
P}^2+{\bf Q}^2-s-t}{{\bf P}^2{\bf Q}^2-st} \ee Since this
contribution is finite, we do not preserve its
$\epsilon$-dependence.


\section{Minimal surfaces with non-planar polygon boundaries}
\setcounter{equation}{0}

In contrast to the situation in AdS spaces, the study of minimal surfaces in flat space
is an old branch of mathematics with close links to the
theory of Riemann surfaces and many impressive results. The first
non-trivial minimal surfaces, helicoid and catenoid, were found by
Meusnier in 1776 \cite{Meu}. Among the next pioneers in the field
were Scherk (1834) and Schwarz (1890). In particular, Schwarz solved
the problem of finding the minimal bounding surface of a skew quadrilateral. For a survey of these old results
and a presentation of the state-of-the-art in the theory of minimal surfaces in
Euclidean space see \cite{Oss,Wolf}.

In string theory the issue of minimal surfaces was first raised by D. Gross
and P. Mende in \cite{GM} in an application to
scattering amplitudes at high energies: the saddle-point approximation to the Veneziano
and Koba-Nielsen formulas for scattering amplitudes is associated with minimal surfaces in
Euclidean (Minkowskian) space of appropriate dimension. They studied the case $n=4$ with the vectors ${\bf p}_a$
lying in a $2+1$-dimensional subspace $R^3_{++-}$.

\subsection{Generalities}

\subsubsection{Ambiguities in the formulation of the problem:}

The problem of finding a minimal area, although it sounds as being a
well-formulated problem, it contains a few ambiguities due to the
necessity of regularizing the infinite area that emerges. While {\it
before} regularization all the formulations are equivalent, this is
not necessarily the case {\it after} the regularization is performed.

The various formulations of this problem use:

\begin{itemize}
\item {\bf Different actions:} Polyakov/Green-Schwarz
($\sigma$-model) and Nambu-Goto. The minimal area formulation is in
terms of the Nambu-Goto action, while motivation for the AdS
geometry (with the fifth coordinate in the target-space associated
with the Liouville field in the first-quantized formalism) is more
transparent in the $\sigma$-model formulation. According to
\cite{DGO}, the Nambu-Goto action with appropriate boundary terms
should be used.
\item
{\bf Different regularizations:} shift and dimensional.
Shift from the AdS boundary is better conceptually, it is compatible with
the RG interpretation of AdS, it preserves supersymmetry and
integrability; an unusual version of dimensional regularization is
used in \cite{AM}. For any regularization, the problem is that
solutions of regularized equations are much more difficult to find
and one needs to get the most from solutions of the {\it
non-regularized} ones: in this respect the approach of
\cite{AM} was successful, but, as often happens in such
circumstances, it is hard to generalize.
\item
{\bf Different descriptions of the target-space:} original
($x^\mu,z$) coordinates and $T$-dual coordinates ($y^\mu,r$) can be
naturally used, the metric in both cases has the Poincare form,
$(dz^2 + dx^\mu dx_\mu)/z^2$ and $(dr^2 + dy^\mu
dy_\mu)/r^2 = (dr^2-dy_0^2+dy_1^2+dy_2^2+dy_3^2)/r^2$, but
boundary conditions are imposed at $z\rightarrow \infty$ and
$r\rightarrow 0$ and they are Dirichlet conditions in the $(r,{\bf
y})$ case. According to \cite{AM}, transition to dual variables also
eliminates boundary terms, which had to be added in the Nambu-Goto
action \cite{DGO}.
\item
{\bf Different coordinates at the boundary:} say, $y_\pm
= y_0\pm y_1, y_2, y_3$ or $y_0,y_1,y_2,y_3$ or
$y_0,\xi_1,\xi_2,y_3$ etc. This is a purely technical issue, but many
papers differ mostly in these choices.
\end{itemize}

\subsubsection*{Literature:} Most considerations in the literature are devoted to a single cusp (see, however,
\cite{DGO}) formed by two null-lines \cite{DGO,Kru,Mak,Buch}. In
\cite{Kru} this was done in rapidity coordinates, and actually
four cusps were implicitly involved. In \cite{AM} this fact was
exploited to construct a solution with rhombic projection of the boundary on the $(y_1,y_2)$ plane.

\subsubsection{The Nambu-Goto action with $y_3=0$ in $y_1,y_2$
projection:}

In this case it is convenient to choose the gauge with the two
world-sheet coordinates identified with the coordinates $(y_1,y_2)$.
Then, the Nambu-Goto action is\be
S_{NG}=\int\int\frac{dy_1dy_2}{r^{2}}\sqrt{H}, \ee \be H=
1-(\partial_1 y)^2 - (\partial_2 y)^2 + (\partial_1
r)^2\big[1-(\partial_2 y)^2\big] + (\partial_2
r)^2\big[1-(\partial_1 y)^2\big] + 2\partial_1r \partial_2
r\partial_1 y\partial_2 y
\ee The equations of motion are \be
\partial_i\frac{\partial_i y}{r^2\sqrt{H}} +
\partial_2\frac{\partial_1 r
(\partial_1r\partial_2y -
\partial_2r\partial_1y)}{r^2\sqrt{H}} -
\partial_1\frac{\partial_2 r (\partial_1r\partial_2y -
\partial_2r\partial_1y)}{r^2\sqrt{H}} = 0 \ee \be
\partial_i\frac{\partial_i r}{r^2\sqrt{H}} +
\partial_2\frac{\partial_1 y
(\partial_1r\partial_2y -
\partial_2r\partial_1y)}{r^2\sqrt{H}} -\partial_1\frac{\partial_2 {\bf
y} (\partial_1r\partial_2y -
\partial_2r\partial_1y)}{r^2\sqrt{H}} +\frac{2}{r^2\sqrt{H}} = 0 \ee

When approaching the boundary given by the segment perpendicular to
a vector $\vec q$, $\vec q\vec y = q_1y_1+q_2y_2=1$, the coordinate $r$
behaves as $r \sim \sqrt{\vec q\vec y -1}$. Poles in $\partial_\bot
r$ are canceled by zeroes of $[1-(\partial_{||}y)^2]$. These zeroes
arise if boundary segments are null-vectors.

\subsubsection{Nambu-Goto vs $\sigma$-model action \label{NGA}}

Formally at the classical level the two formulations should be
equivalent, but with a non-trivial $2d$ metric. In order to put the
$2d$ metric into the conformal gauge, one needs to make a general
coordinate transformation.

It is worth noting, though, that only for the $\sigma$-model action one
can perform a $T$-duality transformation
$(r,{\bf x}) \rightarrow (z,{\bf y})$ with $z=1/r$ and $\partial_i
{\bf x} = z^2\epsilon_{ij}\partial_j {\bf y}$. Indeed, this
transformation does not preserve the shape of the  $2\times 2$
tensor
(induced metric) 
\be G_{ij}(r,{\bf x}) = \frac{\partial_i r\partial_j r +
\partial_i{\bf x}\partial_j{\bf x}} {r^2} \ee
Moreover, it changes the determinant $\det G_{ij}(r,{\bf x})
\neq \det G_{ij}(z,{\bf y})$, thus, the Nambu-Goto action is
formally not $T$-invariant. What is invariant, is the trace:
$\delta^{ij}G_{ij}(r,{\bf x}) = \delta^{ij}G_{ij}(z,{\bf y})$, and
this is enough to guarantee $T$-invariance of the $\sigma$-model action and of the
associated Polyakov and Green-Schwarz actions.

\subsubsection{Equations of motion for the AdS $\sigma$-model}

The equations of motion for the $\sigma$-model action \be S_\sigma =
\int L d^2u, \ \ \ \ L=\frac{(\partial r)^2+(\partial {\bf
y})^2}{r^2} \ee (in appropriate coordinates $(u_1,u_2)$ on
the world sheet) are: \be
\partial \left(\frac{\partial r}{r^2}\right) = -\frac{L}{r} \nn \\
\partial \left(\frac{\partial {\bf y}}{r^2}\right) = 0
\ee and in coordinates $z = 1/r$, ${\bf v} = {\bf
y}/r$ they acquire the form \be
\Delta z = zL, \nn \\
\Delta {\bf v} = {\bf v} L\nn \\
z^2L-(\partial z)^2 = (z\partial {\bf v} - {\bf v}\partial z)^2
\label{eqm} \ee For $L = const$ solutions of the first two equations
are sums of exponentials \be
z=\sum_{a=1}^n z_a e^{\vec k_a \vec u}, \nn \\
{\bf v} = \sum_{a=1}^n {\bf v}_a e^{\vec k_a \vec u}
\label{exposol}
\ee
with ${\vec k}_a^2 = L$.

\subsubsection{Boundary conditions}

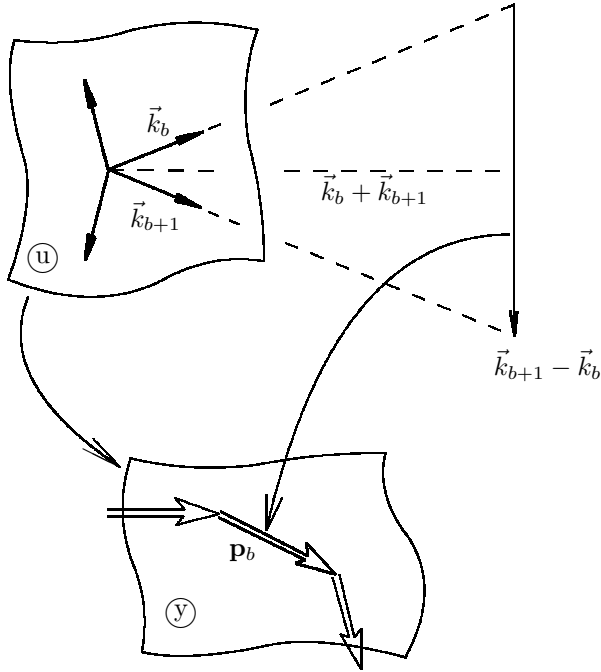
\begin{figure}
\hspace{3cm}
\unitlength 1mm 
\linethickness{0.4pt}
\ifx\plotpoint\undefined\newsavebox{\plotpoint}\fi 
\begin{picture}(90.38,100.277)(0,0)
\put(85.903,100.181){\line(0,-1){44}} \thicklines
\multiput(31.903,78.056)(-.033602151,.130376344){93}{\line(0,1){.130376344}}
\multiput(31.903,77.931)(-.03370787,-.13202247){89}{\line(0,-1){.13202247}}
\multiput(31.908,78.22)(.084266667,.033586667){150}{\line(1,0){.084266667}}
\multiput(44.548,83.258)(-.06364,-.0336){50}{\line(-1,0){.06364}}
\multiput(41.366,81.578)(-.032182,.080364){11}{\line(0,1){.080364}}
\multiput(41.012,82.462)(.1515238,.0336667){21}{\line(1,0){.1515238}}
\multiput(44.194,83.169)(-.108037,-.0327407){27}{\line(-1,0){.108037}}
\multiput(41.277,82.285)(.14375,.03325){8}{\line(1,0){.14375}}
\multiput(42.427,82.551)(-.0505238,-.0337143){21}{\line(-1,0){.0505238}}
\multiput(41.366,81.843)(.0635313,.0331563){32}{\line(1,0){.0635313}}
\multiput(31.997,78.22)(.081313333,-.033593333){150}{\line(1,0){.081313333}}
\multiput(44.194,73.181)(-.05669811,.03335849){53}{\line(-1,0){.05669811}}
\multiput(41.189,74.949)(-.032091,-.088364){11}{\line(0,-1){.088364}}
\multiput(40.836,73.977)(.143625,-.0331667){24}{\line(1,0){.143625}}
\multiput(44.283,73.181)(-.1158276,.0335517){29}{\line(-1,0){.1158276}}
\multiput(40.924,74.154)(.176833,-.0295){6}{\line(1,0){.176833}}
\multiput(41.985,73.977)(-.0420952,.0336667){21}{\line(-1,0){.0420952}}
\multiput(41.101,74.684)(.0635313,-.0331563){32}{\line(1,0){.0635313}}
\multiput(28.726,90.152)(.0325789,-.1907368){19}{\line(0,-1){.1907368}}
\multiput(29.345,86.528)(.1105,.033125){8}{\line(1,0){.1105}}
\multiput(30.229,86.793)(-.0334,.07464444){45}{\line(0,1){.07464444}}
\multiput(28.726,90.152)(.03345946,-.08837838){37}{\line(0,-1){.08837838}}
\multiput(29.964,86.882)(-.0315714,.0378571){14}{\line(0,1){.0378571}}
\put(29.522,87.412){\line(0,-1){.795}}
\multiput(29.522,86.616)(-.0325789,.1488947){19}{\line(0,1){.1488947}}
\multiput(28.903,66.199)(.0331562,.0966563){32}{\line(0,1){.0966563}}
\multiput(30.167,69.111)(-.03369231,-.07546154){39}{\line(0,-1){.07546154}}
\multiput(28.853,66.168)(.033455,.305818){11}{\line(0,1){.305818}}
\multiput(29.221,69.532)(.090818,-.033455){11}{\line(1,0){.090818}}
\multiput(30.22,69.164)(-.098625,.032875){8}{\line(-1,0){.098625}}
\multiput(29.431,69.427)(-.033455,-.224636){11}{\line(0,-1){.224636}}
\thinlines \qbezier(19.551,95.442)(27.487,94.391)(34.792,96.703)
\qbezier(36.369,14.085)(37.525,20.434)(34.981,26.278)
\qbezier(34.792,96.703)(43.148,98.753)(51.715,96.808)
\qbezier(34.981,26.278)(32.727,32.963)(34.866,39.816)
\qbezier(51.715,96.808)(49.35,90.028)(50.559,81.777)
\qbezier(34.866,39.816)(42.323,37.924)(51.4,38.891)
\qbezier(50.559,81.777)(52.714,73.684)(52.556,66.01)
\qbezier(51.4,38.891)(60.303,40.615)(68.743,40.489)
\qbezier(52.556,66.01)(45.514,67.219)(38.471,63.172)
\qbezier(68.743,40.489)(67.414,34.855)(71.865,29.221)
\qbezier(38.471,63.172)(29.694,59.178)(18.605,63.593)
\qbezier(71.865,29.221)(76.259,22.2)(71.403,13.328)
\qbezier(18.605,63.593)(22.126,73.368)(20.812,79.36)
\qbezier(71.403,13.328)(60.65,16.145)(54.059,15.094)
\qbezier(20.812,79.36)(17.711,89.608)(19.446,95.442)
\qbezier(54.059,15.094)(42.786,12.613)(36.369,14.001)
\multiput(53.432,86.858)(.0814035,.0334561){21}{\line(1,0){.0814035}}
\multiput(56.851,88.263)(.0814035,.0334561){21}{\line(1,0){.0814035}}
\multiput(60.27,89.668)(.0814035,.0334561){21}{\line(1,0){.0814035}}
\multiput(63.689,91.073)(.0814035,.0334561){21}{\line(1,0){.0814035}}
\multiput(67.107,92.478)(.0814035,.0334561){21}{\line(1,0){.0814035}}
\multiput(70.526,93.883)(.0814035,.0334561){21}{\line(1,0){.0814035}}
\multiput(73.945,95.289)(.0814035,.0334561){21}{\line(1,0){.0814035}}
\multiput(77.364,96.694)(.0814035,.0334561){21}{\line(1,0){.0814035}}
\multiput(80.783,98.099)(.0814035,.0334561){21}{\line(1,0){.0814035}}
\multiput(84.202,99.504)(.0814035,.0334561){21}{\line(1,0){.0814035}}
\multiput(55.429,68.463)(.0803651,-.0328122){21}{\line(1,0){.0803651}}
\multiput(58.804,67.085)(.0803651,-.0328122){21}{\line(1,0){.0803651}}
\multiput(62.179,65.706)(.0803651,-.0328122){21}{\line(1,0){.0803651}}
\multiput(65.555,64.328)(.0803651,-.0328122){21}{\line(1,0){.0803651}}
\multiput(68.93,62.95)(.0803651,-.0328122){21}{\line(1,0){.0803651}}
\multiput(72.305,61.572)(.0803651,-.0328122){21}{\line(1,0){.0803651}}
\multiput(75.681,60.194)(.0803651,-.0328122){21}{\line(1,0){.0803651}}
\multiput(79.056,58.816)(.0803651,-.0328122){21}{\line(1,0){.0803651}}
\multiput(82.431,57.438)(.0803651,-.0328122){21}{\line(1,0){.0803651}}
\multiput(45.233,72.562)(.0803824,-.0324559){17}{\line(1,0){.0803824}}
\multiput(47.966,71.458)(.0803824,-.0324559){17}{\line(1,0){.0803824}}
\multiput(45.233,83.599)(.0775952,.0325476){21}{\line(1,0){.0775952}}
\put(55.324,78.028){\line(1,0){1.9052}}
\put(59.134,78.028){\line(1,0){1.9052}}
\put(62.944,78.028){\line(1,0){1.9052}}
\put(66.755,78.028){\line(1,0){1.9052}}
\put(70.565,78.028){\line(1,0){1.9052}}
\put(74.376,78.028){\line(1,0){1.9052}}
\put(78.186,78.028){\line(1,0){1.9052}}
\put(81.996,78.028){\line(1,0){1.9052}}
\put(31.884,78.133){\line(1,0){1.9971}}
\put(35.878,78.106){\line(1,0){1.9971}}
\put(39.872,78.08){\line(1,0){1.9971}}
\put(43.866,78.054){\line(1,0){1.9971}}
\put(23.194,66.413){\circle{3.967}}
\put(41.599,19.237){\circle{3.967}}
\put(23.194,66.518){\makebox(0,0)[cc]{u}} \thicklines
\multiput(85.351,59.388)(.0332105,-.177){19}{\line(0,-1){.177}}
\multiput(85.982,56.025)(.0315,.3363){10}{\line(0,1){.3363}}
\thinlines \put(31.849,33.005){\line(1,0){10.196}}
\multiput(42.045,33.005)(-.03321053,.03873684){38}{\line(0,1){.03873684}}
\multiput(40.783,34.477)(.09344444,-.03336508){63}{\line(1,0){.09344444}}
\multiput(46.67,32.375)(-.13138636,-.03345455){44}{\line(-1,0){.13138636}}
\multiput(40.889,30.903)(.0328438,.036125){32}{\line(0,1){.036125}}
\put(41.94,32.059){\line(-1,0){10.091}}
\multiput(62.647,24.386)(.03328333,-.14015){60}{\line(0,-1){.14015}}
\multiput(64.644,15.977)(.0326207,.0398621){29}{\line(0,1){.0398621}}
\put(65.59,17.133){\line(0,-1){5.36}}
\multiput(65.485,11.773)(-.03332927,.05639024){82}{\line(0,1){.05639024}}
\multiput(62.752,16.397)(.0889231,-.0323077){13}{\line(1,0){.0889231}}
\multiput(63.908,15.977)(-.03343939,.12104545){66}{\line(0,1){.12104545}}
\thicklines
\multiput(46.88,32.69)(.066068571,-.033634286){175}{\line(1,0){.066068571}}
\multiput(58.442,26.804)(-.03,.255143){7}{\line(0,1){.255143}}
\multiput(58.232,28.59)(.033525862,-.036241379){116}{\line(0,-1){.036241379}}
\multiput(62.121,24.386)(-.12181818,.03345455){44}{\line(-1,0){.12181818}}
\multiput(56.761,25.858)(.1050769,.0323077){13}{\line(1,0){.1050769}}
\multiput(58.127,26.278)(-.068414201,.033585799){169}{\line(-1,0){.068414201}}
\put(41.622,19.25){\makebox(0,0)[cc]{y}}
\put(49.798,27.277){\makebox(0,0)[cc]{${\bf p}_b$}}
\put(38.649,84.731){\makebox(0,0)[cc]{${\vec k}_b$}}
\put(38.203,71.947){\makebox(0,0)[cc]{${\vec k}_{b+1}$}}
\put(67.339,75.515){\makebox(0,0)[cc]{${\vec k}_b +{\vec k}_{b+1}$}}
\put(90.38,52.176){\makebox(0,0)[]{${\vec k}_{b+1} - {\vec k}_b$}}
\thinlines \qbezier(85.177,69.568)(62.061,69.568)(52.919,30.325)
\qbezier(21.257,61.244)(16.649,49.946)(33.446,38.649)
\put(28.59,41.099){\line(2,-1){4.835}}
\multiput(33.425,38.681)(-.033715043,.035698281){106}{\line(0,1){.035698281}}
\put(52.976,35.002){\line(0,-1){4.73}}
\multiput(52.976,30.272)(.03336876,.06840596){63}{\line(0,1){.06840596}}
\end{picture}
 \caption{{\footnotesize Vectors $\vec k_a$ lie in the
Euclidean $2d$ $u$-plane. They all have the same length $\vec
k_a^2 = L=2$. Far at infinity along the bisector $\vec k_b+\vec
k_{b+1}$ and orthogonal to it is a line, directed along $\vec k_{b+1}-
\vec k_b$. This is mapped by the fields $z$ and ${\bf v}$ onto a
vector ${\bf p}_b$ -- an edge of the polygon $\Pi$ on the
boundary ($z=\infty$) of the AdS target space. }}\label{2dpolyg}
\end{figure}

Since all vectors $\vec k_a$ have equal lengths, the boundary
conditions can also be easily satisfied (see Figure \ref{2dpolyg}):
when $(\vec k_b + \vec k_{b+1})\vec u \longrightarrow \infty$ only
two terms with $a=b-1$ and $a=b$ contribute, and the dependence on
the orthogonal variable $t_b = e^{(\vec k_{b+1}-\vec k_b)\vec u/2}$
gets simple: \be {\bf y} = \frac{{\bf v}_{b+1}t_b + {\bf v}_b
t_b^{-1}} {z_{b+1}t_b + z_b t_b^{-1}} \ee and implies linear
relations between components of the $4d$ vector ${\bf y}$: \be
(z_{b+1}v_b^\nu - z_bv_{b+1}^\nu) y^\mu - (z_{b+1}v_b^\mu -
z_bv_{b+1}^\mu) y^\nu = (v_{b+1}^\mu v_b^\nu - v_b^\mu v_{b+1}^\nu)
\ee As $t_b$ varies from $0$ to $\infty$, the vector ${\bf y}$
changes from $\frac{{\bf v}_b}{z_b}$ to $\frac{{\bf
v}_{b+1}}{z_{b+1}}$, and the boundary conditions imply that this
change is exactly the $b$-th external momentum ${\bf p}_b$: \be
\Delta_b {\bf y} = \frac{{\bf v}_{b+1}}{z_{b+1}} - \frac{{\bf
v}_b}{z_b} = {\bf p}_b \label{boco1} \ee Equivalently, the boundary
conditions are: \be z_b{\bf v}_{b+1} - z_{b+1}{\bf v}_b = z_bz_{b+1}
{\bf p}_b \label{boco} \ee We remind that, in our notation, $b+n\equiv b$.

\subsubsection{The third equation: a problem}

While the first two equations in (\ref{eqm}) are easily satisfied by
the ansatz (\ref{exposol}) with $L=const$, this is not generally
true for the third equation. Indeed, after substitution of
(\ref{exposol}) it becomes \be \sum_{a,b} z_a z_b \Big(L - (\vec
k_a\vec k_b)\Big) E_{a+b} = \sum_{\stackrel{a<b}{c<d}}
({\bf{\cal P}}_{ab}{\bf{\cal P}}_{cd})(\vec k_{ab}\vec k_{cd})
E_{a+b+c+d} \label{eqman} \ee where $\ E_{a_1+\ldots+a_m} = e^{(\vec
k_{a_1}+\ldots+ \vec k_{a_m})\vec u}$, $\ \ \vec k_{ab} = \vec
k_a-\vec k_b\ $ and \be {\bf{\cal P}}_{ab} = z_a{\bf v}_b - z_b{\bf
v}_a = z_az_b\big({\bf p}_a+{\bf p}_{a+1} + \ldots + {\bf
p}_{b-1}\big) = z_az_b\big({\bf p}_a + {\bf P}_{ab}\big)
\label{calP} \ee Note that $\vec k_a$ and $\vec k_{ab}$ are $2d$
vectors on the world sheet, while ${\bf p}_a$ and ${\bf{\cal
P}}_{ab}$ are $4d$ vectors in the ($T$-dualized) target space. Incidentally, ${\bf
P}_{ab}$ are the vectors which enter eq.(\ref{2Fme}).

The problem with eq.(\ref{eqman}) is that the set of exponentials
$E_{a+b+c+d}$ on the r.h.s. is larger then that of $E_{a+b}$ on the
l.h.s. The first term that causes trouble is $(a,b,c,d) =
(a,a,a+1,a-1)$: there is no associated $E = e^{(\vec k_{a+1}+2\vec
k_a +\vec k_{a-1})\vec u}$ on the l.h.s., while on the r.h.s. it
appears with the coefficient \be z_{a+1}z_a^2z_{a-1} \Big[(\vec
k_a-\vec k_{a-1})(\vec k_a-\vec k_{a+1})\Big] (2{\bf p}_{a}{\bf
p}_{a+1})
\ee
The last bracket is nothing but
$t_{a,a+2}=({\bf p}_a+{\bf p}_{a+1})^2 \neq 0$,
while the square bracket with the scalar product of
$2d$ vectors vanishes for all $a$ only for $n=4$ and with the $\vec k$-vectors pointing 
along the diagonals of a rectangle.

Beyond $n=4$ the problem is somewhat reminiscent of Serre relations
in group theory, only here it does not seem
have a simple solution. So, we shall concentrate on the $n=4$ case.

\subsection{The example of quadrilaterals, $n=4$
\label{n4exa}}

\subsubsection{
An anomaly}

$4d$ Lorentz invariance allows one to convert the original geometry,
associated with the four momenta ${\bf p}_a$, to any convenient form
with just two independent parameters $s$ and $t$. To begin with, it
allows to choose $y$-coordinates so that $y_3=0$ and consider
projections of the momenta on the $(y_1,y_2)$ plane, where they form
an ordinary quadrilateral. In order to provide a closed line in the
$y_0$-projection the side lengths of the edges of this quadrilateral should
satisfy an additional constraint \be l_1\pm l_2 \pm l_3 \pm l_4 = 0.
\ee A non-planar quadrilateral in $y$-space arises if the signs are
taken to be $(+-+-)$. Thus for further calculations one can choose
any 2-parametric family of quadrilaterals in $(y_1,y_2)$ plane with
$l_1+l_3=l_2+l_4$. For the two independent parameters $\sqrt{s}$ and
$\sqrt{t}$ one can take the two diagonals (in space-time, not in the
projection on the $(y_1,y_2)$ plane -- this is the same only when
$l_1=l_2$, i.e. for the cases of rhombus and kite).

In \cite{AM}, the shape was chosen to be rhombic, $l_1=l_2=l_3=l_4$,
but we do not impose such restriction in this section. This provides
an additional self-consistency check: the result (regularized
minimal area) should be Lorentz invariant and independent of the
particular shape of the quadrilateral. This is not fully guaranteed,
because any particular solutions of the equations of motion
spontaneously break Lorentz invariance and -- while obviously
restored in the non-regularized problem (indeed, $L=2$ for all 
solutions) -- it can still remain broken after regularization.

\begin{figure}
\epsfxsize 230pt {\epsffile{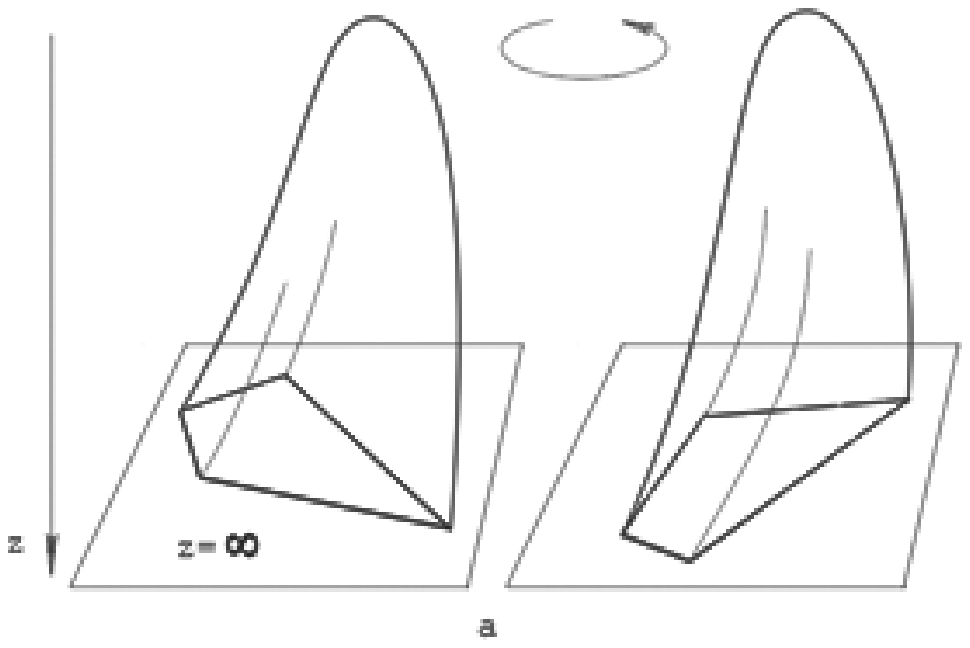}} \epsfxsize 230pt
\hspace{.5cm}{\epsffile{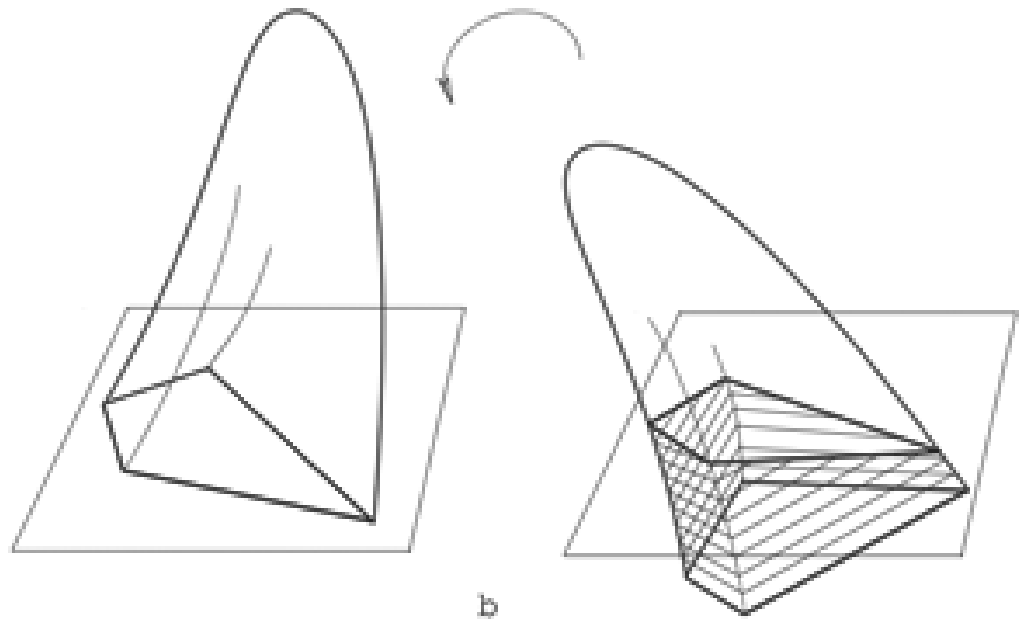}} \caption{\footnotesize Minimal
surfaces, connected (a) by a Lorentz transformation from $SO(3,1)$
and (b) by some non-Lorentzian transformation from $SO(4,2)$. From
the point of view of a remote observer, located at finite $z$,
boundary conditions with three different polygons: $P_1$, its
Lorentz rotated version $P_2$ and an essentially different $P_3$
should be considered equivalent. However, as is clear from the picture,
the actual areas should not coincide: they differ by the area of the shadowed
domain, which may not be negligible. This picture also shows
that the anomaly, considered in this paper, may not be an
artifact of the $\epsilon$-regularization of \cite{AM}: it can be
present for conventional $r^2\to r^2+\epsilon^2$ regularization as
well.} \label{figano}
\end{figure}

Unfortunately this is what eventually happens, as we shall see
below: this self-consistency check actually fails. The reason will
be that the {\it regularized} minimal area, defined according to the
recipe of \cite{AM}, depends on the lifting from the boundary
conditions ($\{{\bf p_a}\}$ or, equivalently, $\{{\bf v_a}\}$) to
the space of solutions (parameterized by ($\{z_a\}$) and there is no
obvious canonical choice of this lifting, i.e. it does not produce a
unique answer. The Lorentz invariance itself can be easily restored
if one asks the lifting to remain intact under Lorentz rotation, but
there is still no distinguished way to obtain an answer for given
$s$ and $t$. As shown in \cite{AM}, a lifting exists that reproduces
the BDS formula, but it is unclear what are the {\it a priori}
reasons to choose this particular lifting and what are the ways to
generalize it to other (non-rhombic) shapes -- other than just an
$SO(4,2)$ rotation of the same {\it ad hoc} prescription. See Figure
\ref{figano} for a pictorial description of this anomaly.

The resolution of the anomaly is well known. When the result is a
non-trivial function on moduli space, one should integrate over it
or, classically, find its extremum. This is exactly what we shall do in Section \ref{7}.

\subsubsection{Equation and solution}

The boundary conditions (\ref{boco}) are
\be
{\cal P}_{12} = z_1{\bf v}_2-z_2{\bf v}_1 = z_1z_2{\bf p}_1, \nn \\
{\cal P}_{23} = z_2{\bf v}_3-z_3{\bf v}_2 = z_2z_3{\bf p}_2, \nn \\
{\cal P}_{34} = z_3{\bf v}_4-z_4{\bf v}_3 = z_3z_4{\bf p}_3, \nn \\
{\cal P}_{41} = z_4{\bf v}_1-z_1{\bf v}_4 = z_4z_1{\bf p}_4 \ee or
\be {\bf p}_a = \frac{{\bf v}_{a+1}}{z_{a+1}} - \frac{{\bf
v}_{a}}{z_{a}} \ee It follows that \be {\cal P}_{13} = z_1{\bf
v}_3-z_3{\bf v}_1 = z_1z_3({\bf p}_1+{\bf p}_2)
\nn \\
{\cal P}_{24} = z_2{\bf v}_4-z_4{\bf v}_2 = z_2z_4({\bf p}_1+{\bf
p}_3) \ee As shown in \cite{AM}, in the case of $n=4$ $L$ can be
constant, as will be verified by the explicit solution below and,
following \cite{AM}, we adjust the scale of $\vec u$ so that $L=2$.
The non-trivial (third) equation of motion in (\ref{eqm}) is \be 2z^2
- (\partial z)^2 = (z\partial {\bf v} - {\bf v}\partial z)^2 \ee In
the form (\ref{eqman}) this equation reads \be \sum_{a,b=1}^4 z_a
z_b\Big(2 - (\vec k_a\vec k_b)\Big)
 E_{a+b} =
\sum_{{a<b}\atop{c<d}} ({\bf{\cal P}}_{ab}{\bf{\cal P}}_{cd})
(\vec k_{ab}\vec k_{cd})
E_{a+b+c+d}
\label{expoeq}
\ee

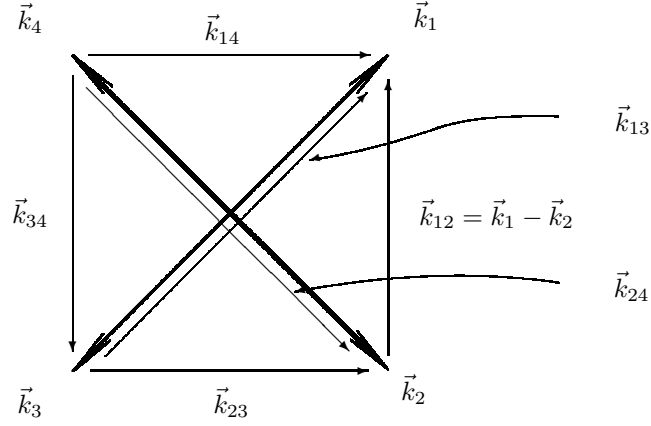
\begin{figure}
\hspace{3cm}
\unitlength 1mm 
\linethickness{0.4pt}
\ifx\plotpoint\undefined\newsavebox{\plotpoint}\fi 
\begin{picture}(101.678,66.299)(0,0)
\thicklines \put(27.156,19.066){\line(1,1){42}}
\multiput(63.936,57.077)(.043747009,.033653846){117}{\line(1,0){.043747009}}
\multiput(69.054,61.015)(-.03362314,-.041219008){121}{\line(0,-1){.041219008}}
\multiput(32.436,56.946)(-.04338843,.03362314){121}{\line(-1,0){.04338843}}
\multiput(27.186,61.015)(.03362314,-.041219008){121}{\line(0,-1){.041219008}}
\multiput(32.173,23.084)(-.04013719,-.033628926){121}{\line(-1,0){.04013719}}
\multiput(27.317,19.015)(.033653846,.042628205){117}{\line(0,1){.042628205}}
\multiput(64.067,23.215)(.042628205,-.033653846){117}{\line(1,0){.042628205}}
\multiput(69.054,19.277)(-.03362314,.042306612){121}{\line(0,1){.042306612}}
\thinlines \put(29.256,61.066){\vector(1,0){37.538}}
\put(69.156,21.166){\vector(0,1){36.619}}
\put(29.606,19.066){\vector(1,0){36.882}}
\put(27.156,58.266){\vector(0,-1){36.619}}
\put(66.298,56.027){\vector(1,1){.07}}\multiput(31.517,20.984)(.03373579049,.03398971872){1031}{\line(0,1){.03398971872}}
\put(29.023,56.684){\vector(1,-1){34.913}} \thicklines
\multiput(27.307,60.992)(.03381827967,-.03373436334){1240}{\line(1,0){.03381827967}}
\multiput(67.993,20.098)(-.03382495565,.0337359426){1169}{\line(-1,0){.03382495565}}
\multiput(28.868,59.535)(.03391735891,-.03373302544){1129}{\line(1,0){.03391735891}}
\multiput(68.201,60.367)(-.03372677292,-.03398822077){1194}{\line(0,-1){.03398822077}}
\multiput(28.556,20.202)(.03372609501,.03399590377){1157}{\line(0,1){.03399590377}}
\put(74.177,66.299){\makebox(0,0)[cc]{${\vec k}_1$}}
\put(72.542,16.649){\makebox(0,0)[cc]{${\vec k}_2$}}
\put(21.555,14.865){\makebox(0,0)[cc]{${\vec k}_3$}}
\put(21.555,66.001){\makebox(0,0)[cc]{${\vec k}_4$}}
\put(47.123,64.515){\makebox(0,0)[cc]{${\vec k}_{14}$}}
\put(83.542,39.987){\makebox(0,0)[cc]{${\vec k}_{12}={\vec k}_1 -
{\vec k}_2$}} \put(21.555,40.136){\makebox(0,0)[cc]{${\vec
k}_{34}$}} \put(48.312,14.865){\makebox(0,0)[cc]{${\vec k}_{23}$}}
\put(101.678,52.474){\makebox(0,0)[cc]{${\vec k}_{13}$}}
\put(101.529,30.325){\makebox(0,0)[cc]{${\vec k}_{24}$}} \thinlines
\qbezier(91.718,52.92)(80.421,53.217)(75.069,51.136)
\put(58.717,47.123){\vector(-4,-1){.07}}\qbezier(75.069,51.136)(66.001,48.089)(58.717,47.123)
\put(56.785,29.582){\vector(-4,-1){.07}}\qbezier(91.867,30.771)(75.812,33.149)(56.785,29.582)
\end{picture}
 \caption{ {\footnotesize Concrete choice (\ref{3.23})
of the $\vec k$-vectors on the $2d$ $u$-plane. } }
\label{kvecs}\label{quadran}
\end{figure}

Let us begin with a special -- $Z_4$-symmetric -- choice of
the $n=4$
$2d$ vectors $\vec k$ with $\vec k^2=L=2$, shown in Figure
\ref{quadran} \be
\vec k_1 = (+1,+1), \ \ \ 
\vec k_2 = (+1,-1), \ \ \ 
\vec k_3  = (-1,-1) = -\vec k_{1}, \ \ \ 
\vec k_4  = (-1,+1) = -\vec k_{2}, \label{3.23}\ee Clearly, in this
case one can relabel indices $3$ and $4$ of the exponentials $E_a$
to $-1$ and $-2$ respectively, what we shall do in some of the
formulas. It is also natural to label $E_{1+4}=E_{1-2}$ and
$E_{1-1}=E_0=1$. In this notation \be z\partial{\bf v} - {\bf
v}\partial z = \sum_{a,b}\vec k_{ab}\otimes {\bf{\cal P}}_{ab}
E_{a+b} = z_1z_2\vec k_{12}\otimes {\bf p}_1\, E_{1+2} + z_1z_3\vec
k_{13}\otimes ({\bf p}_1+{\bf p}_2)\,E_0 - \ee \vspace{-0.5cm}
$$
- z_1z_4\vec k_{14}\otimes {\bf p}_4\, E_{1-2}
+z_2z_3\vec k_{23}\otimes {\bf p}_2\, E_{-1+2} +
z_2z_4\vec k_{24}\otimes ({\bf p}_2+{\bf p}_3)\, E_0
+ z_3z_4\vec k_{34}\otimes {\bf p}_3\, E_{-1-2}
$$
and the equation becomes \be 8\big(z_1z_{3}+z_2z_{4}\big)E_0 +
4\Big(z_1z_2E_{1+2}
 + z_{2}z_3E_{-1+2} + z_{3}z_{4}E_{-1-2} + z_1z_{4}E_{1-2}\Big)
= \nn \\ =
\Big((z_1z_3)^2\vec k_{13}^2({\bf p}_1+{\bf p}_2)^2 +
(z_2z_4)^2\vec k_{24}^2
({\bf p}_2+{\bf p}_3)^2 + 2z_1z_2z_3z_4\big[
(\vec k_{12}\vec k_{34})({\bf p}_1{\bf p}_3) -
  (\vec k_{14}\vec k_{23})({\bf p}_2{\bf p}_4)\big]
  \Big) E_0 + \nn \\
+ 2z_1z_2
\Big(z_1z_3(\vec k_{12}\vec k_{13})
\big[{\bf p}_1({\bf p}_1+{\bf p}_2)\big]+
z_2z_4(\vec k_{12}\vec k_{24})
\big[{\bf p}_1({\bf p}_2+{\bf p}_3)\big]
\Big)E_{1+2} + \nn \\
+ 2z_2z_3
\Big(z_1z_3(\vec k_{23}\vec k_{13})
\big[{\bf p}_2({\bf p}_1+{\bf p}_2)\big] +
z_2z_4(\vec k_{23}\vec k_{24})
\big[{\bf p}_2({\bf p}_2+{\bf p}_3)\big]
\Big)E_{-1+2} + \nn\\
+ 2z_3z_4
\Big(z_1z_3(\vec k_{34}\vec k_{13})
\big[{\bf p}_3({\bf p}_1+{\bf p}_2)\big] +
z_2z_4(\vec k_{34}\vec k_{24})
\big[{\bf p}_3({\bf p}_2+{\bf p}_3)\big]
\Big)E_{-1-2} - \nn\\
- 2z_1z_4
\Big(z_1z_3(\vec k_{14}\vec k_{13})
\big[{\bf p}_4({\bf p}_1+{\bf p}_2)\big] +
z_2z_4(\vec k_{14}\vec k_{24})
\big[{\bf p}_4({\bf p}_2+{\bf p}_3)\big]
\Big)E_{1-2}\ \ \ \
\label{eqm4}
\ee
Many terms are actually absent from the r.h.s.
due to ${\bf p}_a^2=0$
or $\vec k_a^2=2$ or $\vec k_a\vec k_{a+1} = 0$.
Most important, these conditions are enough to
exclude terms like $E_{1+1+2+2}$ or $E_{1+1}$
from the r.h.s., which would have no counterparts
at all on the l.h.s.
More explicitly,

--  terms like $E_{1+1+2+2}$ in the quartic
part (i.e. at the r.h.s.) do not appear
because  ${\bf p}_a^2=0$,

--  terms like $E_{1+1}=E_{(1+2)+(-1+2)}$ in the quartic
part (on the r.h.s.) do not appear because of
$\vec k_a\vec k_{a+1} = 0$,

--  terms like $E_{1+1}=E_{(1+2)+(-1+2)}$ in the quadratic
part (on the l.h.s.) do not appear because of $\vec k_a^2=2$.

\bigskip

\noindent As a result of  these cancelations we are left with five
equations: those for $E_{\pm1\pm2}$ and $E_0=1$ \be
\begin{array}{lrcr}
E_0: &\ \ \ z_1z_3+z_2z_4 =&
(z_1z_3)^2 s + (z_2z_4)^2 t - z_1z_2z_3z_4 u &=
(z_1z_3+z_2z_4)\big(z_1z_3 s + z_2z_4 t\big) \\
E_{1+2}: &\ \ \ z_1z_2 =& z_1z_2\big(z_1z_3 s - z_2z_4(s+u)\big)
&=z_1z_2\big(z_1z_3 s + z_2z_4 t\big) \\
E_{-1+2}: &\ \ \ z_2z_3 =& &z_2z_3\big(z_1z_3 s +z_2z_4 t\big)\\
E_{-1-2}: &\ \ \ z_3z_4 =& z_3z_4\big(-z_1z_3(t+u)+z_2z_4t\big)
&=z_3z_4\big(z_1z_3 s +z_2z_4 t\big)\\
E_{1-2}: &\ \ \ z_1z_4 =& -z_1z_4\big(z_1z_3(t+u)+z_2z_4(u+s)\big)
& = z_1z_4\big(z_1z_3 s +z_2z_4 t\big)
\end{array}
\ee
Here $s=({\bf p}_1+{\bf p}_2)^2=2{\bf p}_1{\bf p}_2$,\ \ \ $t=({\bf p}_2+{\bf p}_3)^2=2{\bf p}_2{\bf p}_3$,\ \ \
$u=({\bf p}_1+{\bf p}_3)^2=2{\bf p}_1{\bf p}_3=2{\bf p}_2{\bf p}_4 = -s-t$.

\bigskip

\noindent
All five equations above coincide and are equivalent to the single relation 
\be z_1z_3 s +z_2z_4 t = 1 \; , \;\; {\rm or} \;\; z_1z_3 t_{13} +z_2z_4 t_{24} = 1 \label{equ4} \ee

\subsection{$SO(4,2)$ symmetry between different choices of $z_a$
\label{linco}}

Eq.(\ref{equ4}) defines the common scale of all factors $z_a$, but
does not fix relations between them. Since all the solutions with
different choices of $\{z_a\}$ have the same Lagrangian value $L=2$,
it is clear that they are related by a symmetry. This symmetry is
nothing but the conformal group $SO(4,2)$. Indeed, our $z$ and ${\bf
v}$ variables are nothing but flat coordinates in
$\mathbb{R}^6_{++++--}$, \be {\bf v} = \frac{{\bf y}}{r} =
\frac{{\bf Y}}{R}, \ \ \ z = \frac{1}{r} = \frac{Y_{-1}+Y_4}{R^2} =
Y_+, \ \ \ \frac{r^2-{\bf y}^2}{r} = Y_{-1}-Y_4 = Y_-, \ee where
$SO(4,2)$ acts linearly and $AdS_5$ is embedded as a quadratic \be
{\bf Y}^2 + Y_+Y_- = Y_{-1}^2+Y_0^2-Y_1^2-Y_2^2-Y_3^2-Y_4^2 = R^2
\label{quadri} \ee The flat metric in $\mathbb{R}^6$ induces the AdS
metric in the Poincare form \be d{\bf Y}^2 + dY_+dY_- = \frac{dr^2 +
d{\bf y}^2}{r^2} \ee

A priori, the ansatz (\ref{exposol}) does not seem to
imply anything nice for the $\vec u$-dependence of the sixth
$\sigma$-model coordinate $Y_-$. However, eq.(\ref{equ4}) is exactly
the condition that $Y_-$ is also a sum of $n=4$ exponentials, \be
Y_- = \sum_{a=1}^{n} w_aE_a \ee Indeed, substituting this expression
together with (\ref{exposol}) into (\ref{quadri}), one obtains
$$
0 = (z_1E_1+z_2E_2+z_3E_3+z_4E_4)(w_1E_1+w_2E_2+w_3E_3+w_4E_4) -
({\bf v}_1E_1+{\bf v}_2E_2+{\bf v}_3E_3+{\bf v}_4E_4)^2 - 1 =$$ $$
\sum_{a=1}^4 \left\{\Big(z_aw_a - {\bf v}_a^2\Big) E_a^2 +
\Big(z_aw_{a+1}+z_{a+1}w_a - 2{\bf v}_a{\bf v}_{a+1}\Big)E_aE_{a+1}
\right\} + \Big(w_1z_3 + w_2z_4 + w_3z_1 + w_4z_2 - 2{\bf v}_1{\bf
v}_3 - 2{\bf v}_2{\bf v}_4 - 1\Big)
$$
The vanishing condition for the first term defines \be w_a =
\frac{{\bf v}_a^2}{z_a} \ee The coefficient in the second term is
then equal to
$$\frac{z_a}{z_{a+1}}\,{\bf v}_{a+1}^2 +
\frac{z_{a+1}}{z_a}\,{\bf v}_a^2 - 2{\bf v}_a{\bf v}_{a+1} =
\frac{\Big(z_a{\bf v}_{a+1} - z_{a+1}{\bf v}_a\Big)^2}{z_az_{a+1}} \
\stackrel{(\ref{boco1})}{=}\ z_az_{a+1}{\bf p}_a\,^2 = 0
$$
and vanishes automatically. Finally, vanishing the last term is
exactly the relation (\ref{equ4}).

Note that the Lorentz transformations $SO(3,1)$ can change ${\bf
p}_a$ and the shape of the boundary in target space, but they are
not enough to relate solutions with different $\{z_a\}$. The problem
is that conformal symmetry and thus the equivalence between different
choices of $\{z_a\}$ is broken by dimensional regularization {\it a
la} \cite{AM}. We return to a discussion of this  {\it anomaly}
problem in Section \ref{quadriepsilon} and especially  in Subsection \ref{7}. The
resolution of the anomaly problem will lead to the existence of preferable
choices for $\{z_a\}$ for given boundary conditions.

\subsection{Solutions for $n=4$}

In this subsection we write down explicitly various solutions for
the $n=4$ case in the rapidity coordinates $\xi_1=\tanh u_1$,
$\xi_2=\tanh u_2$.

\subsubsection{The Alday-Maldacena solution \cite{AM}}

The principal choice of \cite{AM} is
\be z_3=z_1, \ \ \ z_4=z_2 \label{cho21} \ee Then equation
(\ref{equ4}) becomes \be z_1^2s + z_2^2t = 1
\ee but it still leaves a one-parameter freedom in the choice of
$z_1$ and $z_2$. Alday and Maldacena fix it by putting \be z_1^2 =
\frac{1}{2s}, \ \ \ z_2^2 = \frac{1}{2t} \label{cho22} \ee and find
it convenient to reexpress everything in intermediate formulas
through auxiliary parameters $a$ and $b$: \be
s=\frac{A^2}{2(1-b)^2}, \ \ \ t=\frac{A^2}{2(1+b)^2}, \ \ \ z_1 =
\frac{1-b}{A}, \ \ \ z_2 = \frac{1+b}{A}, \ \ \ A = \frac{a}{2\pi}
\label{cho2} \ee This choice of $z_a$ does not actually restrict the
possibility to impose arbitrary boundary conditions and we shall see
in a moment that the generic set of external momenta $\{ {\bf p}_a\}$
can be described with (\ref{cho2}). However, in \cite{AM} the particular
choice is made \be r =
a\frac{\sqrt{(1-\xi_1^2)(1-\xi_2^2)}}{1+b\xi_2\xi_2}, \ \ \ y_0 =
a\frac{\sqrt{1+b^2}\xi_1\xi_2}{1+b\xi_1\xi_2},\ \ \ y_{1} =
a\frac{\xi_{1}}{1+b\xi_1\xi_2},\ \ \ y_{2} =
a\frac{\xi_{2}}{1+b\xi_1\xi_2},\ \ \ y_3 = 0, \label{cho23} \ee
where $\xi_1=\tanh u_1$, $\xi_2=\tanh u_2$ are the rapidity
variables. In these coordinates, the four boundaries of our
quadrilateral are at $\xi_{1,2} = \pm  1$, and the whole solution is
a mapping of a square in the rapidity plane into the target space.
In terms of $y_{1,2}$, the boundaries are at $y_1+by_2=\pm a$ and
$by_1+y_2=\pm a$, i.e. form a {\it rhombus} with diagonals
$\sqrt{s}$ and $\sqrt{t}$. Note that, since $l_1=l_2=l_3=l_4$, the
squares of diagonals in space-time (which are actually $s$ and
$t$) coincide with the squares of their projections onto the
$(y_1,y_2)$ plane.

From (\ref{cho23}) we can deduce the Nambu-Goto action for this
solution: $H = \frac{1+b\xi_1\xi_2}{1-b\xi_1\xi_2}$,\ \ \
$dy_1\wedge dy_2 = \frac{1-b\xi_1\xi_2}{(1+b\xi_1\xi_2)^3}\
d\xi_1\wedge d\xi_2$, the equations of motion are again true, and
\be S_{NG} = \int_0^1\int_0^1
\frac{d\xi_1d\xi_2}{(1-\xi_1^2)(1-\xi_2^2)} \ee obviously coincides
with the $\sigma$-model action.

\subsubsection{Beyond \cite{AM}}

Thus, the Alday-Maldacena solution is fixed by three choices:
(\ref{cho21}), (\ref{cho22}) and (\ref{cho23}). Actually all 
three are ambigous and can be deformed, giving rise to new
solutions: not a big surprise given the $SO(4,2)$ symmetry of the
problem. One could safely take any family of solutions sufficient to
describe arbitrary values of $s$ and $t$ and ignore all the rest --
if $SO(4,2)$ symmetry was not violated by $\epsilon$-regularization.
Since only regularized area makes sense, one should actually analyze
{\it all}  solutions and see what happens -- and the result is
unpleasant: the answer depends on the choice. The answer matches the BDS
formula for the Alday-Maldacena choice, but they do not match in general.
{\it A priori} no way to distinguish the Alday-Maldacena choice
among all others remain unclear. We shall propose a way after regularization
in Section \ref{7}.

Needless to say, it is unclear what should be a counterpart of the
Alday-Maldacena choice for $n>4$, where a variety of methods,
including approximate and numerical, could be used if one knew {\it
what kind} of solution one should concentrate on. It is to
demonstrate all these issues that we proceed with the detailed presentation of other
solutions as well.

\subsubsection{All $z_a$ equal}

Before we proceed with the discussion of the  Alday-Maldacena solution and
its generalizations to arbitrary quadrilaterals in the target space,
we now analyze a much simpler option: when all  four $z_a$ are
the same, and according to (\ref{equ4}) \be z_a =
\frac{1}{\sqrt{s+t}} \label{cho1} \ee
Then (\ref{exposol}) implies that $z = 4\cosh u_1\cosh u_2$ and in
rapidity coordinates, \be
r = a\sqrt{(1-\xi_1^2)(1-\xi_2^2)}, \nn \\
{\bf y} = {\bf \alpha} + {\bf \beta}\xi_1  + {\bf \gamma}\xi_2 +
{\bf\delta}\xi_1\xi_2 \label{cho12} \ee where $a=\sqrt{s+t}$ and the
four $4d$ vectors \be {\bf \alpha} = \frac{a}{4}\Big({\bf v_1} +
{\bf v_2} +
{\bf v_3} + {\bf v_4}\Big), \nn \\
{\bf \beta} = \frac{a}{4}\Big({\bf v_1} + {\bf v_2} -
{\bf v_3} - {\bf v_4}\Big), \nn \\
{\bf \gamma} = \frac{a}{4}\Big({\bf v_1} - {\bf v_2} -
{\bf v_3} + {\bf v_4}\Big), \nn \\
{\bf \delta} = \frac{a}{4}\Big({\bf v_1} - {\bf v_2} + {\bf v_3} -
{\bf v_4}\Big)\ \ee At the four boundaries of the square
$\xi_1,\xi_2=\pm 1\ $, ${\bf y}$ form four segments of straight
lines, which should coincide with the four external momenta: these
are our familiar boundary conditions (\ref{boco1}), \be {\bf
v}_{a+1}-{\bf v}_{a} = \frac{{\bf p}_a}{\sqrt{s+t}} \ee Putting,
say, $\xi_1=1$ one obtains ${\bf y} = ({\bf \alpha}+{\bf\beta}) +
({\bf\gamma}+{\bf\delta})\xi_2$, which varies along the segment
$\xi_2 = [-1,1]$ from $({\bf \alpha}+{\bf\beta}) -
({\bf\gamma}+{\bf\delta})$ to $({\bf \alpha}+{\bf\beta}) +
({\bf\gamma}+{\bf\delta})$, i.e. \be \Delta_1{\bf y} =
2({\bf\gamma}+{\bf\delta}) = -{\bf p_1} \ee Similarly \be
\Delta_2{\bf y} = 2({\bf\beta}+{\bf\delta}) = {\bf p_4},\nn\\
\Delta_{-1}{\bf y} = 2({\bf\gamma}-{\bf\delta}) = {\bf p_3},\nn\\
\Delta_{-2}{\bf y} = 2({\bf\beta}-{\bf\delta}) = -{\bf p_2}
\label{Dy} \ee along the boundaries $\xi_2=1$, $\xi_1=-1$ and $\xi_2=-1$ respectively,
so that equivalently \be
{\bf \beta} = -\frac{1}{4}({\bf p}_2-{\bf p}_4), \nn \\
{\bf\gamma} = -\frac{1}{4}({\bf p}_1-{\bf p}_3), \nn \\
{\bf\delta} = -\frac{1}{4}({\bf p}_1+{\bf p}_3) = \frac{1}{4}({\bf
p}_2+{\bf p}_4) \label{bgd} \ee while ${\bf\alpha}$ is a total shift
of ${\bf y}$ and remains unspecified by the boundary conditions -- like
the weighted common shift of all ${\bf v}$-vectors, ${\bf v}_a
\rightarrow {\bf v}_a+z_a{\bf w}$.

Thus, it is explicitly shown that a solution exists with all $z_a$
 for arbitrary choice of external momenta ${\bf p}_a$. With
the same choice of momenta as in \cite{AM}, \be  {\bf p}_1={2\over
1-b^2}\left(\sqrt{1+b^2},\ 1,\ -b,\ 0\right)\\ {\bf p}_2={2\over
1-b^2}\left({-\sqrt{1+b^2}},\ -b,\ 1,\ 0\right)\\ {\bf p}_3={2\over
1-b^2}\left({\sqrt{1+b^2}},\ -1,\ b,\ 0\right)\\{\bf p}_4={2\over
1-b^2}\left(-\sqrt{1+b^2},\ b,\ -1,\ 0\right)\label{AMcho}\ee one
obtains the solution with \be {\bf v}_1=-{1\over 4}\left(1,\
{1\over\sqrt{1+b^2}},\ {1\over\sqrt{1+b^2}},\ 0\right) \\
{\bf v}_2=-{1\over 4}\left(-1,\ -{1\over \sqrt{1+b^2}},\
{1+2b\over\sqrt{1+b^2}},\  0\right)
\\ {\bf
v}_3=-{1\over 4}\left({1},\ {2b-1\over \sqrt{1+b^2}},\ {2b-1\over
\sqrt{1+b^2}},\ 0\right)\\{\bf v}_4=-{1\over 4}\left({-1},\
{1+2b\over \sqrt{1+b^2}},\ -{1\over \sqrt{1+b^2}},\ 0\right) \ee
This gives \be y_0=a\xi_1\xi_2,\ \ \ \ \ \ \
y_1={a\over\sqrt{1+b^2}}(b+\xi_1-b\xi_2),\ \ \ \ \ \ \
y_2={a\over\sqrt{1+b^2}}(-b+b\xi_1+\xi_2) \ee

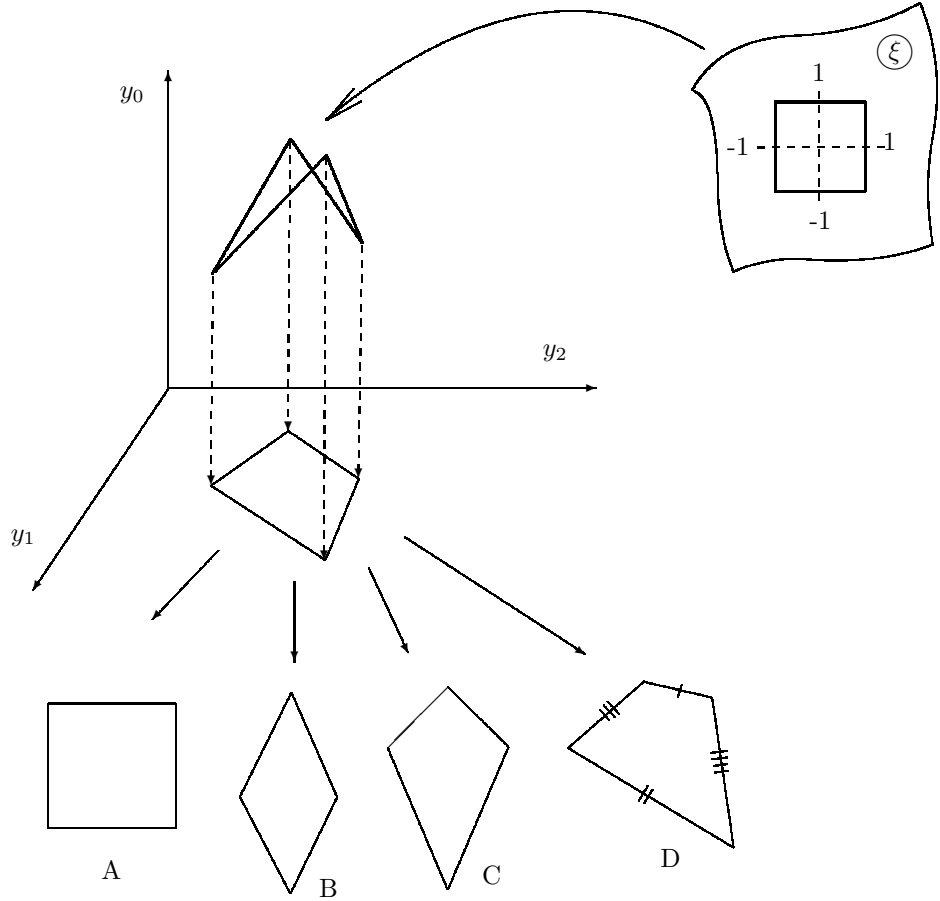
\begin{figure}
\hspace{2cm}
\unitlength 1mm 
\linethickness{0.4pt}
\ifx\plotpoint\undefined\newsavebox{\plotpoint}\fi 
\begin{picture}(132.861,133.807)(0,0)
\put(29.533,75.074){\vector(0,1){42.25}}
\put(29.533,75.074){\vector(1,0){57}}
\put(11.543,48.156){\vector(-2,-3){.07}}\multiput(29.574,75.026)(-.0337028037,-.0502242991){535}{\line(0,-1){.0502242991}}
\thicklines
\multiput(35.408,90.229)(.0336451613,.0581645161){310}{\line(0,1){.0581645161}}
\multiput(50.611,106.139)(-.033690583,-.0352757848){446}{\line(0,-1){.0352757848}}
\multiput(45.838,108.26)(.0337314488,-.0493462898){283}{\line(0,-1){.0493462898}}
\multiput(55.384,94.295)(-.033612676,.082161972){142}{\line(0,1){.082161972}}
\thinlines
\multiput(35.231,62.121)(.047688372,.033711628){215}{\line(1,0){.047688372}}
\multiput(45.484,69.369)(.049571429,-.033671958){189}{\line(1,0){.049571429}}
\multiput(54.853,63.005)(-.033732824,-.082312977){131}{\line(0,-1){.082312977}}
\multiput(50.434,52.222)(-.0517108844,.033670068){294}{\line(-1,0){.0517108844}}
\put(35.231,62.121){\vector(0,-1){.07}}\put(35.515,89.982){\line(0,-1){.9975}}
\put(35.49,87.987){\line(0,-1){.9975}}
\put(35.465,85.992){\line(0,-1){.9975}}
\put(35.439,83.997){\line(0,-1){.9975}}
\put(35.414,82.001){\line(0,-1){.9975}}
\put(35.389,80.006){\line(0,-1){.9975}}
\put(35.363,78.011){\line(0,-1){.9975}}
\put(35.338,76.016){\line(0,-1){.9975}}
\put(35.313,74.021){\line(0,-1){.9975}}
\put(35.288,72.026){\line(0,-1){.9975}}
\put(35.262,70.031){\line(0,-1){.9975}}
\put(35.237,68.036){\line(0,-1){.9975}}
\put(35.212,66.041){\line(0,-1){.9975}}
\put(35.186,64.046){\line(0,-1){.9975}}
\put(45.484,69.369){\vector(0,-1){.07}}\put(45.768,108.013){\line(0,-1){.9927}}
\put(45.75,106.027){\line(0,-1){.9927}}
\put(45.732,104.042){\line(0,-1){.9927}}
\put(45.714,102.057){\line(0,-1){.9927}}
\put(45.695,100.071){\line(0,-1){.9927}}
\put(45.677,98.086){\line(0,-1){.9927}}
\put(45.659,96.101){\line(0,-1){.9927}}
\put(45.641,94.115){\line(0,-1){.9927}}
\put(45.623,92.13){\line(0,-1){.9927}}
\put(45.605,90.145){\line(0,-1){.9927}}
\put(45.587,88.159){\line(0,-1){.9927}}
\put(45.568,86.174){\line(0,-1){.9927}}
\put(45.55,84.189){\line(0,-1){.9927}}
\put(45.532,82.203){\line(0,-1){.9927}}
\put(45.514,80.218){\line(0,-1){.9927}}
\put(45.496,78.233){\line(0,-1){.9927}}
\put(45.478,76.247){\line(0,-1){.9927}}
\put(45.459,74.262){\line(0,-1){.9927}}
\put(45.441,72.277){\line(0,-1){.9927}}
\put(45.423,70.291){\line(0,-1){.9927}}
\put(50.257,52.222){\vector(0,-1){.07}}\put(50.541,105.892){\line(0,-1){.9952}}
\put(50.528,103.901){\line(0,-1){.9952}}
\put(50.515,101.911){\line(0,-1){.9952}}
\put(50.502,99.921){\line(0,-1){.9952}}
\put(50.489,97.93){\line(0,-1){.9952}}
\put(50.476,95.94){\line(0,-1){.9952}}
\put(50.462,93.95){\line(0,-1){.9952}}
\put(50.449,91.959){\line(0,-1){.9952}}
\put(50.436,89.969){\line(0,-1){.9952}}
\put(50.423,87.978){\line(0,-1){.9952}}
\put(50.41,85.988){\line(0,-1){.9952}}
\put(50.397,83.998){\line(0,-1){.9952}}
\put(50.384,82.007){\line(0,-1){.9952}}
\put(50.371,80.017){\line(0,-1){.9952}}
\put(50.358,78.027){\line(0,-1){.9952}}
\put(50.344,76.036){\line(0,-1){.9952}}
\put(50.331,74.046){\line(0,-1){.9952}}
\put(50.318,72.055){\line(0,-1){.9952}}
\put(50.305,70.065){\line(0,-1){.9952}}
\put(50.292,68.075){\line(0,-1){.9952}}
\put(50.279,66.084){\line(0,-1){.9952}}
\put(50.266,64.094){\line(0,-1){.9952}}
\put(50.253,62.104){\line(0,-1){.9952}}
\put(50.24,60.113){\line(0,-1){.9952}}
\put(50.226,58.123){\line(0,-1){.9952}}
\put(50.213,56.133){\line(0,-1){.9952}}
\put(50.2,54.142){\line(0,-1){.9952}}
\put(54.853,63.005){\vector(0,-1){.07}}\put(55.314,94.048){\line(0,-1){.9723}}
\put(55.281,92.103){\line(0,-1){.9723}}
\put(55.248,90.159){\line(0,-1){.9723}}
\put(55.215,88.214){\line(0,-1){.9723}}
\put(55.181,86.27){\line(0,-1){.9723}}
\put(55.148,84.325){\line(0,-1){.9723}}
\put(55.115,82.38){\line(0,-1){.9723}}
\put(55.082,80.436){\line(0,-1){.9723}}
\put(55.049,78.491){\line(0,-1){.9723}}
\put(55.015,76.547){\line(0,-1){.9723}}
\put(54.982,74.602){\line(0,-1){.9723}}
\put(54.949,72.658){\line(0,-1){.9723}}
\put(54.916,70.713){\line(0,-1){.9723}}
\put(54.883,68.768){\line(0,-1){.9723}}
\put(54.849,66.824){\line(0,-1){.9723}}
\put(54.816,64.879){\line(0,-1){.9723}}
\put(13.527,33.136){\line(0,-1){16.608}}
\put(13.527,16.528){\line(1,0){17.028}}
\put(30.555,16.528){\line(0,1){16.608}}
\put(30.555,33.136){\line(-1,0){16.923}}
\multiput(45.901,34.607)(-.033655172,-.068349754){203}{\line(0,-1){.068349754}}
\multiput(39.069,20.732)(.033635,-.06464){200}{\line(0,-1){.06464}}
\multiput(45.796,7.804)(.033706522,.069690217){184}{\line(0,1){.069690217}}
\multiput(51.998,20.627)(-.033685083,.077237569){181}{\line(0,1){.077237569}}
\put(66.713,35.238){\line(-1,-1){7.989}}
\multiput(58.725,27.249)(.033704641,-.079831224){237}{\line(0,-1){.079831224}}
\multiput(66.713,8.329)(.033725,.079270833){240}{\line(0,1){.079270833}}
\multiput(74.807,27.354)(-.034151899,.033708861){237}{\line(-1,0){.034151899}}
\multiput(92.828,36.046)(-.0383522727,-.0336174242){264}{\line(-1,0){.0383522727}}
\multiput(82.703,27.171)(.0559796438,-.0337150127){393}{\line(1,0){.0559796438}}
\multiput(104.703,13.921)(-.03343023,.23255814){86}{\line(0,1){.23255814}}
\multiput(101.828,33.921)(-.14484127,.03373016){63}{\line(-1,0){.14484127}}
\put(27.401,44.278){\vector(-1,-1){.07}}\multiput(36.231,53.527)(-.0337022901,-.0353015267){262}{\line(0,-1){.0353015267}}
\put(46.322,49.323){\vector(0,-1){10.721}}
\put(61.458,39.863){\vector(1,-2){.07}}\multiput(56.202,51.215)(.033692308,-.072769231){156}{\line(0,-1){.072769231}}
\put(85.003,39.653){\vector(3,-2){.07}}\multiput(61.037,55.209)(.0518744589,-.0336709957){462}{\line(1,0){.0518744589}}
\multiput(97.783,35.699)(-.0333333,-.1166667){15}{\line(0,-1){.1166667}}
\multiput(93.283,22.199)(-.03289474,-.05592105){38}{\line(0,-1){.05592105}}
\multiput(93.908,21.699)(-.03308824,-.05514706){34}{\line(0,-1){.05514706}}
\multiput(87.908,33.449)(.03316327,-.03571429){49}{\line(0,-1){.03571429}}
\multiput(87.408,32.949)(.03333333,-.03611111){45}{\line(0,-1){.03611111}}
\multiput(86.908,32.324)(.03289474,-.03618421){38}{\line(0,-1){.03618421}}
\multiput(101.658,26.574)(.28125,.03125){8}{\line(1,0){.28125}}
\multiput(101.783,25.699)(.265625,.03125){8}{\line(1,0){.265625}}
\multiput(102.033,24.824)(.234375,.03125){8}{\line(1,0){.234375}}
\multiput(102.158,23.949)(.234375,.03125){8}{\line(1,0){.234375}}
\put(24.725,114.072){\makebox(0,0)[cc]{$y_0$}}
\put(10.283,55.209){\makebox(0,0)[cc]{$y_1$}}
\put(81.008,79.805){\makebox(0,0)[cc]{$y_2$}}
\put(22,11){\makebox(0,0)[cc]{A}}
\put(50.838,8.622){\makebox(0,0)[cc]{B}}
\put(72.541,10.257){\makebox(0,0)[cc]{C}}
\put(96.326,12.635){\makebox(0,0)[cc]{D}} \thicklines
\put(110.211,113.176){\line(0,-1){11.932}}
\put(110.211,101.243){\line(1,0){12.021}}
\put(122.231,101.243){\line(0,1){11.932}}
\put(122.231,113.176){\line(-1,0){12.109}} \thinlines
\put(107.842,107.095){\line(1,0){.9983}}
\put(109.839,107.105){\line(1,0){.9983}}
\put(111.835,107.116){\line(1,0){.9983}}
\put(113.832,107.126){\line(1,0){.9983}}
\put(115.828,107.137){\line(1,0){.9983}}
\put(117.825,107.147){\line(1,0){.9983}}
\put(119.822,107.157){\line(1,0){.9983}}
\put(121.818,107.168){\line(1,0){.9983}}
\put(123.815,107.178){\line(1,0){.9983}}
\put(116.062,114.608){\line(0,-1){.9723}}
\put(116.062,112.663){\line(0,-1){.9723}}
\put(116.062,110.719){\line(0,-1){.9723}}
\put(116.062,108.774){\line(0,-1){.9723}}
\put(116.062,106.83){\line(0,-1){.9723}}
\put(116.062,104.885){\line(0,-1){.9723}}
\put(116.062,102.941){\line(0,-1){.9723}}
\put(116.062,100.996){\line(0,-1){.9723}}
\qbezier(99.225,114.782)(102.904,121.299)(112.049,121.929)
\qbezier(112.049,121.929)(122.245,122.245)(129.497,126.344)
\qbezier(129.497,126.344)(132.861,118.04)(131.179,108.475)
\qbezier(131.179,108.475)(129.918,101.748)(131.179,94.18)
\qbezier(131.179,94.18)(123.716,91.342)(113.31,92.288)
\qbezier(113.31,92.288)(108.791,92.288)(104.691,90.606)
\qbezier(104.691,90.606)(102.799,94.916)(102.589,102.168)
\qbezier(102.589,102.168)(102.484,113.626)(99.436,114.572)
\put(116.043,117.094){\makebox(0,0)[cc]{1}}
\put(116.253,97.333){\makebox(0,0)[cc]{-1}}
\put(105.322,107.214){\makebox(0,0)[cc]{-1}}
\put(125.503,107.845){\makebox(0,0)[cc]{1}}
\put(126.134,119.827){\circle{4.701}}
\put(126.134,119.827){\makebox(0,0)[cc]{$\xi$}}
\qbezier(100.697,120.248)(78.308,133.807)(50.454,110.788)
\multiput(54.238,114.992)(-.033715043,-.039664756){106}{\line(0,-1){.039664756}}
\multiput(50.664,110.788)(.06166547,.03363571){75}{\line(1,0){.06166547}}
\end{picture}
 \caption{Examples of quadrilaterals on the boundary of
the AdS target space at $r=0$ and also with $y_3=0$, together with their
projections onto the $(y_1,y_2)$ plane:\ \ A. Square.\ \ B.
Rhombus.\ \ C. Kite. \ \ D. generic skew quadrilateral.}
\label{roki}
\end{figure}

Along with these expressions describing the square (at $b=0$) and
rhombus choices \cite{AM} (see Figures \ref{roki}.A,B), one can
consider a kite, Figure \ref{roki}.C, or arbitrary asymmetric skew
quadrilateral satisfying $l_1+l_3=l_2+l_4$, Figure \ref{roki}.D.

\subsubsection{The Alday-Maldacena solution revisited}

Coming back to the Alday-Maldacena solution, it is now clear, that
{\it as for any other choice of $\{z_a\}$} it could describe an
{\it arbitrary} configuration of ${\bf p}_a$, not only rhombic.
Indeed, if we impose (\ref{cho21}) and (\ref{cho2}), we can still
write instead of (\ref{cho23}) \be
r = a\sqrt{(1-\xi_1^2)(1-\xi_2^2)}, \nn \\
{\bf y} = \frac{{\bf \alpha} + {\bf \beta}\xi_1  + {\bf \gamma}\xi_2
+ {\bf\delta}\xi_1\xi_2}{1+b\xi_1\xi_2} \label{cho211} \ee At the
same boundaries $\xi_{1,2}=\pm 1$ we now have straight segments in
 ${\bf y}$ space, parameterized in a slightly more complicated
way. For example at $\xi_1=1$ \be {\bf y} = \frac{({\bf \alpha} + {\bf \beta})  + ({\bf
\gamma} + {\bf\delta})\xi_2}{1+b\xi_2} \ee  It is
indeed a straight line, an intersection of three $3d$ hyperplanes in
$4d$ space given by ${\bf q}{\bf y}=c$ with $\Big(({\bf \gamma} +
{\bf\delta}) -b({\bf \alpha} + {\bf \beta}) \Big) {\bf q} = 0$ and
$c = {\bf q}({\bf \alpha} + {\bf \beta})$. The corresponding vector
\be \Delta_1{\bf y} = \frac{2({\bf \gamma} + {\bf\delta})}{1-b^2} -
\frac{2b({\bf\alpha}+{\bf\beta})}{1-b^2}= -{\bf p}_1 \ee Similarly, the
analogues of (\ref{Dy}) and (\ref{bgd}) are: \be
\Delta_{2}{\bf y} = \frac{2({\bf\beta}+{\bf\delta})}{1-b^2} -
\frac{2b({\bf\alpha}+{\bf\gamma})}{1-b^2}
 = {\bf p_4},\nn\\
\Delta_{-1}{\bf y} = \frac{2({\bf\gamma}-{\bf\delta})}{1-b^2} +
\frac{2b({\bf\alpha}-{\bf\beta})}{1-b^2}
 = {\bf p_3},\nn\\
\Delta_{-2}{\bf y} = \frac{2({\bf\beta}-{\bf\delta})}{1-b^2} +
\frac{2b({\bf\alpha}-{\bf\gamma})}{1-b^2}
 = -{\bf p_2}
\label{DyAM} \ee and \be
{\bf \beta} = -\frac{1-b^2}{4}({\bf p}_2-{\bf p}_4+b({\bf p}_1-{\bf p}_3)), \nn \\
{\bf\gamma} = -\frac{1-b^2}{4}({\bf p}_1-{\bf p}_3+b({\bf p}_2-{\bf p}_4)), \nn \\
b{\bf\alpha}-{\bf\delta} = -\frac{1-b^2}{4}({\bf p}_2+{\bf p}_4) =
\frac{1-b^2}{4}({\bf p}_1+{\bf p}_3) \label{bdgAM} \ee with
 ${\bf\alpha}$ and ${\bf\delta}$ not uniquely specified.

The Alday-Maldacena original choice (\ref{AMcho}) being substituted
into (\ref{bdgAM}) with $b\neq 0$, reproduces (\ref{cho23}), while
substitution of the same (\ref{AMcho}) into (\ref{bgd}) gives a
different-looking but equivalent solution -- before
$\epsilon$-regularization!

\subsubsection{An example of one-cusp solution}

The next examples concern one-cusp solutions (in fact, it is rather
{\it a corner} than {\it a cusp}, the name seems to be due to
historical reasons).

{\bf Choice 3:} Direction towards a cusp is along one of the vectors
$\vec k$, say, $\vec k_1$. In this limit $E_b$ dominates over all
other exponentials. However, one can make use of the freedom to
shift ${\bf v}_a$'s without changing the boundary conditions to put
${\bf v}_1=0$ (this can not be done for all corners/cusps of the
polygon at once, but is allowed in the case of a single
isolated cusp). Then the two adjacent exponentials should be kept in
the formulas for ${\bf v}$ and the {\bf single-cusp} solution
(\ref{exposol}) becomes \be
z = z_1E_1 + O(E_{\pm 2}), \nn \\
{\bf v} = {\bf v}_2 E_2 + {\bf v}_{4} E_{-2} + O(E_{-1})
\label{1cus} \ee Since ${\bf v}_1=0$, the boundary conditions are
now $z_1{\bf v}_2 = z_1z_2{\bf p}_1$ and $-z_1{\bf v}_4 = z_1z_4{\bf
p}_4$. Now using eq.(\ref{equ4}), one obtains $1-z_1z_3s = z_2z_4t =
(2{\bf p}_1{\bf p}_4)z_2z_4 = -2{\bf v}_2{\bf v}_4/z_1^2$. In
the literature, the remaining freedom is typically used to fix
$z_3=0$, which leads to \be\label{cons3} 2{\bf v}_2{\bf v}_{4} =
z_1^2 \ee The solution (\ref{1cus}) still can be chosen in different
ways: depending on the choice of ${\bf v}_2$ and ${\bf v}_{4}$,
which are restricted by a single constraint (\ref{cons3}). In
particular, in \cite{Kru,DGO,Mak,AM} \be
z_1=\sqrt{2}, \ \ \ v_2^+ = 1, \ \ \ v_{4}^- = 1, \nn  \\
z=\sqrt{2}E_1= \sqrt{2}e^{u_1+u_2},\ \ \
v^+ = v^0+v^1 = E_2 = e^{u_1-u_2},\ \ \
v^- = v^0-v^1 = E_{-2}= e^{-u_1+u_2},\ \ \ v^2=v^3=0,
\label{cho31}
\ee
and in \cite{Buch} \be z_1 = \sqrt{2}, \ \ \ v_2^+ = 1,\ \ \ v_{4}^-
= 1, \ \ \
v_2^- = \gamma^2, \ \ \ v_2^2 = \gamma \nn\\
z=\sqrt{2}E_1,\ \ \ v^+  = E_2,\ \ \
v^- = E_{-2}+\gamma^2E_2,\ \ \
v^2=\gamma E_2, \ \ \ v^3=0, \nn \\
{\rm so\ that\ in\ coordinates\ }\  (+,-,2,3) \ \ \ \ \ {\bf p}_1 =
\frac{{\bf v}_2}{z_2} = \frac{(1,\gamma^2,\gamma,0)}{z_2}, \ \ \
{\bf p}_4 = \frac{{\bf v}_{4}}{z_{4}} = \frac{(0,1,0,0)}{z_{4}}
\label{cho32} \ee One can certainly make many other choices (all
equivalent in the above sense, as long as we stay in $AdS_5$).

It is instructive to look once again at the equations
of motion for the solution (\ref{cho31}):
$z = \sqrt{2}E_1 =\sqrt{2}e^{u_1+u_2}$,
$y^\pm = v^\pm/z =E_{-1\pm 2}$, i.e. $y^+ = e^{-2u_2}$,
$y^- = e^{-2u_1}$
and $r=z^{-1} = \sqrt{2y^+y^-}=\sqrt{2}e^{-u_1-u_2}$.
From the two terms on the l.h.s. of the equation
\be
\partial_1(z^2\partial_1  y^\pm) +
\partial_2(z^2\partial_2  y^\pm) = 0
\ee
only one survives for each component of $y$,
because, say, $\partial_2 y^- = 0$.
Further, $\partial_1 y^- = -2y^-$, then multiplication by
$z^2$ (division by $r^2$) converts $y^-$ into $1/y^+$,
which is finally annihilated by the action of the second
$\partial_1$.

\subsection{More quadrilateral solutions:
another kind of deformation and a hidden symmetry
\label{n4exa2}}

\begin{figure}
\hspace{5cm}
\unitlength 1mm 
\linethickness{0.4pt}
\ifx\plotpoint\undefined\newsavebox{\plotpoint}\fi 
\begin{picture}(47.111,24.66)(0,0)
\thicklines
\multiput(8.043,6.983)(.0853955142,.0337192568){443}{\line(1,0){.0853955142}}
\multiput(8.22,6.718)(.0844247446,.0336906258){446}{\line(1,0){.0844247446}}
\multiput(8.043,22.009)(.0855950364,-.0337192568){443}{\line(1,0){.0855950364}}
\multiput(8.043,21.744)(.0863703488,-.03370058){438}{\line(1,0){.0863703488}}
\multiput(10.43,8.397)(-.05208599,-.03314563){56}{\line(-1,0){.05208599}}
\multiput(7.513,6.541)(.1399482,.0331456){24}{\line(1,0){.1399482}}
\multiput(10.253,20.064)(-.04285496,.03348043){66}{\line(-1,0){.04285496}}
\multiput(7.425,22.274)(.1188671,-.0335266){29}{\line(1,0){.1188671}}
\multiput(43.487,21.478)(.1599408,.0336718){21}{\line(1,0){.1599408}}
\multiput(46.846,22.185)(-.05892557,-.03314563){48}{\line(-1,0){.05892557}}
\multiput(43.575,7.336)(.1683588,-.0336718){21}{\line(1,0){.1683588}}
\multiput(47.111,6.629)(-.05836966,.03335409){53}{\line(-1,0){.05836966}}
\thinlines \qbezier(33.941,16.705)(34.648,14.451)(33.941,12.021)
\put(36.239,14.584){\makebox(0,0)[cc]{$\phi$}}
\put(42.957,24.66){\makebox(0,0)[cc]{${\vec k}_1$}}
\put(41.012,5.038){\makebox(0,0)[cc]{${\vec k}_2$}}
\put(13.258,5.657){\makebox(0,0)[cc]{${\vec k}_3$}}
\put(12.286,23.953){\makebox(0,0)[cc]{${\vec k}_4$}}
\end{picture}
 \caption{{\footnotesize Alternative choice of $\vec
k$-vectors on the $u$-plane, which differs from
(\ref{kvecs}) but also provides a solution (\ref{exposol}) with the
{\it same} boundary conditions to the $\sigma$-model equations of
motion. The $\vec k$-vectors lie along diagonals of a {\it rectangular}
and all have the same length, $\vec k_a^2=L=2$. }}\label{15}
\end{figure}
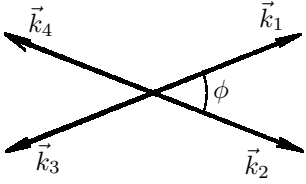

Since the only essential property of quadrilaterals that allowed
(\ref{exposol}) to be an exact solution was $(\vec k_a - \vec
k_{a-1})(\vec k_a - \vec k_{a+1})=0$, it is clear that a further
generalization is possible: the vectors $\vec k_{1,2,3,4}$ can form
diagonals of any rectangular, not necessarily a square, Figure
\ref{15}. One can check, that this is indeed a solution by a
straightforward repetition of the derivation in Section 3: some
coefficients at present depend on the angle between $\vec k_1$ and
$\vec k_2$, which are now not orthogonal, but these angles drop out
of the final relation (\ref{equ4}). What makes this deformation
interesting is that the change of vectors $\vec k_a$ is not a linear
transformation in the space of variables $(z,{\bf v})$, thus
$SO(4,2)$ invariance is not sufficient to explain their existence.
Also, the existence of deformed solutions supports the belief that
simple solutions with $n>4$ can exist: rectangular (rather than
square) quadruples of $\vec k$-vectors naturally arise in
degenerations of regular polygons.

Let us summarize the role of symmetries in the space of
solutions (\ref{exposol}) with $n=4$:

Lorentz symmetry $SO(3,1)$ changes external momenta ${\bf p}_a$
and the shape of the non-planar skew quadrilateral in
target space.
The only invariants are light-likeness of the sides and
the lengths $\sqrt{s}$ and $\sqrt{t}$ of the diagonals.
The shape of the projection on the $(y_1,y_2)$ plane
can be changed from kite or even more generic
configurations to a rhombus.
This can be considered as a change of coefficients
${\bf v}_a$ at fixed $z_a$.

Conformal symmetry $SO(4,2)$ allows to make linear
transformations of both $z_a$ and ${\bf v}_a$ parameters.

None of these target-space symmetries allows to change the
vectors $\vec k_a$. However, at least for $n=4$ such a
change is possible:
all rectangular configurations of four $\vec k_a$ provide
solutions. This can be considered as a certain
rescaling $(u_1,u_2) \rightarrow (\alpha u_1, \beta u_2)$,
which does not look like an obvious
symmetry of the system.

\section{Regularized minimal action \label{quadriepsilon}}
\setcounter{equation}{0}

After regularizing the action in the way proposed in \cite{AM}, one has to evaluate
an integral of the form \be \int L_\epsilon \; z^\epsilon d^2u
\label{Sep} \ee where $L_\epsilon$ is a certain modification of either
the $\sigma$-model or the Nambu-Goto action. An attractive feature
of this formula is that the finite contribution comes from the {\it
second-order} term in the expansion in powers of $\epsilon$, so that it
has a chance to acquire a biliner form, which is needed to reproduce
the double contour integral formula (\ref{loopin}). For the same
reason, however, different formulations of the minimal area problem,
while equivalent for $\epsilon=0$, will not necessarily lead to
the same answer for $\epsilon$-finite terms in (\ref{Sep}). In this
section we examine the dependence of (\ref{Sep}) on the choice of
$z_a$ and minimize it with respect to them.

\subsection{A puzzle}

An apparent problem with equation (\ref{Sep}), arising already in the
case of $n=4$, is the possibility to choose all $z_a=1$, which makes
the answer fully independent of external momenta. To be more
precise, the prescription of \cite{AM} just to make the only
replacement $r \rightarrow r\sqrt{1+\epsilon/2}$ in the solution of
equations of motion {\it at zero} $\epsilon$ and insert it back into
the action (\ref{Sep}), implies the substitution \be \int
\frac{(\partial r)^2 + (\partial {\bf y})^2}{r^{2+\epsilon}} \,d^2u
\longrightarrow \frac{1}{(1+\epsilon/2)^{\epsilon/2}} \int
\frac{(1+\epsilon/2)(\partial r)^2 + (\partial {\bf y})^2}
{(1+\epsilon/2)r^2}
\frac{d^2u}{r^\epsilon} = \nn \\
= \frac{1}{(1+\epsilon/2)^{\epsilon/2}}
\int \left(\frac{(\partial r)^2}{r^2} +
\frac{1}{1+\epsilon/2}\left(L-\frac{(\partial r)^2}{r^2}\right)
\right)\frac{d^2u}{r^\epsilon} =
\frac{1}{(1+\epsilon/2)^{1+\epsilon/2}}
\int \left(L + \frac{\epsilon}{2}\frac{(\partial z)^2}{z^2}
\right) 
z^\epsilon d^2u \label{Sep1} \ee where $(\partial {\bf y})^2$ is
expressed through $r$ and its derivatives from $L=\frac{(\partial
r)^2 + (\partial {\bf y})^2}{r^2} = 2$. The factor
$(1+\epsilon/2)^{-(1+\epsilon/2)} = 1 - \frac{\epsilon}{2} +
O(\epsilon^3)$. Under naive application of the prescription of
\cite{AM}, expression (\ref{Sep1}) is a concrete realization of
(\ref{Sep}),
and it suffers from the same problem: it depends only on $z$ and
becomes trivial (independent of external momenta) with an
allowed choice of all $z_a$ equal. Note that the same argument does not
work, at least in such a simple form, for the Nambu-Goto action,
because the $y$-variables do not disappear from the final expression above.

\subsection{Equations of motion}

We forget for a while the prescription (\ref{Sep1}) and start
directly from the deformed $\sigma$-model action, \be \int
\frac{(\partial r)^2 + (\partial {\bf y})^2}{r^{2+\epsilon}} \,d^2u \ee We
keep the notation $L$ and $z$ for the old quantities,
$L=((\partial r)^2 + (\partial {\bf y})^2)/r^2$ and $z=r^{-1}$,
while the new $\epsilon$-dependent quantities will be marked by
tildes: $\tilde L=((\partial r)^2 + (\partial
{\bf y})^2)/r^{2+\epsilon}$ and $\tilde z=r^{-1-\epsilon}$.

The equations of motion now read
\be
\partial\left(\frac{\partial r}{r^{2+\epsilon}}\right)
+ \frac{(2+\epsilon)}{2}\frac{L}{r^{1+\epsilon}} = 0, \nn \\
\partial\left(\frac{\partial {\bf y}}{r^{2+\epsilon}}\right) = 0
\ee
or, in terms of ${\bf v}={\bf y}/r$,
\be
\Delta \tilde z = (1+\epsilon)
\Big(1+\frac{\epsilon}{2}\Big)\tilde z L, \nn \\
\partial\Big((1+\epsilon)\tilde z\partial {\bf v} -
{\bf v}\partial\tilde z\Big)=0, \nn \\
(1+\epsilon)^2\tilde z^2 L = 
\left\{(\partial \tilde z)^2 + \Big((1+\epsilon)\tilde z\partial {\bf v} -
{\bf v}\partial\tilde z\Big)^2 \right\} \label{eqepsi} \ee In contrast to the
 case of $\epsilon = 0$, the mixing term in the second
equation now survives, and the equation for ${\bf v}$ does not look like the
first equation for $\tilde z$. To avoid confusion, we emphasize that in the
above equations $L = \tilde L_{\epsilon = 0}$.

\subsection{$\epsilon$-deformed one-cusp solution of \cite{Kru,AM,KRTT}}

This solution is a direct generalization of (\ref{cho31}). It
possesses an immediate generalization to (\ref{cho32}), but not to
the generic one-cusp limit.

If the only non-vanishing components of $z$ and ${\bf v}$ are $z_1$
and $v^\pm_{\pm 2}$, then one can write \be
z = z_1E_1, \nn \\
v^\pm = v^\pm_{\pm 2} E_{\pm 2}^\gamma \label{cho3ep} \ee where
$\gamma$ is some $\epsilon$-dependent power, which still needs to be
determined. The crucial problem with the generic single-cusp limit
is that, for $\epsilon \neq 0$, one can not safely add the term
${\bf v}_1 E_1^\gamma$ to ${\bf v}$: with the value of $\gamma$
required, it then contributes an undesired term $E_{1+1+1+1}$ to the
equations. Therefore, we consider restricted ansatz (\ref{cho3ep})
without reliable justification (no symmetry transformation is
immediately available at $\epsilon \neq 0$ to bring any one-cusp
limit to this form).

Substituting (\ref{cho3ep}) into (\ref{eqepsi}) one obtains \be
(1+\epsilon)^2 k_1^2 (z_1E_1)^{1+\epsilon} =
(1+\epsilon)(1+\epsilon/2)L (z_1E_1)^{1+\epsilon}, \nn \\
\Big(\gamma\vec k_2 + (1+\epsilon)\vec k_1\Big)
\Big(\gamma \vec k_2 - \vec k_1\Big)v_2^+E_2^\gamma E_1^{1+\epsilon}
 = 0, \nn \\
\Big(-\gamma\vec k_2 + (1+\epsilon)\vec k_1\Big)
\Big(-\gamma \vec k_2 - \vec k_1\Big)v_{-2}^-
E_{-2}^\gamma E_1^{1+\epsilon}
 = 0, \nn \\
(1+\epsilon)^2(z_1E_1)^{2(1+\epsilon)}\left(-L + \vec k_1^2
+ \Big(\gamma \vec k_2 - \vec k_1\Big)v_2^+ \Big(-\gamma \vec k_2 -
\vec k_1\Big)v_{-2}^-\right) = 0 \ee Given $\vec k_1^2=\vec k_2^2=2$
and $\vec k_1\vec k_2=0$, these equations imply \be
L=2\frac{1+\epsilon}{1+\epsilon/2}, \nn \\
\gamma^2 = 1+\epsilon, \nn \\
L=2 +2(1-\gamma^2)v_2^+v_2^- = 2-\epsilon v_2^+v_2^-
\ee
or
\be
v_2^+v_2^- = \frac{2}{\epsilon}
\left(1-\frac{1+\epsilon}{1+\epsilon/2}\right) =
-\frac{1}{1+\epsilon/2}
\ee

We repeat that this nice exact solution is a deformation of a very
special type of a cusp solution at $\epsilon = 0$ -- with ${\bf v}$
growing slower than $z$, ${\bf v}_1=0$ at $z_1\neq 0$, and it is
hard to extract any information from it, which can be justly used in
application to solution from more general classes. As we saw, the
prescription $r \rightarrow r\sqrt{1+ \epsilon/2}$ or rather ${\bf
v} \rightarrow {\bf v}/\sqrt{1+ \epsilon/2}$, while successfully
applied in \cite{AM}, cannot work equally well for all
$4d$-equivalent solutions with different sets of $\{z_a\}$.

\subsection{Alternative  one-cusp solution}

If one considers the generic one-cusp limit with ${\bf v}$ growing
at the same rate as $z$, then the exponential form of asymptotics
(\ref{exposol}) is no longer true, and this once again demonstrates that
the equivalence between different solutions is violated as a result of the
$\epsilon$-regularization. This change of asymptotical behavior is a
characteristic feature of Whitham deformations: when equations are
infinitesimally deformed, the change of solutions is not quite
infinitesimal -- it is at any given value of the arguments $\vec u$, but
for a given $\epsilon$ and sufficiently large $\vec k\vec u$ the
deformation can be as big as one wishes, the asymptotics is
changed, or, in other words, the large $\vec k \vec u$ and small
$\epsilon$ limits do not commute.

To be concrete, consider the
one-cusp limit $E_b \rightarrow \infty$ with all
non-vanishing coefficients $z_b$ and ${\bf v}_b$.
Before $\epsilon$-regularization solution (\ref{exposol})
in this limit becomes simply
\be
z = z_bE_b + O(E_{b,b\pm 1}), \nn \\
{\bf v} = {\bf v}_bE_b + O(E_{b,b\pm 1}) \ee Let us simply neglect
all $O(E_{b,b\pm 1})$ terms, i.e. demand that $z_b$ and ${\bf v}_b$
are the only non-vanishing coefficients. Then \be z\partial {\bf v}
- {\bf v} \partial z = 0 \label{difvan} \ee and let us look for a
solution to $\epsilon$-deformed equations (\ref{eqepsi}) with
exactly the same property (\ref{difvan}). This is not going to be a
generic solution, but still it provides some useful information.

Since (\ref{difvan}) is nothing but \be (1+\epsilon)\tilde z\partial
{\bf v} - {\bf v}\partial\tilde z = 0, \ee this restriction
drastically simplifies (\ref{eqepsi}) and reduces it to \be
\Delta\tilde z = (1+\epsilon)(1+\epsilon/2)\tilde z L, \nn \\
(\partial \tilde z)^2 = (1+\epsilon)^2\tilde z^2 L
\ee
It follows that
\be
\tilde z\Delta\tilde z = \frac{1+\epsilon/2}{1+\epsilon}
(\partial\tilde z)^2
\ee
or
\be
\partial (\partial \log z) = \sigma (\partial \log z)^2
\label{logzsig} \ee with $\sigma = -\epsilon/2$ (there would be an
additional factor of $(1+\epsilon)^{-1}$ in $\sigma$ if
(\ref{logzsig}) was written in terms of $\tilde z$). This equation
is easily converted into the Laplace equation \be \Delta
z^{-1/\sigma} = 0 \ee with the real part of any complex analytic
function as generic solution. Since we are interested in a solution
which behaves as $\log z = \vec k\vec u + O(\sigma)$ as $\sigma
\rightarrow 0$, one can easily solve (\ref{logzsig}) iteratively and
obtain \be z = z_b{\cal E}_b = z_b\frac{1}{(1-\sigma\vec k_b\vec
u)^{1/\sigma}} \label{zepsi} \ee  The corresponding $L$ is no longer a
constant at $\sigma\neq 0$: \be L = \frac{\vec k_b^2}{(1-\sigma\vec
k_b\vec u)^2},\ \ \ \ \ \vec k^2 = 2 \label{Sepsi} \ee Of course,
for a given $\vec u$, ${\cal E}_b = (1-\sigma\vec k_b\vec
u)^{-1/\sigma} \rightarrow E_b = e^{\vec k_b\vec u}$ as $\sigma
\rightarrow 0$, but at given $\sigma\neq 0$ the asymptotic
behavior at $\vec k_b\vec u \rightarrow \infty$ is completely
different: ${\cal E}_b$ grows/falls faster and reaches infinity/zero
at {\it finite} values of $\vec k_b\vec u = \pm\sigma^{-1}$. Such a
drastic change of the asymptotic behavior is a well-known phenomenon
in Whitham theory. As a side remark, note that for infrared
regularization $\epsilon$ should be negative and $\sigma =
-\epsilon/2$ positive.

Unfortunately, this alternative one-cusp solution has the same
drawbacks as the previous one: it does not provide us with a
complete polygon solution. Indeed, one may try now to substitute
(\ref{exposol}) at $\epsilon \neq 0$ with ${\cal E}$ in place of all
$E$. However, such a substitution does not provide an exact solution
to equations (\ref{eqepsi}). Instead \be z\partial {\bf v} - {\bf
v}\partial z = \sum_{a,b}
k_b{\cal P}_{ab}{\cal E}_a{\cal E}_b', \ \ \ \ \  
\partial(z\partial {\bf v} - {\bf v}\partial z) = \sigma(1+\sigma)
\sum_{a,b} \frac{{\cal P}_{ab} \left((\vec k_a - \vec k_b)\vec
u\right) \Big(2-\sigma(\vec k_a+\vec k_b)\vec u\Big)} {(1-\sigma
\vec k_a\vec u)^{2+1/\sigma} (1-\sigma \vec k_b\vec u)^{2+1/\sigma}}
\ee Moreover, $L$ then turns into \be L =
\frac{\sum_{a,b}z_az_b({\cal E}_a''{\cal E}_b + {\cal E}_a{\cal
E}_b'') + \epsilon \vec k_a\vec k_b{\cal E}_a' {\cal E}_b'}
{(1+\frac{\epsilon}{2})\sum_{a,b} z_az_b {\cal E}_a{\cal E}_b} \ee
and is still independent of the external momenta ${\bf p}_a$.

\subsection{The $\sigma$-model action \label{inteva}}

We now return to the area integral (\ref{Sep1}) for the solution
(\ref{exposol}) with $n=4$ and then consider the analogous integral
for the deformation of the Nambu-Goto action a la \cite{AM}.

In order to calculate the regularized $\sigma$-model action
(\ref{Sep1}), one may use the formula \be
{\cal R}_\epsilon\equiv \frac{1}{(1+\epsilon/2)^{1+\epsilon/2}} \int \left(L +
\frac{\epsilon}{2}\frac{(\partial z)^2}{z^2}
\right) 
z^\epsilon d^2u = \frac{1}{(1+\epsilon/2)^{1+\epsilon/2}}
\left(2-\frac{1}{2(1-\epsilon)}\sum_{a,b} (\vec k_a\vec k_b)z_az_b
\frac{\partial^2 }{\partial z_a\partial z_b}\right) \int z^\epsilon
d^2u \label{SepS1mt} \ee derived in Appendix A. In \cite{AM} the
same quantity was parameterized by two functions $I_1$ and $I_2$ \be
{\cal R}_\epsilon \ =\ 2\int z^\epsilon (1+\epsilon I_1 + \epsilon^2
I_2 + \ldots)d^2u \ee According to (\ref{SepS1mt}), integrals with
$I_1$ and $I_2$ will be immediately known (note that they themselves
depend on $\epsilon$!), once one evaluates \be {\cal J}\{z_a\} = \int
z^\epsilon d^2u \ee This integral is calculated in Appendix A (the
calculation generalizes the calculation of \cite[Appendix
B]{AM} to arbitrary values of $z_a$). The answer is (see
Appendix A for details) \be
{\cal R}_\epsilon \ = \ {\cal K}_\epsilon \left\{1 +
\frac{\epsilon}{4}\log(z_1z_2z_3z_4) +
\frac{\epsilon^2}{8}\log(z_1z_3)\log(z_2z_4)\right\} \label{Sep3mt}
\ee with \be {\cal K}_\epsilon = \tilde K_\epsilon
\left(1+\frac{\epsilon}{2}+\frac{\epsilon^2}{2}\right)  =
\frac{8}{\epsilon^2|\sin\phi|}
\left(1+\epsilon^2\Big(\frac{1}{4}-\frac{\pi^2}{12}\Big)\right) \ee
where $\phi$ is the angle between the vectors $\vec k_1$ and $\vec
k_2$ (we consider here the general case of a rectangle).
Eq.(\ref{Sep3mt}) is our final formula for the regularized "minimal
area" for $n=4$ in the $\sigma$-model approach.

\bigskip

For the Alday-Maldacena choice $z_1=z_3=1-b$
and $z_2=z_4=1+b$ and $\phi=\frac{\pi}{2}$ one obtains from
(\ref{Sep3mt}):
\be
{\cal R}_\epsilon^{AM} \ =  \frac{1}{2}\,{\cal  K}_\epsilon
\left\{ (1-b)^\epsilon + (1+b)^\epsilon - \frac{\epsilon^2}{2}
\Big(\!\log\frac{1-b}{1+b}\,\Big)^2\right\} \ee According to
\cite{AM} this expression should be multiplied by \be
\frac{\sqrt{\lambda_Dc_D}}{2\pi a^\epsilon} = \frac{\sqrt{\lambda
\mu^{2\epsilon}} (2\pi)^\epsilon \sqrt{1+\epsilon}} {2\pi
a^\epsilon\sqrt{1-\frac{\pi^2\epsilon^2}{12}}} \ee to give \be {\rm
Area}_\epsilon =  2^{1+2\epsilon}\frac{\tilde {\cal
K}_\epsilon}{\pi\epsilon^2} \left\{\sqrt{\frac{\lambda
\mu^{2\epsilon}}{(-s)^\epsilon}} + \sqrt{\frac{\lambda
\mu^{2\epsilon}}{(-t)^\epsilon}} -
\frac{\epsilon^2}{8}\left(\log\frac{s}{t}\right)^2\right\} \ee where
\be \tilde {\cal K}_\epsilon = 1+ \frac{\epsilon}{2}(1-\log 2) +
\frac{\epsilon^2}{8} \left(1-\frac{\pi^2}{3} - 2\log 2 + (\log
2)^2\right) \ee Note that our normalization of $z$ contains an extra
factor of $4$ as compared to \cite{AM} (the sum of four exponents is
$4$ times a product of two cosines) and this contributes a factor
$4^\epsilon$ in the answer. An extra factor of $2$ is due to our choice of the 
constant $L=2$, instead of the $L=1$ of \cite{AM}.

Thus, (\ref{Sep3mt}) is in full agreement with \cite{AM}, but one
can use it equally well for other choices of $\{z_a\}$, and the
answer obviously depends on this choice. Even more striking, it
strongly depends on the angle $\phi$ between the vectors $\vec k_1$
and $\vec k_2$, which, along with $\{z_a\}$, are the moduli in the
solution space.

\subsection{The Alday-Maldacena solution as a minimum in the
moduli space\label{7}}

Since solutions with different $\{z_a\}$ are not equivalent, the
regularized action depends on these parameters. In such a case, one
should naturally look for a new extremum: the minimum of
(\ref{Sep3mt}) in the moduli space of solutions. Since the moduli
space is the hypersurface (\ref{equ4})
\be
z_1z_3 s + z_2z_4 t = 1
\label{constra}
\ee
in the space of $z_a$-variables, we should look
for a minimum of (\ref{Sep3mt}) with respect to $z_a$ under the
constraint (\ref{constra}). This gives
\be
\frac{1}{\epsilon
z_1}\Big(1 + \frac{\epsilon}{2}\log(z_2z_4)\Big) =
\lambda z_3 s, \nn \\
\frac{1}{\epsilon z_2}\Big(1 + \frac{\epsilon}{2}\log(z_1z_3)\Big) =
\lambda z_4 t, \nn \\
\frac{1}{\epsilon z_3}\Big(1 + \frac{\epsilon}{2}\log(z_2z_4)\Big) =
\lambda z_1 s, \nn \\
\frac{1}{\epsilon z_4}\Big(1 + \frac{\epsilon}{2}\log(z_1z_3)\Big) =
\lambda z_2 t,
\label{fineq}
\ee
where $\lambda$ is the Lagrange multiplier.
A possible solution of this system is:
\be
z_1 = z_3 = \frac{1}{\sqrt{2s}}\left(1-
\frac{\epsilon}{8}\log\frac{s}{t} + O(\epsilon^2)\right), \nn \\
z_2 = z_4 = \frac{1}{\sqrt{2t}}\left(1+
\frac{\epsilon}{8}\log\frac{s}{t} +O(\epsilon^2)\right)\ \
\label{finsol}
\ee
i.e. we obtain exactly the Alday-Maldacena choice
(\ref{cho22}) for $z_a$, with 
corrections of order $\epsilon$. The latter could lead to a different 
finite term in (\ref{Sep3mt}), however they exactly cancel each
other, since as one may check
\be
\frac{\epsilon}{4}\log (z_1z_2z_3z_4) =
-\frac{\epsilon}{4}\log(4st) + O(\epsilon^3)
\label{corref}
\ee

Of course, (\ref{finsol}) is not the generic solution of
(\ref{fineq}). In fact, from (\ref{Sep3mt}) it follows that the
$\epsilon$-regularization did not fully break the non-Lorentz
part of $SO(4,2)$: the answer depends on the products $z_1z_3$ and
$z_2z_4$ and is invariant under rescalings $z_1 \rightarrow \alpha
z_1$, $z_3 \rightarrow z_3/\alpha$ and $z_2 \rightarrow \beta z_1$,
$z_4 \rightarrow z_4/\beta$. However, exactly because (\ref{Sep3mt})
depends only on the two products, the other solutions, obtained by
such rescalings, do not affect the vanishing of the correction in
(\ref{corref}).

The regularized action (\ref{Sep3mt}) depends also on the
angle $\phi$ through a factor $|\sin\phi|^{-1}$ in
${\cal K}_\epsilon$,
and minimum is obviously located at $\phi=\pi/2$.

\subsection{The Nambu-Goto action}

Nambu-Goto (NG) action seems a more difficult issue to address in
the approach of \cite{AM}, though it is much better from geometrical
and, perhaps, even conceptual points of view \cite{DGO}.

First of all, as mentioned in Section \ref{NGA}, the NG Lagrangian is not
invariant under $T$-duality. This means that one needs to borrow the
beautiful formulation of the minimal area problem with boundary formed
by external momenta from the $\sigma$-model formalism and then use
it as a starting point for NG calculations. In principle, this is
not a drawback, especially if one believes that the origin of the
AdS/CFT (string/gauge) duality is in Polyakov's formalism with the
Liouville field playing the role of the $5-th$ dimension: then the
$\sigma$-model-like formulation is the starting point in any case
and the NG-like formulation in terms of minimal areas is a derivable
concept.

Second, not all the  $\sigma$-model solutions (\ref{exposol})
necessarily satisfy the Nambu-Goto equations of motion
(Alday-Maldacena solution does). Description of moduli space of
Nambu-Goto solutions is an open problem even for $n=4$.

Third, if this moduli space is also large, as in the $\sigma$-model
case, further calculations can be more difficult.  For a given
solution one can not immediately get rid of the $y$-fields when the
action is evaluated with the help of the Alday-Maldacena
prescription (to substitute $r$ by $r\sqrt{1+\epsilon/2}$), as we
did in (\ref{Sep1}): some direct calculation involving all the
fields $z$ and $v$ should be performed. It is an intriguing
question, if the result will be the same as in the $\sigma$-model
case, and -- even if not -- if the Alday-Maldacena solution is still
a minimum in the moduli space.

\section{Conclusion and prospects}
\setcounter{equation}{0}

To summarize, the seemingly {\it ad hoc}
choice of $z_a$-parameter values made in \cite{AM} indeed corresponds
to a minimum of the regularized area in the moduli space of
all possible solutions of the AdS $\sigma$-model with given
boundary conditions and the one-to-one correspondence between
minimal surfaces and boundary conditions is partly restored after
regularization. Of course, whether or not the minimal surface is regularization
independent remains an open question.
In general, in order to find the minimum one needs to know the area as
a function on the moduli space, i.e. analysis of some particular
solution is not sufficient. In practice, however, it appeared sufficient
to restrict consideration to the $SO(4,2)$-orbit of a particular
solution, despite the fact that this symmetry does not act transitively
on the entire moduli space (for instance, it does not change the angle $\phi$
between the $\vec k$-vectors). Since the $SO(4,2)$ symmetry is
broken by regularization, the $z_a$-dependence of the regularized
action is not automatically given by symmetry arguments
and should be determined by straightforward calculation.
This implies that to address the
$n>4$ problem one will need to construct the full family
of solutions and evaluate their regularized action.

\bigskip

Modulo the above comments, the reasoning of \cite{AM}
reduces in the framework of the AdS/CFT correspondence
the problem of $n$-point amplitudes of $N=4$ SYM at strong coupling
to a couple of well-defined problems in the field of integrable
systems:

1) Find solutions of the $2d$ integrable $SO(4,2)$ sigma-model
allowing for growing asymptotics on the world sheet, and

2) Find their Whitham deformations, induced by an
$\epsilon$-regularization, which breaks the integrability of
the $\sigma$-model.

The regularized minimal area is then defined by a minimum of
some still-to-be-determined function on the moduli space of solutions.

\bigskip

Using the results of \cite{BHT} one may write the BDS conjectured formula, given in the introduction,
in terms of the regularized Polyakov's \cite{Pol1} double contour integral (\ref{loopin})
over the auxiliary polygon $\Pi$ in momentum space,
formed by the momenta of the scattering process under study, i.e.
\be
{\rm Amplitude\ in\ perturbative}\ N=4\ {\rm SYM}\ \  \sim \ \
\exp \left(\frac{\gamma(\lambda)}{4} \oint_\Pi\oint_\Pi
\frac{dy^\mu dy_\mu'}{(y-y')^{2+\epsilon}}\right),
\ee
Then the AdS/CFT duality in the sector of $n$-point
amplitudes, may equivalently be stated in purely {\it geometrical} terms, namely:
why does this integral coincide with the area
of the minimal surface $\Sigma$ defined by above Whitham-deformed
$\sigma$-model solutions:
\be
\oint_\Pi\oint_\Pi \frac{dy^\mu dy_\mu'}{(y-y')^{2+\epsilon}} =
{\rm Area}_\epsilon(\Sigma), \ \ \ \ \  \ \Pi = \partial\Sigma
\label{duali}
\ee
This was explicitly verified in \cite{AM} and also here for $n=4$.
The exact relation between  integrable structures on the two sides of this
formula \cite{BA,Zar,Gor} remains to be understood.
We emphasize that the result of \cite{AM} is entirely based
on the {\it deviations} from ordinary integrability. It suggests the
crucial role of a very different -- {\it Whitham} --
integrability \cite{Whith,RG},
which did not yet attract much attention in
the studies of  the AdS/CFT correspondence.

\section{Appendix: Evaluation of the regularized area}
\def\theequation{A.\arabic{equation}}
\setcounter{equation}{0}

In this Appendix, we explain how to calculate the regularized area
in the $\sigma$-model case (\ref{Sep1}). The calculation is a
straightforward extension of the calculation in \cite{AM} to the case of
generic $z_a$.

For $z = \sum_a z_ae^{\vec k_a\vec u}$ we have: \be
\left(\frac{\partial }{\partial \vec u} - \sum_a \vec k_a
z_a\frac{\partial }{\partial z_a}\right)z = 0 \ \ \ \Longrightarrow
\ \ \ \ \frac{(\partial z)^2}{z^2} = \sum_{a,b} (\vec k_a\vec
k_b)\frac{z_az_b}{z^2} \frac{\partial z}{\partial z_a}\frac{\partial
z}{\partial z_b}
\ee
Next, from 
\be \frac{\partial^2z}{\partial z_a\partial z_b} = 0 \ \ \
\Longrightarrow \ \ \ \frac{\partial^2 z^\epsilon}{\partial
z_a\partial z_b} = \epsilon(\epsilon-1) \frac{z^\epsilon}{z^2}
\frac{\partial z}{\partial z_a}\frac{\partial z}{\partial z_b} \ee
so that \be \frac{(\partial z)^2}{z^2}\, z^\epsilon =
-\frac{1}{\epsilon(1-\epsilon)}\sum_{a,b} (\vec k_a\vec k_b)z_az_b
\frac{\partial^2 z^\epsilon}{\partial z_a\partial z_b}\ 
\ee Thus substituting (\ref{exposol}) into (\ref{Sep1}) we obtain:
\be {\cal R}_\epsilon \equiv \frac{1}{(1+\epsilon/2)^{1+\epsilon/2}} \int \left(L +
\frac{\epsilon}{2}\frac{(\partial z)^2}{z^2}
\right) 
z^\epsilon d^2u = \frac{1}{(1+\epsilon/2)^{1+\epsilon/2}}
\left(2-\frac{1}{2(1-\epsilon)}\sum_{a,b} (\vec k_a\vec k_b)z_az_b
\frac{\partial^2 }{\partial z_a\partial z_b}\right) \int z^\epsilon
d^2u \label{SepS1} \ee In \cite{AM} the same quantity was
parameterized by two functions $I_1$ and $I_2$: \be {\cal R}_\epsilon \
=\ 2\int z^\epsilon (1+\epsilon I_1 + \epsilon^2 I_2 + \ldots)d^2u
\ee and looking at the l.h.s. of (\ref{SepS1}) with $L=2$ and
$(1+\epsilon/2)^{-(1+\epsilon/2)} = 1- \epsilon/2 + O(\epsilon^3)$,
one immediately sees that \be 2I_2 + I_1 = -\frac{1}{2} \ee
According to (\ref{SepS1}) integrals with $I_1$ and $I_2$ will be
immediately known (note that they themselves depend on $\epsilon$!),
if one evaluates \be {\cal J}_n\{z_a\} = \int z^\epsilon d^2u \ee If
$n=4$ and $\vec k_3=-\vec k_1$ and $\vec k_4=-\vec k_2$, as is the
case for our solutions, then we can use $\tilde u_1 = \vec k_1 \vec u$
and $\tilde u_2 = \vec k_2\vec u$ as new coordinates on
the world sheet, and they can be further shifted by
$\frac{1}{2}\log(z_1/z_3)$ and $\frac{1}{2}\log(z/2/z_4)$
respectively in order to give: \be {\cal J}_4\{z_a\} =
\frac{2^\epsilon}{|\vec k_1\times\vec k_2|} \int
\Big(\sqrt{z_1z_3}\cosh \tilde u_1 + \sqrt{z_2z_4}\cosh\tilde u_2
\Big)^\epsilon
d^2\tilde u = \nn \\
= \frac{2^{1+\epsilon}}{|\vec k_1\times\vec k_2|}
\int \left\{\Big(\sqrt{z_1z_3}+\sqrt{z_2z_4}\,\Big)
\cosh \hat u_1\cosh \hat u_2
+\Big(\sqrt{z_1z_3}-\sqrt{z_2z_4}\,\Big) \sinh \hat u_1\sinh \hat
u_2 \right\}^\epsilon d^2\hat u \ee In the second line the variables
are rotated once again, $\hat u_1 = \frac{\tilde u_1+\tilde
u_2}{2}$, $\hat u_2 = \frac{\tilde u_1-\tilde u_2}{2}$, and the
resulting integral can be evaluated, say, by expanding in powers of
the second item, as was done in \cite{AM}: \be \int \Big(A\cosh \hat
u_1\cosh \hat u_2 + B \sinh \hat u_1\sinh \hat u_2 \Big)^\epsilon
d^2\hat u = \frac{A^\epsilon}{\Gamma(-\epsilon)} \sum_{k=0}^\infty
\frac{\Gamma(2k-\epsilon)}{(2k)!} \left(-\frac{B}{A}\right)^{2k}
\left(\int \tanh^{2k}\! \hat u \cosh^\epsilon\! \hat u \,d\hat
u\right)^2
\label{ABint} \ee The last integral converges for $\epsilon < 0$ and
is given by a $B$-function formula with $\xi = \tanh\hat u$: \be
\int \tanh^{2k}\! \hat u \cosh^\epsilon\! \hat u\, d\hat u =
\int_0^1 (\xi^2)^{k-1/2}(1-\xi^2)^{-1-\epsilon/2} d(\xi^2) =
\frac{\Gamma\left(k+\frac{1}{2}\right)
\Gamma\left(-\frac{\epsilon}{2}\right)}
{\Gamma\left(k+\frac{1-\epsilon}{2}\right)} \label{Bfu} \ee It
remains to substitute (\ref{Bfu}) into (\ref{ABint}) and make use of
the doubling formula \be \Gamma(2k-\epsilon) =
\frac{2^{2k-\epsilon-1}}{\sqrt{\pi}}
\Gamma\left(k-\frac{\epsilon}{2}\right)
\Gamma\left(k+\frac{1-\epsilon}{2}\right) \ee to obtain \cite{AM}:
\be (\ref{ABint})\ = \
\frac{A^\epsilon\Gamma^2(-\frac{\epsilon}{2})}
{2^{1+\epsilon}\Gamma(-\epsilon)}\sum_{k=0}^\infty
\frac{\Gamma\left(k+\frac{1}{2}\right)
\Gamma\left(k-\frac{\epsilon}{2}\right)} {k!\ \Gamma\left(k +
\frac{1-\epsilon}{2}\right)} \left(-\frac{B}{A}\right)^{2k} =
\frac{\pi A^\epsilon\Gamma^2\big(-\frac{\epsilon}{2}\big)}
{\Gamma^2\big(\frac{1-\epsilon}{2}\big)}
\phantom{.}_2F_1\left(\frac{1}{2},-\frac{\epsilon}{2};
\frac{1-\epsilon}{2}; \,\left(\frac{B}{A}\right)^2\,\right)
\label{JFint} \ee At the last step one applies doubling to
$\Gamma(-\epsilon)$ and includes a factor
$\frac{\Gamma(\frac{1}{2})\Gamma(-\frac{\epsilon}{2})}
{\Gamma(\frac{1-\epsilon}{2})}$ from the definition of the
$F$-function.

Now we return to (\ref{SepS1}) and substitute the
evaluated expression (\ref{JFint}) for ${\cal J}_4$. The differential
equation for the hypergeometric function $\phantom{.}_2F_1$ can be
used to evaluate the derivatives. Alternatively, one can expand
(\ref{JFint}) and keep the first relevant powers of $\epsilon$
already at this stage, as was done in \cite{AM} with the help of the
asymptotic formula \be
\phantom{.}_2F_1\left(\frac{1}{2},-\frac{\epsilon}{2};
\frac{1-\epsilon}{2}; \,C\,\right) = 1 + \frac{\epsilon}{2}\log(1-C)
+ \frac{\epsilon^2}{2} \log(1-\sqrt{C})\log(1+\sqrt{C}) +
O(\epsilon^3) \ee so that \be
(\ref{JFint}) \ = \ K_\epsilon
\left(1+\frac{\epsilon}{2} \log(A^2-B^2) +
\frac{\epsilon^2}{2}\log(A-B)\log(A+B)
+ O(\epsilon^3)\right) = \nn \\
= K_\epsilon
\left(1+
\frac{\epsilon}{4}\log(16z_1z_2z_3z_4)
+ \frac{\epsilon^2}{8}\log(4z_1z_3)\log(4z_2z_4)
+ O(\epsilon^3)\right) = \nn \\
= 2^\epsilon K_\epsilon
\left(1+
\frac{\epsilon}{4}\log(z_1z_2z_3z_4)
+ \frac{\epsilon^2}{8}\log(z_1z_3)\log(z_2z_4)
+ O(\epsilon^3)\right) = \nn \\
= 2^\epsilon K_\epsilon\left\{1+
\frac{\epsilon}{4}\log(z_1z_2z_3z_4)
+ \frac{\epsilon^2}{32}\Big(\log(z_1z_2z_3z_4)\Big)^2 -
\frac{\epsilon^2}{32}\Big(\log\frac{z_1z_3}{z_2z_4}\Big)^2
+ O(\epsilon^3)\right\} \ee where \be K_\epsilon = \frac{\pi
\Gamma^2\big(-\frac{\epsilon}{2}\big)}
{\Gamma^2\big(\frac{1-\epsilon}{2}\big)} =
\frac{4^{1-\epsilon}}{\epsilon^2}\left(
\frac{\Gamma^2(1-\frac{\epsilon}{2})}{\Gamma(1-\epsilon)}\right)^2
=\frac{4^{1-\epsilon}}{\epsilon^2} \left(1 +
\frac{\epsilon^2}{2}\Big[\big(\Gamma'(1)\big)^2 - \Gamma''(1)\Big] +
O(\epsilon^3)\right) =
\frac{4^{1-\epsilon}(1-\frac{\pi^2\epsilon^2}{12})}{\epsilon^2} \ee
because $\Gamma(1+z) = 1-\gamma z +\left(\frac{\pi^2}{12}+
\frac{\gamma^2}{2}\right)z^2 + \ldots$

If the angle between vectors $\vec k_1$ and $\vec k_2$ is $\phi$,
then we obtain, up to the terms of order $O(\epsilon)$ \be {\cal R}_\epsilon \
= \ \tilde K_\epsilon \left(1-\frac{1}{4(1-\epsilon)}\sum_{a,b}
(\vec k_a\vec k_b)z_az_b \frac{\partial^2 }{\partial z_a\partial
z_b}\right) \left(1+
\frac{\epsilon}{4}\log(z_1z_2z_3z_4)
+ \frac{\epsilon^2}{8}\log(z_1z_3)\log(z_2z_4) \right) \label{Sep2}
\ee with \be \tilde K_\epsilon =
\left(1-\frac{\epsilon}{2}\right)\frac{2^{1+\epsilon}}{2|\sin\phi|}
\cdot 2^\epsilon K_\epsilon\cdot 2 =
\frac{8\left(1-\frac{\epsilon}{2}-\frac{\pi^2\epsilon^2}{12}\right)}
{\epsilon^2|\sin\phi|} \ee where the last factor $2$ comes from $L=2$.

The action of the differential operator on the first logarithmic
term is simple, since this logarithm is just a sum $\sum_a \log
z_a$, only four terms with $a=b$ and $\vec k_a^2=2$ contribute: \be
\sum_{a,b} (\vec k_a\vec k_b)z_az_b \frac{\partial^2 }{\partial
z_a\partial z_b}\log(z_1z_2z_3z_4) =-2\cdot(1+1+1+1) = -8 \ee The
action on the second logarithmic term is more interesting. For the
same reason $\partial^2/\partial z_1\partial z_3$ and
$\partial^2/\partial z_2\partial z_4$ annihilate this term, while
the four operators $\partial^2/\partial z_a\partial z_{a+1}$ are
multiplied by $(\vec k_a\vec k_{a+1}) = -2(-)^a\cos\phi$ so that
they do not contribute at all when $\vec k_a$ are directed along
diagonals of a {\it square} and cancel among each other even if
$\phi\neq \pi/2$. Keeping this in mind we obtain: \be \sum_{a,b}
(\vec k_a\vec k_b)z_az_b \frac{\partial^2 }{\partial z_a\partial
z_b} \log(z_1z_3)\log(z_2z_4) = 2(-1-1)\Big(\log(z_2z_4) +
\log(z_1z_3)\Big)
+ 4\cos\phi\Big(1-1+1-1\Big) =\nn \\
= -4\log(z_1z_2z_3z_4) \ee Therefore (\ref{Sep2}) becomes \be
{\cal R}_\epsilon \ = \ \tilde K_\epsilon \left\{\left(1+
\frac{\epsilon}{4}\log(z_1z_2z_3z_4)
+ \frac{\epsilon^2}{8}\log(z_1z_3)\log(z_2z_4) \right) +
\frac{8\epsilon}{16(1-\epsilon)} +
\frac{4\epsilon^2}{32}\log(z_1z_2z_3z_4)\right\} = \nn \\
= {\cal K}_\epsilon \left\{1 + \frac{\epsilon}{4}\log(z_1z_2z_3z_4)
+ \frac{\epsilon^2}{8}\log(z_1z_3)\log(z_2z_4)\right\} \label{Sep3}
\ee with \be {\cal K}_\epsilon = \tilde K_\epsilon
\left(1+\frac{\epsilon}{2}+\frac{\epsilon^2}{2}\right) =
\frac{8\left(1-\frac{\epsilon}{2}-\frac{\pi^2\epsilon^2}{12}\right)}
{\epsilon^2\sin\phi}
\left(1+\frac{\epsilon}{2}+\frac{\epsilon^2}{2}\right) =
\frac{8}{\epsilon^2|\sin\phi|}
\left(1+\epsilon^2\Big(\frac{1}{4}-\frac{\pi^2}{12}\Big)\right) \ee
Equation (\ref{Sep3}) is our final formula for the regularized "minimal area"
for $n=4$ in the $\sigma$-model approach.

\bigskip

For the Alday-Maldacena choice of $z_a$ variables $z_1=z_3=1-b$ and
$z_2=z_4=1+b$ and $\phi=\frac{\pi}{2}$
we obtain from (\ref{Sep3}):
$$
{\cal R}_\epsilon \ = \ {\cal  K}_\epsilon
\left\{1+\frac{\epsilon}{2}\log(1-b^2) +
\frac{\epsilon^2}{4}\left[\Big(\log(1-b)\Big)^2
+\Big(\log(1+b)\Big)^2 - \Big(\!\log\frac{1-b}{1+b}\,\Big)^2\right]
\right\} =
%
$$ \vspace{-0.3cm}
\be = \frac{1}{2}\,{\cal  K}_\epsilon
\left\{ (1-b)^\epsilon + (1+b)^\epsilon - \frac{\epsilon^2}{2}
\Big(\!\log\frac{1-b}{1+b}\,\Big)^2\right\} \ee According to
\cite{AM} this expression should be multiplied by \be
\frac{\sqrt{\lambda_Dc_D}}{2\pi a^\epsilon} = \frac{\sqrt{\lambda
\mu^{2\epsilon}} (2\pi)^\epsilon \sqrt{1+\epsilon}} {2\pi
a^\epsilon\sqrt{1-\frac{\pi^2\epsilon^2}{12}}} \ee to give: \be {\rm
Area}_\epsilon = 2^{1+2\epsilon} \frac{\tilde {\cal
K}_\epsilon}{\pi\epsilon^2}
 \sqrt{\lambda \mu^{2\epsilon}}
\left\{\left[\frac{2\pi(1-b)}{2^{3/2}\,a}\right]^\epsilon +
\left[\frac{2\pi(1+b)}{2^{3/2}\,a}\right]^\epsilon -
\frac{\epsilon^2}{2} \Big(\!\log\frac{1-b}{1+b}\,\Big)^2\right\} =
\nn \\ = 2^{1+2\epsilon}\frac{\tilde {\cal
K}_\epsilon}{\pi\epsilon^2} \left\{\sqrt{\frac{\lambda
\mu^{2\epsilon}}{(-s)^\epsilon}} + \sqrt{\frac{\lambda
\mu^{2\epsilon}}{(-t)^\epsilon}} -
\frac{\epsilon^2}{8}\left(\log\frac{s}{t}\right)^2\right\} \ee where
\be \tilde {\cal K}_\epsilon = 2^{-\epsilon/2}
\left(1+\epsilon^2\Big(\frac{1}{4}-\frac{\pi^2}{12}\Big)\right)
\sqrt{\frac{1+\epsilon}{1-\frac{\pi^2\epsilon^2}{12}}} = 1+
\frac{\epsilon}{2}(1-\log 2) + \frac{\epsilon^2}{8}
\left(1-\frac{\pi^2}{3} - 2\log 2 + (\log 2)^2\right) \ee

\section*{Acknowledgements}

We are indebted to T.Mironova for help with the Figures.
Work is supported in part by the INTERREG-IIIA Greece-Cyprus
program, as well as by the European Union contract MRTN-CT-512194.
A.Mironov and A.Morozov acknowledge the kind hospitality and support
of the Institute of Plasma Physics and the Department of Physics of
the University of Crete, where the final part of this work was done.
Also, their work is partly supported by the Russian Federal Nuclear
Energy Agency, by the joint grant 06-01-92059-CE,  by NWO project
047.011.2004.026, by INTAS grant 05-1000008-7865, by
ANR-05-BLAN-0029-01 project and by the Russian President's Grant of
Support for the Scientific Schools NSh-8004.2006.2, by RFBR grants
07-02-00878 (A.Mir.) and 07-02-00645 (A.Mor.).

\end{document}